\documentclass[11pt,a4paper]{article}

\usepackage{mathtools}
\usepackage{authblk} 
\usepackage{algpseudocode,algorithmicx,algorithm}

\usepackage{mathrsfs}
\usepackage{latexsym,bm}
\usepackage{amsmath,amsfonts,amsmath,amssymb,amsthm}
\usepackage{extarrows}

\usepackage{graphicx,subfigure,epstopdf,float}
\usepackage{enumerate,cases,multirow}
\usepackage{makecell}
\usepackage{caption}

\usepackage{longtable,colortbl,arydshln,threeparttable}
\definecolor{mygray}{gray}{.9}

\usepackage{indentfirst}
\usepackage[top=25mm,bottom=20mm,left=25mm,right=20mm]{geometry}
\usepackage{cite}

\usepackage{listings}

\usepackage{makeidx}        
\usepackage{booktabs}
\usepackage[bookmarksnumbered,colorlinks,citecolor=red,linkcolor=red,hyperindex,linktocpage=true]{hyperref}

\newcommand{\ket}[1]{| #1 \rangle} 
\newcommand{\bra}[1]{\langle #1 |} 

\newcommand{\bb}{\boldsymbol}
\newcommand{\vect}{\boldsymbol}
\newcommand{\littleo}{{\scriptstyle \mathcal{O}}}
\def \d {\mathrm{d}}
\def \e {\mathrm{e}}
\def \i {\mathrm{i}}
\def \D {\Delta}

\newcounter{parentalgorithm}

\makeatother

\newtheorem{theorem}{Theorem}[section]
\newtheorem{lemma}{Lemma}[section]

\newtheorem{example}{\bf Example}[section]
\theoremstyle{remark}
\newtheorem{remark}{\bf Remark}[section]

\numberwithin{equation}{section}

\begin{document}

\title{Time complexity analysis of quantum algorithms via linear representations for nonlinear ordinary and partial differential equations}
\author[1,2,3]{Shi Jin \thanks{shijin-m@sjtu.edu.cn}}
\author[2,3,4]{Nana Liu \thanks{nana.liu@quantumlah.org}}
\author[1]{Yue Yu \footnote{Corresponding author.} \thanks{terenceyuyue@sjtu.edu.cn}}
\affil[1]{School of Mathematical Sciences, Institute of Natural Sciences, MOE-LSC, Shanghai Jiao Tong University, Shanghai, 200240, P. R. China.}
\affil[2]{Institute of Natural Sciences, Shanghai Jiao Tong University, Shanghai 200240, China.}
\affil[3]{Ministry of Education, Key Laboratory in Scientific and Engineering Computing, Shanghai Jiao Tong University,
Shanghai 200240, China}
\affil[4]{University of Michigan-Shanghai Jiao Tong University Joint Institute, Shanghai 200240, China}

\maketitle

\begin{abstract}
  We construct quantum algorithms to compute the solution and/or physical observables of nonlinear ordinary differential equations (ODEs) and nonlinear Hamilton-Jacobi equations (HJE) via linear representations or exact mappings between nonlinear ODEs/HJE and linear partial differential equations (the Liouville equation and the Koopman-von Neumann equation). The connection between the linear representations and the original nonlinear system is established through the Dirac delta function or the level set mechanism. We compare the quantum linear systems algorithms based methods and the quantum simulation methods arising from different numerical approximations, including the finite difference discretisations and the Fourier spectral discretisations for the two different linear representations, with the result showing that the quantum simulation methods usually give the best performance in  time complexity. We also propose the Schr\"odinger framework to solve the Liouville equation for the HJE with the Hamiltonian formulation of classical mechanics, since it can be recast as the semiclassical limit of the Wigner transform of the Schr\"odinger equation. Comparsion between the Schr\"odinger and the Liouville framework will also be made.
\end{abstract}



\tableofcontents

\section{Introduction}

Some of the most important problems in physics, chemistry, engineering, biology and finance are modelled by nonlinear ordinary and partial differential equations (ODEs and PDEs).
Prominent examples include  climate modelling, aircraft design, molecular dynamics and drug design, deep learning neural networks and mean-field games in mathematical finance. Among the most important examples of  nonlinear ODEs include Newton's equations in molecular dynamics, while examples of nonlinear PDEs include the Euler and Navier-Stokes equations in fluid dynamics, the Boltzmann equations in rarified gas, and the Hamilton-Jacobi equations in geometric optics, front propagation, mean-field games and optimal control. In spite of tremendous progresses in developing classical algorithms for solving these equations, there remain major challenges that are difficult to be handled by classical algorithms, for examples the curse-of-dimensionality, multiple scales, strong nonlinearity, and large numbers of degree of freedoms. On the other hand, quantum
algorithms, due to their potential polynomial and even exponential advantages, could  be the game changer to deal with some of these difficulties.  However, since quantum algorithms are based on the principle of quantum mechanics, which is fundamentally linear (as far as we know), so far the development  of quantum algorithms have been mostly confined to linear problems, with the most notable in  linear algebra  \cite{Nielsen2010,Lipton2010,LiuN2016onequmode,LiuN2019clinet,Deutsch1992rapid,
Shor1997prime,HHL2009,Ambainis2012VTAA,Childs2017QLSA,Costa2021QLSA,Subasi2019AQC,Gilyen2019QSVD,Lin2022Notes, fang2022time}. For linear ODEs and PDEs, once they are numerically discretised, they became linear algebra problems which can then be handled by quantum linear algebra solvers   (e.g. \cite{Clader2013preconditioned,Childs2021high,Costa2019Wave,Linden2020heat,Engel2019qVlasov,Cao2013Poisson,DLW2022forwarding}).

Since most natural phenomena are nonlinear,
the ability of quantum computing to solve nonlinear problems will
significantly extend the horizon of quantum computing.
The most natural idea of handling nonlinear problems by a quantum computer is to represent the nonlinear problem in a linear way, so  quantum algorithms for linear problems can be used.  There are two approaches in recent literatures. One is to {\it approximate} the nonlinear problem  through linearisation of the nonlinearity, or truncation of the equation, which is referred to as  {\it linear approximation methods}.  The second is coined   as {\it linear representation methods} \cite{JinLiu2022nonlinear}, which tries to find a map from the linear to the nonlinear systems, and usually yields a system in the phase space which is {\it equivalent} (without any approximation) to the original system.

In the linear approximation approach, in which one {\it linearises} the nonlinear system, modeling errors are inevitable, hence the approach may only be valid for a short time,  weak or special nonlinearities, and consequently after a long time solutions  may lose significant nonlinear features.  A more appealing approach uses the Carlemann linearisation in \cite{Liu2021nonlinear} or the techniques in \cite{Lloyd2020nonlinear, Leyton2008nonlinear}, but they
are restricted to polynomial nonlinearities, and need to truncate the system since only finite number of equations is used. Such a truncation may lead to loss of some important nonlinear features of the original system.
It  is similar to moment closure technique in  kinetic theory, in which one attempts to close the moment system with finite number of moments but often ends up with a closed system that has mathematical stability problems \cite{Grad1963asymptotic, Bobylev1982chapman} or physical realizibility issues \cite{Levermore1996moment}. This is similarly true for methods in \cite{Lloyd2020nonlinear, Leyton2008nonlinear}, see \cite{Lin2022KvN}.

For the linear representation methods, there are two current approaches. One is
the Koopman-von Neumann (KvN) approach \cite{Joseph2020KvN} (see also \cite{Dobin2021plasma}), which was introduced for nonlinear ODEs. The other is the level set method, first introduced in \cite{JinLiu2022nonlinear}, applicable to both nonlinear ODEs, and nonlinear PDEs -- more specifically the Hamilton-Jacobi equations and scalar nonlinear hyperbolic equations. Both approaches introduce equations in the phase space but the extra dimensionality that is difficult for classical algorithms can be significantly eased by quantum algorithms, hence quantum advantages in most numerical parameters can still be achieved, even including the measurements of physical observables as analysed in \cite{JinLiu2022nonlinear}.

The goal of this paper is to compare the two linear representation methods~--~the Liouville equation in the level set approach and the KvN equation, their variants and their different numerical approximations.  The KvN is, so to  speak, the square of the Liouville equation, and, when the solution is smooth, they are equivalent if the force is divergence free (for example in the case of Hamilton system).  While both equations are of linear transport nature, which can be solved similarly by classical algorithms, the KvN equation, due to its unitary structure, can be directly solved  by quantum simulations, while the Liouville equation  was solved by quantum linear system solvers in \cite{JinLiu2022nonlinear}. Despite the absence of the unitary structure, we still proposed a ``quantum simulation'' algorithm for the Liouville representation in Appendix~\ref{subsect:spectralLiouvillerep} by using the dimensional splitting Trotter based approximation. The basic idea is to transform the asymmetric evolution in each direction into a symmetric one, which requires only a simple variable substitution with the transformation matrix being diagonal. However, different from the traditional time-marching Hamiltonian simulation, non-unitary procedures for the variable substitution are involved, which may lead to exponential increase of the cost arising from multiple copies of initial quantum states at every time step as pointed out in Remark~\ref{rem:multiplecopies}.

The connection between the Liouville or KvN equation with the original nonlinear system can be made through the Dirac delta-function $\delta (x)$, which is naturally defined in the weak sense for the Liouville equation which solves for the probability distribution of particles. However, to connect the KvN model -- which computes, so to speak, the square root of the probability density distribution, one needs to use $\sqrt{\delta}$, which is not well-defined  mathematically, even in the weak sense, so one needs to be more careful in interpreting its solution (as will be discussed in Section \ref{sec: KvN}) and the consequent numerical  convergence in a suitable solution space. In addition, since the KvN is not in conservation form,
it also requires higher smoothness for the force field than the Liouville approach.

Another alternative approach to the Liouville equation (in the classical Hamiltonian formalism) is to use the linear Schr\"odinger equation, which approximates the Liouville equation in the classical limit by sending the ``Planck constant'' to zero. Here the Planck constant is an artificial small parameters, which can be chosen to depend on $\varepsilon$, the numerical precision. The Schr\"odinger equation can be solved by quantum simulation techniques \cite{Jin2022quantumSchrodinger,Nielsen2010}, or by quantum linear algebraic solvers after spatial and temporal discretisations.  We will compare all these different models and their different approximations: spectral methods vs. finite difference methods; and  quantum simulation vs. quantum linear algebra solvers, and identify the pros and cons of each of these solvers in order to identify the best possible linear representation method.

Our results on time complexities for the computation of physical observables are summarised in Tab.~\ref{tab:summaryObs}.
For nonlinear ODEs, the quantum simulation method has less computational cost than the quantum linear systems algorithm (QLSA) based approaches for both Liouville and KvN representations, and the cost of computing the physical observables for the Liouville representation has a multiplicative factor squared times as larger as the one for the KvN representation (if the number of copies needed of the initial state is neglected).

For nonlinear Hamilton-Jacobi equations (in the classical Hamiltonian formalism), the Schr\"odinger framework/representation, using quantum simulation, has advantages over other approaches in time complexity, for both $d$ (the space dimension) and $\varepsilon$. However, we would like to point out that the solution to the Schr\"odinger equation has oscillations of frequency of $\mathcal{O}(1/\sqrt{\varepsilon})$, as shown in Fig. \ref{fig:Pden}.  If one wants high resolution results~--~for example oscillations free, then the Liouville representation with spectral approximation and QLSA has edges on $d$, and $\varepsilon$ for smooth solution, while the Liouville replantation with finite difference approximation and QLSA has edges in $\varepsilon$ if solution is less smooth (in Sobolev space $H^l$ for $l \le 4$).

For general scalar hyperbolic equation, one can still use the Liouville representation \cite{JinLiu2022nonlinear} but the Schr\"odinger representation is not available. It is also unclear how to devise the KvN approach, as will be discussed in Section \ref{Sec: Hyp}.

We also note that we do not compare these quantum alrogithms with the corresponding classical algorithms, which were partly made, for example, in \cite{JinLiu2022nonlinear}. In particular, for nonlinear ODEs, these algorithms could be more expensive than the classical solvers \cite{JinLiu2022nonlinear}.


It should be pointed out that the optimal QLSA with query complexity $Q = \mathcal{O} ( s \kappa \log(1/\varepsilon) )$ is presented in \cite{Costa2021QLSA}, where $s$ is the sparsity, meaning it has at most $s$ nonzero entries per row and column, $\kappa$ is the condition number of the coefficient matrix. On the other hand, the gate complexity may be quantified by $\mathcal{O}( Q \text{poly} (\log Q, \log N) )$, which is larger than the query complexity only by logarithmic factors \cite{Childs2017QLSA,Lin-Tong-2020,Costa2021QLSA}. For simplicity, we use $\widetilde{\mathcal{O}}(Q)$ to denote the case where all logarithmic factors are suppressed.

The outline of the paper is as follows. In Section \ref{sect:reviewNonlinearODEs} we give a more detailed review of the several techniques for solving the nonlinear differential equations. In Section \ref{sect:linearrep} we summarize the general framework for computing the nonlinear ODEs by using the linear representation approaches, and analyse in detail the time complexity of the QLSA based methods and the quantum simulation methods. We propose a simple algorithm to compute the observables of integral form and figure out the multiplicative factor for the sampling procedure.
In Section \ref{sect:semiclassical} we propose and analyse several quantum algorithms for solving the nonlinear Hamilton-Jacobi PDEs. By using the level set mechanism proposed in \cite{JinLiu2022nonlinear}, we map the nonlinear PDEs of $(d+1)$-dimension to a linear $(2d+1)$-dimensional Liouville equation, referred to as the Liouville representation for nonlinear PDEs, and repeat the analysis of the quantum algorithms for the associated Liouville equation as in the previous section. In Section \ref{sect:Schrodinger}, we propose a Schr\"odinger framework to solve the nonlinear Hamilton-Jacobi PDEs (in the classical Hamiltonian formalism) since the Schr\"odinger equation can be transformed into the quantum Liouville equation via the Wigner transform, which in turn leads to the Liouville equation when taking the semiclassical limit. We present the quantum interpretation of the classical time-splitting Fourier spectral method proposed for the Schr\"odinger equation, and give a detailed discussion about the associated sampling law and the gate complexity for the computation of the expectation of observables. In Section \ref{Sec: Hyp} we briefly study scalar nonlinear hyperbolic equations.  Discussion and summary are given in Section \ref{summary}.

\section{Review of quantum nonlinear differential equation solvers} \label{sect:reviewNonlinearODEs}

In \cite{Leyton2008nonlinear} a quantum algorithm, perhaps the first nonlinear differential equation solver, is described to solve sparse systems of nonlinear differential equations whose nonlinear terms are quadratic polynomials:
\[\frac{\d \bb{z}(t)}{\d t}  = F(\bb{z}), \qquad f_\alpha = \sum\limits_{k,l} a_{kl}^{(\alpha)}z_kz_l, \]
where $\bb{z}(t) = [z_1(t), z_2(t), \cdots,z_n(t)]^T$ and $F = [f_1(\bb{z}), f_2(\bb{z}), \cdots, f_n(\bb{z})]^T$. In the algorithm at least two copies of the state vector are required at every iteration step, which means the complexity scales exponentially with the number of steps to represent the nonlinearity throughout the evolution. Note that polynomials of degree higher than two (or more general nonlinearities) can be reduced to the quadratic case by introducing additional variables. However, the approach is only effective for low-order polynomials and is unlikely to be suitable for practical ODE solvers.  Ref.~\cite{Liu2021nonlinear} presents a quantum Carleman linearisation algorithm for a class of quadratic nonlinear differential equations
\[\frac{\d u}{\d t} = F_2 u^{\otimes 2} + F_1 u + F_0(t),  \qquad  u^{\otimes 2} = [u_iu_j].\]
Compared to the approach of \cite{Leyton2008nonlinear}, their algorithm improves the complexity from an exponential dependence on the evolution time to a nearly quadratic dependence. This algorithm is efficient in the regime $R<1$, where the quantity $R$ characterizes the relative strength
of the nonlinear and dissipative linear term. As an alternative, Ref.~\cite{XWG2021homotopy} uses a different truncation method~-~the homotopy perturbation method~-~to transform the problem into a series of nonlinear ODEs with a special structure, which are further embedded into linear ODEs with a technique similar to the Carleman linearisation. The dependence on the error is exponentially better than the quantum Carleman linearisation algorithm, but the ratio of nonlinearity to dissipation is required to be smaller. Moreover, the algorithm is only discussed for homogeneous differential equations. In \cite{Krovi2022improved}, the author presents substantially generalised and improved quantum algorithms over
prior work for inhomogeneous linear ODEs and applies it to nonlinear ODEs using Carleman linearisation.  Ref.~\cite{Engel2021embedding} describes a strategy of embedding the nonlinear system into an infinite-dimensional linear system with several specific mappings presented, which is then truncated to finite dimension. In particular, the linear Carleman embedding is also included in this approach and connections between
several embedding techniques are discussed.
Ref.~\cite{Lloyd2020nonlinear} describes a quantum algorithm, referred to as the quantum nonlinear Schr\"odinger linearisation technique, to solve the following nonlinear ODEs
\[\frac{\d f}{\d t} + f(x) x = b(t),\]
where $x \in \mathbb{C}^d$ and $f(x)$ is a $d\times d$ matrix that is an order $m$ polynomial function of the vectors $x$ and $x^\dag$.
After technical treatment, $f(x)$ can be assumed in the form of $f(x) = x^{\dag \otimes m} F x^{\otimes m}$ for a suitable tensor $F$. In the approach, the forward Euler discretisation $x(t+\Delta t) = x - \Delta t f(x) x$ is approximated by $N$-body Schr\"odinger evolution:
\[x - \Delta t f(x) x \approx \e^{-i H \Delta t} x^{\otimes N}, \qquad H = -i \begin{pmatrix} N \\ m\end{pmatrix}^{-1} \sum\limits_{j_1 \cdots j_m} F_{j_1 \cdots j_m}, \]
with
 \[F_{j_1 \cdots j_m} = I \otimes \cdots \otimes F \otimes I \otimes \cdots \otimes F \otimes \cdots \otimes I,\]
 where the right-hand side has $N$ matrices and $F$s are placed in the $j_l$-th positions.

 The above methods can be categorised as linear approximation approaches as in the introduction, in which the nonlinear term is linearised, so the
approach may only be valid for a short time, for weak nonlinearities and consequently may lose significant nonlinear
features of the problem after a long time \cite{JinLiu2022nonlinear}. The linear approximation approach is very different from the linear representation method for the ODE, where the original nonlinear ODE is transformed into a linear counterpart by using the level set mechanism {\it without any modelling errors}, and for {\it general nonlinearity}.
On the other hand, when reduced to the linear representations, the impact of the nonlinearity will be reflected in the CFL condition (or the numerical stability condition) as well as the accumulation of the local truncation error for the upwind discretizations. For example, the CFL condition in Theorem \ref{thm:FDLiouvilleRep} and the truncation error in Remark \ref{rem:kvntruc}.

\section{Linear representation methods for nonlinear ODEs: Liouville vs. KvN representations} \label{sect:linearrep}

In this section we consider the following nonlinear ODEs
\[
\frac{\d q_j}{\d t} = F_j(q_1,\cdots, q_d), \quad q_j(0) =  q_{0,j}, \quad j = 1,\cdots,d,
\]
where $q_{0,j}$ are initial data and $F_j$ are real-valued, which can be written in vector form as
\begin{equation}\label{ODEsFx}
\frac{\d q(t)}{\d t} = F(q(t)), \quad q(0) = q_0 , \quad q = [q_1,\cdots,q_d]^T.
\end{equation}

\subsection{Linear representation methods}

\subsubsection{The Liouville representation} \label{subsubsec:Liouville}

For $x = (x_1,\cdots,x_d)$, let $\delta(x) = \Pi_{i=1}^d \delta(x_i)$  be the Dirac delta distribution. The Liouville equation corresponding to \eqref{ODEsFx} can be derived by considering
 a function $\rho(t, x): \mathbb{R}^+ \times \mathbb{R}^d \to \mathbb{R}$, defined by
\begin{equation}\label{rhodelta}
\rho(t,x) = \delta( x-q(t)),
\end{equation}
which represents the probability distribution in space $x$ that corresponds to the solution $x =q$.
By the properties of the delta function, one obtains the solution of \eqref{ODEsFx} by taking the moment:
\begin{equation}\label{xrho}
q(t) = \int x\delta( x-q(t)) \d x = \int x\rho(t,x) \d x.
\end{equation}
Other quantity of interest $G(q(t))$ can be obtained by
the moment
\begin{equation}\label{xrho-G}
G(q(t)) = \int G(x)\delta( x-q(t)) \d x = \int G(x)\rho(t,x) \d x.
\end{equation}
To this end, we can characterize the dynamics of $\rho(t,x)$ and find the solution $q(t)$ via \eqref{xrho}.
One can check that $\rho$ satisfies, in the weak sense, the linear $(d + 1)$-dimensional PDE \cite{Dobin2021plasma,JinLiu2022nonlinear}
\begin{equation}\label{LiouvilleRepresentation}
\begin{cases}
\partial_t \rho(t,x) + \nabla_x \cdot [F(x) \rho(t,x)  ] = 0,\\
\rho_0(x):=\rho(0,x) = \delta(x-q_0).
\end{cases}
\end{equation}
Since the initial data involves a delta function,
one can  consider the following problem with the smoothed initial data
\begin{equation}\label{LiouvilleApp}
\begin{cases}
\partial_ t\rho^{\omega}(t,x) + \nabla \cdot [F(x) \rho^{\omega}(t,x)  ] = 0,\\
\rho^{\omega}_0(x):=\rho^{\omega}(0,x) = \delta_\omega(x-q_0)
\end{cases}
\end{equation}
when solving \eqref{LiouvilleRepresentation} by a classical or quantum algorithm, where $\omega$ is a smoothing parameter of the delta function \cite{JinLiu2022nonlinear}. For example, in one-dimensional case, one can choose
\[\delta_\omega = \begin{cases}
\frac{1}{\omega}\beta(x/\omega), \quad  & |x| \le \omega, \\
0, \quad & |x| > \omega,
\end{cases}\]
where typical choices of $\beta$ include
\[\beta(x) = 1 - |x| \qquad \mbox{and} \qquad \beta(x) = \frac{1}{2} ( 1 + \cos(\pi x) ).\]
Here $\omega = mh$ and $m$ is the number of mesh points within the support of $\delta_\omega$. For $d$ dimensions, one defines
\begin{equation*}
\delta_\omega(x) = \Pi_{i=1}^d \delta_\omega(x_i), \quad x = (x_1,\cdots,x_d).
\end{equation*}
In addition, the periodic boundary conditions can be imposed since $\rho(0,x)$ or $\rho^\omega(0,x)$ has compact support and the solutions to problems \eqref{LiouvilleRepresentation} and \eqref{LiouvilleApp} propagate with finite speed.

To compare with the KvN approach to be introduced next, note
the equation in \eqref{LiouvilleRepresentation} is usually transformed to the (some classical) analogue of the Schr\"odinger equation
\begin{equation}\label{LiouvilleODE}
\i \partial_t \rho = L \rho,
\end{equation}
where $L$ is referred to as the Liouville operator, satisfying
\[L \rho :=  -\i \sum\limits_j \Big( F_j \frac{\partial }{\partial x_j}  + \frac{\partial F_j}{\partial x_j} \Big)\rho.\]
The operator $L$ is generally not a Hermitian operator, and thus cannot be directly simulated by quantum Hamiltonian unless ${\rm div} F = 0$ (In this case one has
\[L \rho :=  -\i \sum\limits_j \Big( F_j \frac{\partial }{\partial x_j}  + \frac{\partial F_j}{\partial x_j} \Big)\rho
= -\i \sum\limits_j \Big( F_j \frac{\partial }{\partial x_j}  + \frac{1}{2}\frac{\partial F_j}{\partial x_j} \Big)\rho,\]
which is the symmetric KvN operator defined later).

\subsubsection{The Koopman-von Neumann representation}\label{sec: KvN}

The starting point of the Koopman-von Neumann (KvN) approach to classical mechanics is the introduction of a Hilbert space of complex and square integrable functions $\psi$ referred to as the KvN wave functions such that $\rho := |\psi|^2$ can be interpreted as the probability density of finding a particle at the point of the phase space.
To this end, it is desirable to find a complex-valued function $\psi$ which satisfies a dynamical behavior similar to that of the Schr\"odinger equation, and such that $\rho = |\psi|^2 = \psi^\dag \psi$ is a solution of \eqref{LiouvilleODE}.

 Formally, one can verify that $\psi$ satisfies  \cite{Bogdanov-Bogdanova-2014,Bogdanov-Bogdanova-2019,Joseph2020KvN,Lin2022KvN}
\[\i \partial_t \psi = \mathcal{H}_{\text{KvN}} \psi,\]
where the KvN operator is
\[\mathcal{H}_{\text{KvN}}  = -\i \sum\limits_j \Big( F_j \frac{\partial}{\partial x_j} + \frac{1}{2} \frac{\partial F_j}{\partial x_j}  \Big).\]
It can be written in the following symmetric form
\[\mathcal{H}_{\text{KvN}}\psi = \frac{1}{2} \sum\limits_j \Big(  F_j(x) P_j \psi + P_j  (F_j(x)\psi) \Big)
 =: \sum\limits_{j=1}^d H_j \psi,\]
where
\[H_j \psi = \frac{1}{2}  \Big( F_j(x) P_j \psi + P_j  (F_j(x)\psi) \Big), \quad P_j = -\i \frac{\partial}{\partial x_j}.\]
By introducing the position operator $\hat{x}_j$ and the momentum operator $\hat{P}_j$, the KvN representation can be rewritten as
\begin{equation}\label{KvNOperator}
\i \partial_t \ket{\psi} =  \frac{1}{2} \sum\limits_j (  F_j(\hat{x}) \hat{P}_j  + \hat{P}_j  F_j(\hat{x}) ) \ket{\psi}
 =: \sum\limits_{j=1}^d \hat{H}_j \ket{\psi},
\end{equation}
where the notation $F_j(\hat{x})$ denotes a nonlinear map of the position operator which resembles the nonlinear flow, potentially through series expansions.
Unlike the Liouville operator, the KvN operator is Hermitian and thus allows for quantum Hamiltonian simulations.

To obtain the quantity of interest $G(q(t))$ one uses
\begin{equation}\label{xrho-psi}
G(q(t)) = \int G(x)\delta( x-q(t)) \d x = \int G(x)|\psi(t,x)|^2 \d x.
\end{equation}

The setting of Eq.~\eqref{rhodelta} may be essential, since it allows to determine the solution of the ODEs via \eqref{xrho}. To be consistent, one may need to set $\psi(t,x) = \delta^{1/2} (x-q(t))$, the square root of the delta function.
 Namely, one needs to solve
 \begin{equation}\label{KvN-IP}
\begin{cases}
\i \partial_t \psi = \mathcal{H}_{\text{KvN}} \psi,\\
\psi(0,x) = \delta^{1/2}(x-q_0).
\end{cases}
\end{equation}
The solution to the above problem, in general, {\it cannot be defined mathematically} since one cannot define $\delta^{1/2}$, let alone its derivative, {\it even in the weak sense}.  However,  by connecting it with \eqref{LiouvilleRepresentation} by $\rho=|\psi|^2$
one can make sense of the solution to \eqref{KvN-IP} since $|\psi|^2$ satisfies the Liouville equation
\eqref{LiouvilleRepresentation} which can be defined in the weak sense \cite{LiPa, GMMP}. This connection is quite important, especially if one discretises \eqref{KvN-IP} and hopes the numerical solution will converge, as will be elaborated more later on.

At the discrete level, as was done in \cite{Dobin2021plasma}, one could instead consider  smoothing the $\delta$-function  and obtain
\begin{equation}\label{KvNSmooth}
\begin{cases}
\i \partial_t \psi^\omega = \mathcal{H}_{\text{KvN}} \psi^\omega,\\
\psi^{\omega}(0,x) = \delta_\omega^{1/2}(x-q_0),
\end{cases}
\end{equation}
which satisfies
\begin{equation}\label{rhopsi2}
\rho^\omega = |\psi^\omega|^2 = (\psi^\omega)^\dag \psi^\omega.
\end{equation}

 Although for every fixed $\omega$, the problem \eqref{KvNSmooth} is well defined, it will not be possible to define the $\omega \to 0$ limit which is needed to make mathematical sence of the KvN equation with initial data $\sqrt{\delta(x-q_0)}$. Furthermore, when discretising \eqref{KvNSmooth} numerically, one will not be able to prove the convergence of the numerical approximation (when $\omega$, mesh size and time step all go to zero) toward the  solution of the KvN since the solution of KvN is not suitably defined. In such a case, a different choice of $\omega$ or mesh size could pick up different numerical solutions, which is a well-known phenomenon when numerically approximating the (weak) solution of hyperbolic conservation laws \cite{LeVeque2002FVM}. These solutions may not be unique, unless stronger conditions, such as the entropy condition, are satisfied by the numerical approximations.

Even if one disregards the $\sqrt{\delta}$ problem, when using the Liouville equation \eqref{LiouvilleRepresentation}, one only needs $F$ to be Lipschitz continuous to define its solution, like for the original ODEs \eqref{ODEsFx}. For the KvN equation, since it is not in conservation form, one needs ${\text {div}} F$ to be Lipschitz continuous. Thus this is more restrictive than the case of the Liouville representation.

Unlike in the Liouville representation, in the KvN  framework one {\it cannot} get the ensemble average defined as $\frac{1}{M_0}\sum_{j=1}^{M_0} a_j $ for $M_0$ different initial data \cite{JinLiu2022nonlinear}, where $a(x)$ is a quantity of interest (say physical observables). Rather it gives
$(\frac{1}{M_0}\sum_{j=1}^{M_0}\sqrt{a_j})^2$, {\it Thus one loses quantum advantage in $M_0$ if one is interested in the ensemble average with $M_0 \gg 1$ initial data}. These two kinds of ensemble average observables are relevant to the quantum sampling setting, referring to \cite{HW20adaptive,CLL2022log}.

\subsubsection{Computation of the quantity of interest and error analysis} \label{subsubsect:obsLinear}

After regularising the $\delta$ function by $\delta_\omega$,
one now needs to approximate the physical quantity of interest, which are
\begin{itemize}
  \item For the Liouville approximation:
  \begin{equation}\label{phiObs}
  \langle G_{\rho^\omega}(t) \rangle = \int G(x) \rho^\omega (t,x) \d x,
  \end{equation}
  where $\rho^\omega$ is the solution of \eqref{LiouvilleApp};
  \item For the KvN approximation:
  \begin{equation}\label{psiObs}
  \langle O_{\psi^\omega}(t) \rangle = \int G(x) |\psi^\omega (t,x)|^2 \d x,
  \end{equation}
  where $\psi^\omega$ is the solution of \eqref{KvNSmooth} satisfying $\rho^\omega = |\psi^\omega|^2$ for every fixed $\omega>0$,
\end{itemize}
by some quadrature rules.

By choosing specific $G$ one recovers physical quantities of interest (usually observables). For example, if $G(x)=x$, \eqref{phiObs} and \eqref{psiObs} gives $q$, the solution of the original equation \eqref{ODEsFx}.
For other choices of $G$ one could recover other physical quantities such as momentum and energy for PDE problems such as the Hamilton-Jacobi equations. See \cite{JinLiu2022nonlinear} and Eq.~\eqref{obser}.

For this quantity, we have two ways to approximate it.
For the Liouville approximation, one can compute the integral by using the numerical quadrature rule
\begin{align}\label{Grho}
 \langle G(t_n) \rangle
 & = \int_{[0,1]^d} G(x) \rho(t_n,x) \d x  \approx \int_{[0,1]^d} G(x) \rho^\omega(t_n,x) \d x \nonumber \\
 & =: \langle G_{\rho^\omega}(t_n) \rangle  \approx \frac{1}{M^d}\sum_{\vect{j}} G_{\vect{j}} \rho^{\omega}_{ \vect{j},n}
  =: \langle G_{\rho^\omega,n}\rangle,
\end{align}
 where, $G_{\vect{j}} = \bb{\omega}_{\bb{j}} G(x_{\bb{j}})$ with $\bb{\omega}_{\bb{j}}$ being the weight,  $\bb{j} = (j_1,\cdots,j_d)$ and $M$ is the number of points in each dimension of the $d$-dimensional space. Throughout the paper, we only consider the trapezoidal rule: with $w = [\frac{1}{2}, 1, \cdots, 1, \frac{1}{2}]^T$, the weight vector can be arranged as $\sum\nolimits_{\bb{j}} \bb{\omega}_{\bb{j}} \ket{\bb{j}} = w \otimes \cdots \otimes w$. Accordingly, the solution vector is denoted as
 \[\bb{\rho}_n^\omega = \sum\nolimits_{\bb{j}} \bb{\rho}^\omega_{\bb{j},n} \ket{\bb{j}} = \sum\limits_{j_1,\cdots,j_d}\bb{\rho}^\omega_{j_1,\cdots,j_d,n} \ket{j_1} \otimes \cdots \otimes \ket{j_d}.\]
That is, the $n_{\bb{j}}$-th entry of $\bb{\rho}_n^\omega$ is $\bb{\rho}^\omega_{\bb{j},n}$, with the global index given by
\begin{equation}\label{globalindex}
n_{\bb{j}}: = j_12^{d-1} + \cdots + j_d2^0.
\end{equation}
 Note that for periodic boundary conditions, $\rho$ can be assumed to be periodic since the solution to the Liouville equation is essentially zero outside a compact support. In such a case, the trapezoidal rule is of spectral accuracy \cite{Kurganov2009}.

 \begin{lemma}[Error of the Liouville representation] \label{lem:errLiouville}

 Let $\rho^\omega$ be the analytical solution of the Liouville representation \eqref{LiouvilleRepresentation} with the smoothed initial data, and $\rho^\omega_h$ the numerical solution of $\rho^\omega$. Then
\begin{equation}\label{rhoErr}
e_{\rho} := |\langle  G(t_n)  \rangle-\langle G_{\rho^\omega_h,n} \rangle| \le
C( \omega \e^{t_n \|\text{div} F\|_\infty} + d\Delta x^{\ell}/\omega^{\ell+1} + e_{\rho,h}),
\end{equation}
where $ \|\text{div} F\|_\infty = \text{sup}_q |\text{div} F (q)|$, $\ell$ is the Sobolev regularity of $\rho^\omega$ (namely $\rho^\omega \in C^\ell$), and $e_{\rho,h} = |\bb{\rho}^\omega_n - \bb{\rho}^\omega_{h,n}|$ is the (relative) discretisation error for the linear Liouville equation, given by for examples:
\begin{itemize}
  \item For the first-order upwind finite difference scheme, one has
  \begin{equation}\label{UW-error}
      e_{\rho,h}  \le C \Big( \frac{\Delta t}{\omega} + \frac{d\Delta x}{\omega^2} \Big)=  \mathcal{O} \Big( \frac{d\Delta x}{\omega^2} \Big)
\end{equation}
  with the  CFL condition $ d\lambda = \mathcal{O}(1)$ (for numerical stability), where $\lambda = \Delta t/\Delta x$.
  \item For the Fourier spectral discretisation, one has
  \begin{equation}\label{Spectral-error}
 e_{\rho,h} \le C \Big( \frac{\Delta t^\alpha}{\omega^\alpha} + \frac{d\Delta x^\ell}{\omega^{\ell+1}} \Big),
  \end{equation}
   where $\alpha$ depends on the accuracy of the temporal discretisation.
\end{itemize}
 \end{lemma}

\begin{proof}
The error can be split as
\begin{align}
e_{\rho}
& = | \langle G(t_n) \rangle - \langle G_{\rho^\omega_h,n}\rangle | = | \langle G_\rho(t_n) \rangle - \langle G_{\rho^\omega_h,n}\rangle | \nonumber\\
&  \le | \langle G_\rho(t_n) \rangle - \langle G_{\rho^\omega} (t_n)\rangle |
  + |\langle G_{\rho^\omega} (t_n)\rangle - \langle G_{\rho^\omega,n} \rangle |
  + |\langle G_{\rho^\omega,n} \rangle - \langle G_{\rho_h^\omega,n}\rangle | \nonumber\\
& =: I_1 + I_2 + I_3. \label{Liouvilledecomp}
\end{align}
For $I_1$, one can apply the the method of characteristics as done in \cite{Raviart1985particle,JinLiu2022nonlinear}. To do so, we introduce the characteristics of \eqref{LiouvilleRepresentation} as
\[\begin{cases}
\dfrac{\d X}{\d t} = F(X), \quad  X \in \mathbb{R}^d, \\
X(s) = x,
\end{cases}\]
with the solution denoted by $X(t; x, s)$. Let the Jacobian determination of the map from $x$ to $X$ be
\[J(t; x,s)= \text{det} \Big( \frac{\partial X_i}{\partial x_j}(t; x,s)\Big).\]
Then one has
\[ J(t; x, s)>0,\qquad
  J(t; x,s) = \exp \Big( \int_s^t \nabla \cdot F(X(\sigma; x,s)) \d\sigma \Big).  \]
By the method of characteristics (see Eq.~(1.11) of \cite{Raviart1985particle} or Appendix J of \cite{JinLiu2022nonlinear}), the solution to \eqref{LiouvilleRepresentation} can be given by
\[\rho(t,x) = \rho_0( X(0; x, t) )J(0; x,t) = \rho_0( X(0; x, t) ) \exp \Big( -\int_0^t \nabla \cdot F(X(\sigma; x,t)) \d\sigma \Big).\]
Similarly, the solution to \eqref{LiouvilleApp} is
\begin{equation*}
\rho^\omega(t,x) = \rho_0^\omega( X(0; x, t) )J(0; x,t) = \rho_0^\omega( X(0; x, t) ) \exp \Big( -\int_0^t \nabla \cdot F(X(\sigma; x,t)) \d\sigma \Big).
\end{equation*}
Therefore,
\begin{align*}
I_1
& = | \langle G_\rho(t_n) \rangle - \langle G_{\rho^\omega} (t_n)\rangle |
  = \Big| \int_{[0,1]^d} G(x) ( \rho(t_n,x)- \rho^\omega(t_n,x)) \d x \Big| \\
& = \Big| \int_{[0,1]^d} G(x) \Big( \rho_0( X(0; x, t_n) )- \rho_0^\omega( X(0; x, t_n) ) \Big) J(0; x,t_n) \d x \Big| \\
& \le C \omega \max_{x\in [0,1]^d} J(0; x,t_n) \le C \omega \e^{t_n \|\text{div} F\|_\infty},
\end{align*}
where $ \|\text{div} F\|_\infty = \text{sup}_q |\text{div} F (q)|$.

The second term is just the error of the quadrature rule, hence $I_2 \le C d \Delta x^\ell/\omega^{\ell+1}$,
where the $1/\omega^{\ell+1}$ factor comes from the $\ell$-th derivative of $\delta_w$. Obviously, $I_3 \le C e_{\rho,h}$.
The final result comes from the standard error analysis for the linear hyperbolic equation \cite{LeVeque2002FVM}.
\end{proof}

\begin{remark}
Note that for the upwind scheme, due to the CFL condition
$\Delta t=\mathcal{O}(\Delta x/d)$, one has
\[\mathcal{O}(\Delta t/\omega)=\mathcal{O}(\Delta x/d\omega) = \littleo( d\Delta x/\omega^2),\]
hence the second equality in \eqref{UW-error} holds. For the spectral discretisation in Subsect.~\ref{subsubsect:spectralLiouoville}, we require that
\[d \D t^\alpha/\omega^\alpha \sim d\Delta x^{\ell}/\omega^{\ell+1} \sim \varepsilon,\]
see \eqref{meshSpectralLiouville} for example, where $d$ for time comes from the dimension splitting in \eqref{Wdelta}. This means
\[ \D t^\alpha/\omega^\alpha \le d \D t^\alpha/\omega^\alpha \lesssim d\Delta x^{\ell}/\omega^{\ell+1}.   \]
That is, we can still combine the two terms on the right hand side of \eqref{Spectral-error}. We leave it as is in what follows.
\end{remark}

For the KvN approximation, the quadrature rule gives
\begin{align}\label{Opsi}
 \langle O_{\psi^\omega} (t_n)\rangle
 & = \int_{[0,1]^d} G(x) |\psi^\omega(t_n,x)|^2 \d x
  \approx \frac{1}{M^d}\sum_{\vect{j}} G_{\vect{j}} |\psi^{\omega}_{ \vect{j},n}|^2 \nonumber \\
 & = \frac{1}{M^{d/2}}(\bb{\psi}_n^\omega)^\dag G_M \bb{\psi}_n^\omega
  =: \langle O_{\psi^\omega,n}\rangle,
\end{align}
where the elements of the vector $\bb{\psi}_n^\omega$ are arranged as $\bb{\psi}_n^\omega = \sum_{\bb{j}} \psi^{\omega}_{ \vect{j},n} \ket{\bb{j}}$, and $G_M = \text{diag}( \bb{g} )$ is a diagonal matrix with $\bb{g} = \sum_{\bb{j}} G_{\vect{j}}/M^{d/2}  \ket{\bb{j}}$ satisfying $\|\bb{g}\| \sim 1$.

 \begin{lemma}[Error of KvN representation] \label{lem:errKvN}

 Let $\psi^\omega$ be the analytical solution of the KvN representation \eqref{KvNSmooth}, and $\psi^\omega_h$ the numerical solution of $\psi^\omega$. Then
\begin{equation}\label{psiErr}
e_{\psi} := |\langle  G(t_n)  \rangle-\langle O_{\psi^\omega_h,n} \rangle| \le
C( \omega \e^{t_n \|\text{div} F\|_\infty} + d\Delta x^{\ell}/\omega^{\ell+1} + e_{\psi,h}),
\end{equation}
where $\ell$ is the regularity of $\rho^\omega$, and $e_{\psi,h} = | |\bb{\psi}_n^\omega|^2 - |\bb{\psi}_{h,n}^\omega|^2 |$ is the (relative) discretisation error for the KvN equation \eqref{KvNOperator}.

\begin{itemize}
  \item For the first-order upwind finite difference scheme, one has
  \[e_{\psi,h}  \le C \Big(  \frac{d\Delta x}{\omega^2} \Big),\]
  with the CFL condition $ d\lambda = \mathcal{O}(1)$, where $\lambda = \Delta t/\Delta x$.
  \item For the Fourier spectral discretisation, one has
  \[e_{\psi,h} \le C \Big( \frac{\Delta t^ \alpha}{\omega^\alpha} + \frac{d\Delta x^\ell}{\omega^{\ell+1}} \Big),\]
  where $\alpha$ depends on the accuracy of the temporal discretisation.
\end{itemize}
 \end{lemma}
\begin{proof}
Noting that $\rho^\omega=|\psi^\omega|^{2}$, we can split the error as
\begin{align*}
e_\psi
& = |\langle G(t_n)\rangle-\langle O_{\psi_h^\omega, n}\rangle| \\
& \le |\langle G_\rho(t_n)\rangle-\langle O_{\psi^\omega}(t_n)\rangle| + |\langle O_{\psi^\omega}(t_n)\rangle-\langle O_{\psi^\omega, n}\rangle|+|\langle O_{\psi^\omega, n}\rangle-\langle O_{\psi_h^\omega, n}\rangle| \\
& = |\langle G_{\rho}(t_n)\rangle-\langle G_{\rho^\omega}(t_n)\rangle| +|\langle O_{\psi^\omega}(t_n)\rangle-\langle O_{\psi^\omega, n}\rangle|+|\langle O_{\psi^\omega, n}\rangle-\langle O_{\psi_h^\omega, n}\rangle| \\
& =: I_1+I_2+I_3 .
\end{align*}
Here, $I_1$ is exactly the first term in \eqref{Liouvilledecomp} for the Liouville representation, so $I_1 \leq C \omega \e^{t_n \|\text{div} F\|_\infty}$. The second term is the (relative) error of the quadrature rule. One again has $I_2 \leq C d \Delta x^\ell / \omega^{\ell+1}$ since $\rho^\omega=|\psi^\omega|^2$, where the $1 / \omega^{\ell+1}$ factor comes from the $\ell$-th derivative of $\delta_w$. For the last term, one has
\begin{align*}
I_3
& = |\langle O_{\psi^\omega, n}\rangle-\langle O_{\psi_h^\omega, n}\rangle|
  =|(\bb{\psi}_n^\omega)^{\dagger} G_{M} \bb{\psi}_n^\omega-(\bb{\psi}_{h, n}^\omega)^{\dagger} G_{M} \bb{\psi}_{h, n}^\omega| \\
& = \frac{1}{M^d} \Big|\sum_{j} G_{\bb{j}}| \psi_{\bb{j}, n}^\omega|^2-\sum_{j} G_{\bb{j}} |(\psi_h^\omega)_{\bb{j}, n}|^2 \Big|
  \lesssim  | |\bb{\psi}_n^\omega|^2-|\bb{\psi}_{h, n}^\omega|^2 |.
\end{align*}
This completes the proof.
\end{proof}


\begin{remark}
When discretising the Liouville equation by the upwind scheme, since it is in conservative form, one can show that the $l_1$ norm at later time is bounded by the norm at $t=0$ when the CFL condition is satisfied (namely it is $l_1$ contracting). In addition, the accumulation of the local truncation error which grows {\it linearly} in $t$.
However, this is not true for the KvN representation. If one uses the upwind scheme to discretize the transport (spatial derivative) term,
since the $1/2$ term is a forcing term, it will contribute to the $l_1$ error an exponentially growing term like $\e^{t_n \|\text{div} F\|_\infty}$. When $t_n \le T = \mathcal{O}(1)$ this is not a big issue. But it is worth pointing  out the difference with the Liouville representation here, since if one wants to compute the long time solution, the $l_1$ contraction of the upwind scheme for the Liouville representation gives much smaller error. The above discussion will be further elaborated later.

For simplicity, we only consider $t_n \le T = \mathcal{O}(1)$ in this article.
\end{remark}

\subsection{Finite difference discretisation for the Liouville representation}

\subsubsection{The QLSA for the finite difference discretisation}

The Liouville representation can be rewritten as
\begin{equation}\label{LiouvilleExplicit}
\begin{cases}
\frac{\partial w}{\partial t} + \sum\limits_{i=1}^d \frac{\partial}{\partial x_i} ( F_i(x) w(t,x) ) = 0,\\
w_0(x) = \delta_{\omega}(x-x_0),
\end{cases}
\end{equation}
where we have assumed the smoothed initial data $\delta_\omega$.

Denote $\bb{e}_i = [0,\cdots, 1, \cdots, 0]$ to be the unit vector with the $i$-th entry being 1. Let $\bb{j} = (j_1,\cdots,j_i, \cdots, j_d)$ and $x_i = j_i \D x$ are the $i$-th components of $x_{\bb{j}}$. The first-order upwind discretisation at $(t_n,x_{\bb{j}})$ takes the following form \cite[Appendix K]{JinLiu2022nonlinear}:
\begin{align*}
& \partial_t w  \longrightarrow  \frac{w_{\bb{j}}^{n+1} - w_{\bb{j}}^n}{\D t},   \nonumber \\
& \frac{\partial}{\partial x_i} ( F_i(x) w(t,x) )
   \longrightarrow
  \frac{1}{\Delta x} \Big( \Big\{F_i(x_{j_i+1/2})\Big\}_{\bb{j}}^-  w^n_{\bb{j}+\bb{e}_i} - \Big\{F_i(x_{j_i-1/2})\Big\}_{\bb{j}}^+  w^n_{\bb{j}-\bb{e}_i} \Big) \nonumber \\
& \quad \hspace{3cm} + \frac{1}{\Delta x} \Big(\Big\{F_i(x_{j_i+1/2})\Big\}_{\bb{j}}^+ - \Big\{F_i(x_{j_i-1/2})\Big\}_{\bb{j}}^- \Big) w_{\bb{j}}^n, 
\end{align*}
where
\begin{equation*} 
\alpha^+ = \max \{ \alpha, 0\} = \frac{\alpha + |\alpha|}{2}\ge 0, \quad
\alpha^- = \min \{ \alpha, 0\} = \frac{\alpha-|\alpha|}{2} \le 0,
\end{equation*}
and
\[\Big\{F_i(x_{j_i\pm 1/2})\Big\}_{\bb{j}} = \frac{1}{2}(F_i(x_{\bb{j}\pm \bb{e}_i}) + F_i(x_{\bb{j}}) )
= \frac{1}{2}(F_i(x_{j_1, \cdots, j_i \pm 1, \cdots, j_d}) + F_i(x_{j_1, \cdots, j_i, \cdots, j_d}) ) .\]
For convenience we introduce the following notation
\[a_{\bb{j}}^{i,\pm} = \Big\{F_i(x_{j_i+1/2}) \Big\}_{\bb{j}}^\pm, \qquad
b_{\bb{j}}^{i,\pm} = \Big\{F_i(x_{j_i-1/2})\Big\}_{\bb{j}}^\pm.\]
The discrete scheme can be written as
\begin{equation} \label{LiouvilleRepdiffscheme}
w_{\bb{j}}^{n+1} -  \Big[ 1 - \lambda \sum\limits_{i=1}^d (a_{\bb{j}}^{i,+}    - b_{\bb{j}}^{i,-}) \Big] w^n_{\bb{j}}
 + \lambda \sum\limits_{i=1}^d \Big[ a_{\bb{j}}^{i,-}  w^n_{\bb{j}+\bb{e}_i}  - b_{\bb{j}}^{i,+}   w^n_{\bb{j}-\bb{e}_i} \Big]   = 0,
\end{equation}
where $\lambda = \D t/\D x$. In matrix form one has
\[\bb{w}^{n+1} - B \bb{w}^n = \bb{f}^{n+1}, \quad n = 0,1,\cdots, N_t-1,\]
with $\bb{f}^i$ being the terms resulting from the initial and boundary conditions, where the nodal values at $t=t_n$ are arranged as $\bb{w}^n = \sum_{\bb{j}}w_{\bb{j}}^n \ket{\bb{j}}$. With the help of the global index \eqref{globalindex}, the non-zero entries of $B$ can be given by
\[B_{n_{\bb{j}}, n_{\bb{j}} } = 1 - \lambda \sum\limits_{i=1}^d (a_{\bb{j}}^{i,+}    - b_{\bb{j}}^{i,-}), \qquad
B_{n_{\bb{j}}, n_{\bb{j}+\bb{e}_i} } =  -\lambda a_{\bb{j}}^{i,-},  \qquad
B_{n_{\bb{j}}, n_{\bb{j}-\bb{e}_i} } =  \lambda b_{\bb{j}}^{i,+}.\]
By introducing the notation $\bb{w} = [\bb{w}^1; \cdots ;\bb{w}^{N_t}]$, where ``;" indicates the straightening of $\{\bb{w}^i\}_{i\ge 1}$ into a column vector, one obtains the following linear system
\begin{equation}\label{systemQdiffLiouvilleRepresenation}
L \bb{w} = F,
\end{equation}
where
\[L =
\begin{bmatrix}
I  &            &           &            \\
-B & I     &           &            \\
        &\ddots      & \ddots    &    \\
        &            & -B   & I     \\
\end{bmatrix}
, \qquad
F =
\begin{bmatrix}
\bb{f}^1 \\
\bb{f}^2  \\
\vdots\\
\bb{f}^{N_t} \\
\end{bmatrix}.
\]
For periodic boundary conditions, one has $\bb{f}^1 = B\bb{w}^0$ and $\bb{f}^i = \bb{0}$ for $i\ge 2$.

\begin{theorem} \label{thm:FDLiouvilleRep}
Suppose that $\lambda = \Delta t/\Delta x$ satisfies the following CFL condition
\[ \lambda   \sum\limits_{i=1}^d \sup_x |F_i(x)| \le 1.\]
\begin{enumerate}[(1)]
  \item The condition number and the sparsity of $L$ satisfy $\kappa = \mathcal{O}(1/\Delta t)$ and $s = \mathcal{O}(d)$.
  \item For fixed spatial step $\Delta x$, let $\Delta t = \mathcal{O}(\Delta x/d)$ and $\omega = (d\Delta x)^{1/3}$. Given the error tolerance $\varepsilon$, the gate complexity of the QLSA (for the problem in Eq.~\eqref{systemQdiffLiouvilleRepresenation}) is
\[N_{\text{Gates}} = \widetilde{\mathcal{O}}\Big( \frac{d^3}{\varepsilon^3} \log \frac{1}{\varepsilon} \Big).\]
\end{enumerate}
\end{theorem}
\begin{proof}

1) We claim that $\|B\|_2 \le 1 + \Delta t \|{\rm div} F\|_{\infty}$. To this end, one can show that the summation of the absolute values of  row  or column entries is not greater than $1 + \Delta t \|{\rm div} F\|_{\infty}$.

Let $R_i$ be the sum of absolute values of row entries. Since the $i$-th equation in $Ax = b$ is actually the $i$-th row of $A$ multiplied by $x$, the entries of the $i$-th row of $A$ are completely determined by the $i$-th equation. On the other hand, the grid values corresponding to the initial or boundary conditions in the $i$-th equation will increase the values of the summation of the absolute values of row entries since they will be moved to the right-hand side. That is, $R_i\le \tilde{R}_i$, where $\tilde{R}_i$ includes the contribution from the initial-boundary values.

With the CFL condition, one has
\[c: = 1 - \lambda \sum\limits_{i=1}^d (a_{\bb{j}}^{i,+}    - b_{\bb{j}}^{i,-}) \ge 0.\]
The argument is as follows. Without loss of generality, we set $d = 1$ and obtain
\[c = 1 - \frac{\lambda}{4}( F_{j+1} - F_{j-1} + | F_{j+1} + F_j| + |F_{j-1} + F_j|) = : 1 - \frac{\lambda}{4} c_F,\]
where $F_j = F(x_j)$. This reduces to verify that $c_F$ contains only four terms of the grid values of $F$:
\begin{itemize}
  \item For $F_{j-1} + F_j\ge 0$, one has
  \[c_F = F_{j+1} - F_{j-1} + | F_{j+1} + F_j| + F_{j-1} + F_j = F_{j+1} + F_j + | F_{j+1} + F_j|, \]
  as required.
  \item For $F_{j-1} + F_j< 0$, one has
  \begin{align*}
  c_F
  & = F_{j+1} - F_{j-1} + | F_{j+1} + F_j| - F_{j-1} - F_j \\
  & = \begin{cases}
 F_{j+1} - F_{j-1} + F_{j+1} - F_{j-1}, \qquad & \mbox{if $F_{j+1} + F_j \ge 0$}, \\
  - F_{j-1} - F_j - F_{j-1} - F_j, \qquad & \mbox{if $F_{j+1} + F_j < 0$},
  \end{cases}
  \end{align*}
  as required.
\end{itemize}

The above argument implies that
\begin{align*}
R_i
& \le \tilde{R}_i \le 1 - \lambda \sum\limits_{i=1}^d (a_{\bb{j}}^{i,+}    - b_{\bb{j}}^{i,-})
 + \lambda \sum\limits_{i=1}^d ( b_{\bb{j}}^{i,+} - a_{\bb{j}}^{i,-}) \\
& = 1 - \lambda \sum\limits_{i=1}^d (a_{\bb{j}}^{i,+} + a_{\bb{j}}^{i,-} - b_{\bb{j}}^{i,+} -  b_{\bb{j}}^{i,-} )
 = 1 - \lambda \sum\limits_{i=1}^d (a_{\bb{j}}^i  - b_{\bb{j}}^i) \\
& = 1 - \lambda \sum\limits_{i=1}^d \Big(\Big\{F_i(x_{i+1/2})\Big\}_{\bb{j}}  - \Big\{F_i(x_{i-1/2})\Big\}_{\bb{j}} \Big).
\end{align*}
Since $F$ is smooth, we obtain from the mean value theorem that
\begin{align*}
R_i
& \le  1 - \lambda \sum\limits_{i=1}^d \Big(\Big\{F_i(x_{i+1/2})\Big\}_{\bb{j}}  - \Big\{F_i(x_{i-1/2})\Big\}_{\bb{j}} \Big) \\
& = 1 - \lambda \sum\limits_{i=1}^d \partial_{x_i}F_i(\xi_i) \D x  \qquad   (x_{i-1/2}< \xi_i \le x_{i+1/2}) \\
& = 1 - \Delta t \sum\limits_{i=1}^d \partial_{x_i}F_i(\xi_i) \le 1 + \Delta t \|{\rm div} F\|_{\infty},
\end{align*}
where $F_i(\xi_i):= F_i(x_{j_1}, \cdots, \xi_i, \cdots, x_{j_d})$.

Let $C_j$ be the absolute column under discussion. The $j$-th column of $A$ is exactly the collection of the coefficients of $x_j$ in each equation of $Ax=b$, hence the absolute column sum is simply the sum of the absolute values of the coefficients with respect to $x_j$. Consider the variable $w_{\bb{j}}^n$ in \eqref{LiouvilleRepdiffscheme}. Notice that the other subscripts are only shifted left and right once in some direction, and thus the other elements of the corresponding column are only changed by the upper index $i$. This again implies that
\begin{align*}
C_j
\le 1 - \lambda \sum\limits_{i=1}^d (a_{\bb{j}}^{i,+}    - b_{\bb{j}}^{i,-})
 + \lambda \sum\limits_{i=1}^d ( b_{\bb{j}}^{i,+} - a_{\bb{j}}^{i,-}) \le 1 + \Delta t \|{\rm div} F\|_{\infty}.
\end{align*}

2) By definition, $\sigma_{\min}(L) = 1/\sigma_{\max}(L^{-1})$.
After simple algebra, one has
\[
{ \scriptsize
L^{-1} =
\begin{bmatrix}
I         &              &                 &          &            \\
B        & I       &                 &          &        \\
B^2      &  \ddots      &  \ddots         &          &\\
   \vdots       &  \ddots      &  \ddots         &  \ddots  &   \\
B^{N_t-1} &   \cdots     &   B^2     &  B &  I     \\
\end{bmatrix}
= \begin{bmatrix}
I          &              &                 &          &            \\
                & I       &                 &          &        \\
                &              &  \ddots         &          &\\
                &              &                 &  \ddots  &   \\
                &              &                 &          &  I      \\
\end{bmatrix}+
\begin{bmatrix}
                 &              &                 &          &            \\
B        &              &                 &          &        \\
                 &  \ddots      &                 &          &\\
                 &              &  \ddots         &          &   \\
                 &              &                 &  B &         \\
\end{bmatrix} + \cdots,}
\]
which gives
\begin{align*}
\sigma_{\max}(L^{-1})= \|L^{-1}\|_2 \le \|I\|_2 + \|B\|_2 + \|B\|_2^2 +\cdots +\|B\|_2^{N_t-1}.
\end{align*}
According to the previous analysis, one has
\[ \|B\|_2 \le 1 + \Delta t \|{\rm div} F\|_{\infty} \le c = \begin{cases}
1, \quad & \|{\rm div} F\|_{\infty} = 0, \\
1 + \Delta t, \quad & 0<\|{\rm div} F\|_{\infty}<1, \\
1 + \Delta t \|{\rm div} F\|_{\infty}, \quad & \|{\rm div} F\|_{\infty} \ge 1,
\end{cases}  \]
and hence
\[\sigma_{\max}(L^{-1}) \le 1 + c + c^2 + \cdots c^{N_t-1} = \frac{c^{N_t}-1}{c-1}. \]
Noting that $(1+x/n)^n \le \e^x$ holds for any real number, we then have
\[\sigma_{\max}(L^{-1}) \le \frac{1}{\Delta t} \times \begin{cases}
1, \quad & \|{\rm div} F\|_{\infty} = 0 \\
\e, \quad & 0<\|{\rm div} F\|_{\infty}<1 \\
\exp(\|{\rm div} F\|_{\infty}), \quad & \|{\rm div} F\|_{\infty} \ge 1
\end{cases}, \]
which can be simply written as
\[\sigma_{\max}(L^{-1}) \le \exp(\|{\rm div} F\|_{\infty} + 1) \frac{1}{\Delta t},\]
hence
\[\sigma_{\min}(L) \ge  \frac{\Delta t}{ \exp(\|{\rm div} F\|_{\infty} + 1) }.\]
By the Gershgorin-type theorem for singular values \cite{Qi-1984,Horn2013},
\[\sigma_{\max}(L) \le 1 + \|B\|_2 \le 2 + \Delta t \|{\rm div} F\|_{\infty}.\]
which gives
\[\kappa(L)
\le ( 2 + \Delta t \|{\rm div} F\|_{\infty} )\exp(\|{\rm div} F\|_{\infty} + 1)  \frac{1}{\Delta t}
\lesssim \exp(\|{\rm div} F\|_{\infty})  \frac{1}{\Delta t}.\]

3)  In view of the CFL condition, we set $\Delta t = \mathcal{O}(\Delta x /d)$. According to Lemma \ref{lem:errLiouville}, the classical error of the numerical approximation (if the first order upwind scheme is used to discretize the Liouville equation) to the observables is $\mathcal{O}(\omega + \Delta t/\omega + d \Delta x/\omega^2)$. One can choose $\omega$ such that $\omega \sim d \Delta x/\omega^2$ or $\omega = (d\Delta x)^{1/3}$ so the numerical error becomes
$O( (d\Delta x)^{1/3})$.
To reach the precision $\mathcal{O}(\varepsilon)$, we choose $\Delta x \sim \varepsilon^3/d$, and hence $\Delta t/\omega \sim \varepsilon^2/d^2 \le \varepsilon$. This naturally leads to the query complexity \cite{Costa2021QLSA}
\[Q = \mathcal{O}\Big(s \kappa \log \frac{1}{\varepsilon} \Big)
    = \mathcal{O}\Big( \frac{d^3}{\varepsilon^3} \log \frac{1}{\varepsilon} \Big).\]
The gate complexity is larger than the query complexity only by logarithmic factors.
\end{proof}

\begin{remark}\label{Rmk2.3}
Since the Liouville equation is in conservative form, one can get the $l_1$ contracting of the upwind scheme. Without loss of generality we set $d=1$. The upwind scheme in \eqref{LiouvilleRepdiffscheme} is then given by
\[w_j^{n+1} -  ( 1 - \lambda (a_j^+ - b_j^-) ) w^n_j  + \lambda ( a_j^-  w^n_{j+1}  - b_j^+   w^n_{j-1} )   = 0,\]
where $a_j^{\pm} = F(x_{j+1/2})^\pm$ and $b_j^{\pm} = F(x_{j-1/2})^\pm$. Since $b_j^{\pm} = a_{j-1}^\pm$, the scheme can be rewritten as
\[
w_j^{n+1} -  ( 1 - \lambda (a_j^+ - a_{j-1}^-) ) w^n_j
 + \lambda ( a_j^-  w^n_{j+1}  - a_{j-1}^+   w^n_{j-1} ) = 0.
\]
Under the CFL condition given in Theorem \ref{thm:FDLiouvilleRep}, one has
\begin{align*}
\|\bb{w}^{n+1}\|_1
& = \sum\limits_j |w_j^{n+1}| \le \sum\limits_j \Big[ (1 - \lambda (a_j^+ - a_{j-1}^-))  |w^n_j|
 - \lambda a_j^-  |w^n_{j+1}|  + \lambda a_{j-1}^+   |w^n_{j-1}| \Big] \\
& = \sum\limits_j \Big[ (1 - \lambda (a_j^+ - a_{j-1}^-))  |w^n_j|
 - \lambda a_{j-1}^-  |w^n_j|  + \lambda a_j^+   |w^n_j| \Big] \\
& = \sum\limits_j |w_j^n| = \|\bb{w}^n\|_1,
\end{align*}
as required.
\end{remark}

\subsubsection{The algorithm for the computation of the observable} \label{subsubsect:obsLiouvilleRep}

The quantum algorithm to approximate physical observables is presented in \cite{JinLiu2022nonlinear} with a detailed analysis on the gate complexity, where the observable is computed by using the amplitude estimation algorithm with block-encoding techniques augmented by amplitude amplification.  Amplitude amplification was used to achieve optimal scaling of the query complexity with respect to the error $\varepsilon$ while measuring an expectation value. However, since our purpose here is only to compare the strengths of the Liouville approach versus the Koopman-von Neumann approach, for simplicity we can instead compute the observable with a more straightforward means, without using amplitude amplification. In this paper, we first obtain the quantum state proportional to the solution of the problem, either with QLSA or with quantum simulation, then compute the observable afterwards.

\subsubsection*{The expectation of the observable}

In order to measure observables, we need to express them as Hermitian operators.
Let $\bb{w}_{\bb{j},n} :=\bb{\rho}^\omega_{\bb{j},n}$ be the solution of the upwind finite difference method (with the smoothed initial data $\delta_\omega$). One has
\[\ket{\psi} = \frac{1}{N_{\psi}} \sum\limits_{\bb{j},n} \bb{w}_{\bb{j},n} \ket{\bb{j}} \ket{n}
,\]
where the normalization constant $N_{\psi} = \|\bb{w}\|$.
With $G_{\bb{j}}$ in \eqref{Grho}, we define the state
\[|G_n\rangle := \frac{1}{N_G}\sum_{\vect{j}}G^\dagger_{\vect{j}} |\vect{j}\rangle |n\rangle,\]
where $N_G= (\sum_{\vect{j}}|G^\dagger_{\vect{j}}|^2)^{1/2}$ is the normalisation constant.
We assume we are given the density matrix $\mathcal{G} :=|G_n\rangle \langle G_n|$ and we define
$\Upsilon := \langle \psi |\mathcal{G} | \psi \rangle$. Simple algebra yields
 \begin{equation}\label{eq:gequation}
  \langle G(t_n) \rangle \approx \langle G_{\rho^\omega,n}\rangle  = n_{\psi}n_G|\sqrt{\Upsilon}|,
 \end{equation}
where $n_G = N_G/M^{d/2} = \mathcal{O}(1)$ is known and $n_{\psi} = N_\psi /M^{d/2}$ may be unknown.
We further define \begin{equation}\label{ObsLiouvilleOLSA}
\langle O \rangle = \langle G_{\rho^\omega,n}\rangle^2 = (n_Gn_{\psi})^2 \Upsilon := \langle \psi | O | \psi \rangle, \qquad
O = (n_Gn_{\psi})^2 \mathcal{G}.
\end{equation}
Then one only needs to estimate it to precision $\varepsilon$ since
\[| \langle G_{\rho^\omega,n}\rangle- \langle G_{\rho^\omega,n}\rangle_{\text{app}}|
 = \frac{1}{| \langle G_{\rho^\omega,n}\rangle + \langle G_{\rho^\omega,n}\rangle_{\text{app}}|}
 |\langle O \rangle - \langle O \rangle_{\text{app}}| \]
and $\langle G_{\rho^\omega,n}\rangle$ and $\langle G_{\rho^\omega,n}\rangle_{\text{app}}$ can be considered as $\mathcal{O}(1)$, where the subscript ``app'' refers to the approximations.

\begin{figure}[H]
  \centering
  \includegraphics[scale=0.5]{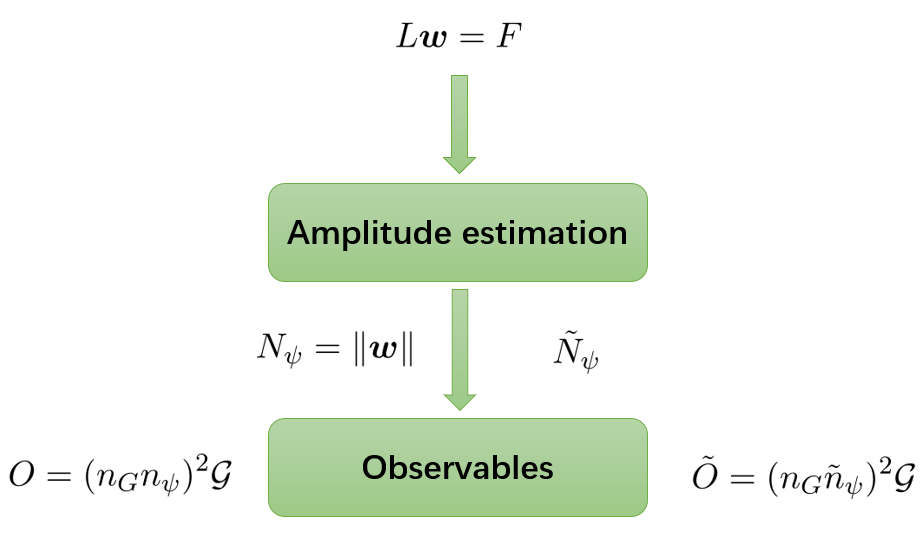}\\
  \caption{Construction of the observables when $\|\bb{w}\|$ is unknown.}\label{fig:Obs}
\end{figure}

The problem then reduces to approximating the normalisation constant $N_\psi = \|\bb{w}\|$ to a desired precision, where $\bb{w} = L^{-1} F$. This can be referred to as the amplitude estimation or linear equation norm estimation in quantum computing \cite{Linden2020heat,Chakraborty2019blockEncode}. As shown in Fig.~\ref{fig:Obs}, for the QLSA of the upwind discretisation, we first compute an approximation $\tilde{N}_\psi$ of $N_\psi$ by using amplitude estimation, and then construct the approximate observable with $n_\psi$ replaced by $\tilde{n}_\psi$.

\subsubsection*{The general sampling law}

Since the measurement outcome is probabilistic in general, we have to evaluate the expectation value via sampling. Let $O$ be an observable with $\mu:=\langle O \rangle = \langle \psi| O | \psi \rangle$ being the expectation value, where $\ket{\psi}$ is a quantum state. Suppose that we conduct $n$ experiments with the outcomes labelled $\mu_1, \cdots, \mu_n$. By the law of large numbers
\[{\rm P}_{\text{r}}\Big( \Big|\frac{\mu_1+\cdots + \mu_n}{n} -  \mu\Big| < \varepsilon \Big) \ge 1 - \frac{\text{Var}(O)}{n\varepsilon^2},\]
where $\text{Var}(O)$ is the variance. For a given lower bound $p$, the number of samples required to estimate $\langle O \rangle$ to additive precision $\varepsilon$ satisfies
\[1 - \frac{\text{Var}(O)}{n\varepsilon^2} \ge p \quad \Longrightarrow \quad
n \ge \frac{1}{1-p} \frac{\text{Var}(O)}{\varepsilon^2}.\]
This implies a multiplicative factor $\text{Var}(O)/\varepsilon^2$ in the total gate complexity \cite{Joseph2020KvN,Lin2022Notes}, which is referred to as the ``general sampling law'' in this article.
We remark that in many cases of interest, the number of repetitions can be reduced to $\mathcal{O}(1/\varepsilon)$, up to polylogarithmic factors. For example, quantum algorithms based on amplitude amplification and estimation are able to compute numerical approximations to sums and integrals with a quadratic speedup over classical probabilistic algorithms, so that the number of repetitions of the quantum simulation follows the ``quantum sampling law'' $\mathcal{O}(1/\varepsilon)$, up to polylogarithmic factors \cite{Joseph2020KvN}.
When $O = U$ is a unitary gate acting on the space of $\ket{\psi}$, one can apply the Hadamard test to create a random variable whose expected value is the expected real part $\text{Re} \langle \psi| U | \psi \rangle$ or imaginary part $\text{Im} \langle \psi| U | \psi \rangle$ \cite{AJL2009HadamardTest}. The above sampling law is still valid. Such a technique is also employed in \cite{DLT2022groud,LT2022Heisenberg} for ground-state preparation and energy estimation, where $U = \e^{-\i \tau H}$ is the time evolution operator.
In this paper, we only assume the general sampling law $\mathcal{O}(\text{Var}(O)/\varepsilon^2)$. Noting that
\[ {\rm Var}(O) = (n_Gn_{\psi})^4 {\rm Var} (\mathcal{G}) \lesssim
n_{\psi}^4 , \]
we may need to include this multiplicative factor $n_{\psi}^4$ in the query complexity. Below we will show that in fact we can replace this factor instead by $n_{\psi_0}$ where $n_{\psi_0} = N_{\psi_0}/M^{d/2} = \|\bb{w}^0\|/M^{d/2}$ as defined in \cite{JinLiu2022nonlinear}.

For the QLSA of the upwind discretisation, referring to the linear system \eqref{systemQdiffLiouvilleRepresenation}, and noting that $\|B\|\lesssim 1 + \D t$ and $\|L^{-1}\| = \sigma_{\max}(L^{-1}) \lesssim N_t$, we have
  \[N_\psi = \|\bb{w}\| = \|L^{-1}F\| \le N_t \|F\| \lesssim N_t \|\bb{w}^0\|, \qquad
\mbox{and} \qquad n_\psi \lesssim N_t n_{\psi_0},\]
which gives the multiplicative factor $N_t^4 n_{\psi_0}^4$. However this  $N_t^4$ factor can be removed as addressed in \cite{Berry-2014,Lin2022Notes} by adding $N_t$ copies of the final state $\bb{w}^{N_t}$. That is, we add the following additional equations
\begin{equation}\label{dilation}
\bb{w}^{n+1} - \bb{w}^n = 0, \quad n = N_t, \cdots, 2N_t
\end{equation}
in \eqref{systemQdiffLiouvilleRepresenation}, which is referred to as the dilation procedure. For simplicity, we assume that
$\|\bb{w}^1\| = \cdots = \|\bb{w}^{N_t}\| = \|\bb{w}^0\| = N_{\psi_0}$. Let the padded state vector be
\begin{equation} \label{paddingw}
\widehat{\bb{w}}
 = [\widehat{\bb{w}}^1; \cdots; \widehat{\bb{w}}^{N_t};\widehat{\bb{w}}^{N_t},\cdots, \widehat{\bb{w}}^{N_t}  ]
  = \ket{0} \otimes \bb{x} + \ket{1} \otimes \bb{y},
\end{equation}
where $\ket{0} = [1,0]^T$, $\ket{1} = [0,1]^T$, and the unnormalized vectors are
\[\bb{x} = [\widehat{\bb{w}}^1; \cdots; \widehat{\bb{w}}^{N_t}], \qquad
\bb{y} = [\widehat{\bb{w}}^{N_t},\cdots, \widehat{\bb{w}}^{N_t}  ],\]
satisfying
\[\|\widehat{\bb{w}}^{N_t}\|^2 = \frac{1}{2N_t} = \frac{1}{2N_t N_{\psi_0}^2} \|\bb{w}^{N_t}\|^2.\]
Let us block the matrix $O$ as $(O_{ij})$ according to the structure of $\bb{w}$. One easily finds that $O_{ij} = \bb{O}$ are zeros matrices when $(i,j) \ne (N_t,N_t)$. Then,
\begin{align}
\langle O \rangle
& = \langle \psi | O | \psi \rangle = \frac{1}{N_\psi^2 } \langle \bb{w} | O | \bb{w} \rangle
  = \frac{1}{N_\psi^2 } (\bb{w}^{N_t})^\dag O_{N_t,N_t} \bb{w}^{N_t} \label{expectationNt}\\
& = \frac{2N_t N_{\psi_0}^2}{N_\psi^2 } (\widehat{\bb{w}}^{N_t})^\dag O_{N_t,N_t} \widehat{\bb{w}}^{N_t}
  = \frac{2N_{\psi_0}^2}{N_\psi^2 } (\bb{y})^\dag O_{\bb{y}} \bb{y}^{N_t} \nonumber \\
& = \frac{ N_{\psi_0}^2}{M^{d/2} } \frac{M^{d/2}}{N_\psi^2 } \langle \widehat{\bb{w}} | \widehat{O} | \widehat{\bb{w}} \rangle
  = n_{\psi_0}^2 \frac{1}{n_\psi^2 } \langle \widehat{\bb{w}} | \widehat{O} | \widehat{\bb{w}} \rangle, \nonumber
\end{align}
where $O_{\bb{y}} = \text{diag}(O_{N_t,N_t}, \cdots, O_{N_t,N_t})$, and $\widehat{O} = \text{diag}(\bb{O}, \cdots, \bb{O}, O_{\bb{y}})$. It is evident that $\text{Var}(\widehat{O}) \sim \text{Var}(O)$, and hence $\text{Var}(\widehat{O}/n_{\psi}^2) \lesssim 1$, which implies that the new multiplicative factor is $n_{\psi_0}^4$, as expected. It's worth pointing out that the solution vector in \eqref{paddingw} only requires one ancilla qubit.

\subsubsection*{Boosting the success probability of the final state projection}

It can be seen from \eqref{expectationNt} that the observable is an expectation value taken with respect to the value of the state at the final time. When the solution decays exponentially in time, the success probability of projecting the history state onto the final state is exponentially small.
One can raise the success probability via the amplitude amplification as in \cite{BerryChilds2017ODE,Childs-Liu-2020}. This implies a
multiplicative factor $g = \max_{t \in [0,T]} \| \bb{w}(t) \| /  \| \bb{w}(T) \|$  in the time complexity, which characterises the decay of the final state relative to the initial state.

In conclusion, the additional multiplicative factor in the time complexity is $g n_{\psi_0}^4/\varepsilon^2$ for the computation of the observable. In the subsequent discussion, we will ignore the parameter $g$ for convenience. For the range of $n_{\psi_0}$, please refer to Lemma 14 in \cite{JinLiu2022nonlinear} for some discussions.

\subsubsection{The gate complexity of computing the observable}

The corresponding result for the quantum state is given in Theorem \ref{thm:FDLiouvilleRep}. We now consider the computation of the observable. In this case, we have to discuss the additional cost in estimating the norm of $\bb{w}$.
Let $\langle \tilde{O} \rangle_{\text{app}}$ be the approximate value of $\langle \tilde{O} \rangle$. Then the total error $\text{Err}$ satisfies
\begin{equation}\label{ErrObsLiouville}
\text{Err}:= |\langle O \rangle - \langle \tilde{O} \rangle_{\text{app}} | \le |\langle O \rangle - \langle \tilde{O} \rangle |
+ |\langle \tilde{O} \rangle - \langle \tilde{O} \rangle_{\text{app}}| =: I_1 + I_2,
\end{equation}
where the first term is for the estimation of $N_\psi = \|\bb{w}\|$, and the second one is for the sampling.

We first need to evaluate $N_\psi = \|\bb{w}\|$, with the result described as follows. One can refer to \cite[Theorem 16]{Linden2020heat} and \cite[Corollary 32]{Chakraborty2019blockEncode} for details.

\begin{lemma}[Estimation of $\|A^{-1}b\|$] \label{lem:linearNorm}
Let $Ax = b$ for an $N\times N$ matrix with sparsity $s$ and condition number $\kappa$. Then there exists a quantum algorithm that outputs $\tilde\alpha$ such that
\[|\tilde\alpha - \|x\| | \le \eta \|x\|\]
with probability at least 0.99, in time
\[\mathcal{O}\Big( (T_U + T_b) \frac{\kappa}{\eta}  \log^3 \kappa \log \log \frac{\kappa}{\eta} \Big), \]
where $T_b$ is the time of constructing the state $\ket{b} = \frac{1}{\|b\|}\sum b_i \ket{i}$, and
\[T_U = \log N \Big( \log N +  \log^{2.5} \frac{s \kappa \log(\kappa/\eta) }{\eta}  \Big) \log^2 \frac{\kappa}{\eta}.\]
\end{lemma}

The cost $T_b$ is neglected throughout the paper. With the help of the above result, we are able to bound the gate complexity of computing the observable for the QLSA of the Liouville representation.

\begin{theorem} \label{thm:obsFDLouvilleRep}
Suppose the condition of Theorem \ref{thm:FDLiouvilleRep} is satisfied and $\sup_x |F_i(x) | = \mathcal{O}(1)$ for $i = 1,\cdots, d$. Given the error tolerance $\varepsilon$, if the QLSA for the upwind finite difference discretisation is used, then the observable of the Liouville representation \eqref{LiouvilleRepresentation} can be computed with gate complexity given by
\[N_{\text{Gates}}( \langle O \rangle) = \widetilde{\mathcal{O}}\Big( \frac{n_L^4 d^3}{\varepsilon^5} \log \frac{1}{\varepsilon}\Big),\]
where $n_L = \|(\bb{\rho}^\omega)^0\|/M^{d/2}$.
\end{theorem}
\begin{proof}
(1) For the error $I_1$ in \eqref{ErrObsLiouville}, let $\alpha$ and $\tilde{\alpha}$ be the exact and approximate norms of $\bb{w}$, respectively. Denote $\tilde{n}_{\psi} = \tilde{N}_\psi/M^{d/2} = \tilde{\alpha}/M^{d/2}$ and $n_{\psi} = N_\psi/M^{d/2} = \alpha/M^{d/2}$, where $|\tilde{\alpha} - \alpha| \le \eta \alpha$. Then
\[| (\tilde{n}_{\psi})^2 - (n_{\psi})^2 |
\le \eta n_{\psi} (n_{\psi} + \tilde{n}_{\psi})
\le \eta (2 + \eta) (n_{\psi})^2,\]
and the error
\begin{align*}
I_1 = | (n_G \tilde{n}_{\psi})^2 \Upsilon - (n_Gn_{\psi})^2 \Upsilon |
  \le  \eta (1 + \eta) (n_Gn_{\psi})^2 \Upsilon = \eta (2 + \eta) \langle O \rangle.
\end{align*}
This suggests to take $\eta = \mathcal{O}(\varepsilon)$ since $\langle O \rangle = \mathcal{O}(1)$.

For the error $I_2$, by the general sampling law, we can obtain an approximation $\langle \tilde{O} \rangle_{\text{app}}$ to precision $\varepsilon$ by repeating the quantum algorithm $k = \mathcal{O}(n_{\psi_0}^4/\varepsilon^2)$ times, where $n_{\psi_0} = n_L := \|(\bb{\rho}^\omega)^0\|/M^{d/2}$.

(2) According to Theorem \ref{thm:FDLiouvilleRep}, one has
\[M = 1/\Delta x \sim d/\varepsilon^3, \qquad
\Delta t \sim \Delta x /d , \qquad \kappa \sim 1/\Delta t \sim d^2/\varepsilon^3, \qquad
s \sim d.\]

(3) Let $T_1$ be the gate complexity of obtaining the estimation of $N_\psi$. By Lemma \ref{lem:linearNorm},
\[T_1 = \widetilde{\mathcal{O}}( \kappa/\eta ) = \widetilde{\mathcal{O}}( d^2/\varepsilon^4).\]
Let $T_2$ be the gate complexity of the QLSA. Then,
\[T_2 = \widetilde{\mathcal{O}}\Big( s \kappa \log \frac{1}{\varepsilon}\Big) = \widetilde{\mathcal{O}}\Big( \frac{d^3}{\varepsilon^3} \log \frac{1}{\varepsilon}\Big). \]
The overall gate complexity is
\[T = T_1 + k T_2 = \widetilde{\mathcal{O}}\Big( \frac{n_{\psi_0}^4 d^3}{\varepsilon^5} \log \frac{1}{\varepsilon}\Big).\]
The proof is completed.
\end{proof}

\begin{remark} \label{rem:obsk}
 As observed in the proof, the overall complexity $T$ of the algorithm is dominated by the complexity $kT_2$ of sampling. This implies, when computing the observables, we just need to multiply the original gate complexity under an appropriate mesh strategy by the sampling factor $k = \mathcal{O}({\rm Var}(O)/\varepsilon^2)$. We further remark that the classical cost contains exponential terms in dimension like $d^d$ and $(1/\varepsilon)^d$, which is absent in applications where $n_L$ does not grow so quickly.
\end{remark}

\begin{remark}
Despite the absence of the unitary structure, we can still propose a ``quantum simulation'' algorithm for the Liouville representation as shown in Appendix~\ref{subsect:spectralLiouvillerep} by using the dimensional splitting Trotter based approximation. The basic idea of the algorithm is to transform the asymmetric evolution in each direction into a symmetric one, which requires only a simple variable substitution with the transformation matrix being diagonal. However, unlike the traditional time-marching Hamiltonian simulation, non-unitary procedures for the variable substitution are involved, which leads to exponential increase of the cost arising from multiple copies of initial quantum states at every time step as pointed out in Remark~\ref{rem:multiplecopies}.
\end{remark}

\subsection{Finite difference discretisation for the KvN representation}

\subsubsection{The QLSA for the finite difference discretisation}

We consider the upwind finite difference discretisation for the KvN representation, which can be written as
\begin{equation}\label{KvNExplicit}
\begin{cases}
\partial_t u + \sum\limits_{i=1}^d F_i \frac{\partial u}{\partial x_i}  + \frac{1}{2} ( {\rm div} F) u = 0, \qquad u = \psi^\omega,\\
u(x,0) = \psi_0^\omega.
\end{cases}
\end{equation}
The scheme reads
\begin{align}
u_{\bb{j}}^{n+1}
 - \Big[ 1 - \frac{1}{2}\Delta t ({\rm div} F)_{\bb{j}}
 - \lambda \sum\limits_{\ell=1}^d ( b_{\bb{j}}^{\ell,+} - b_{\bb{j}}^{\ell,-}) \Big] u_{\bb{j}}^n
 + \lambda \sum\limits_{\ell=1}^d \Big[ b_{\bb{j}}^{\ell,-} u_{\bb{j}
 + \bb{e}_\ell}^n - b_{\bb{j}}^{\ell,+} u_{\bb{j} - \bb{e}_\ell}^n\Big] = 0, \label{FDMKvN}
\end{align}
where
\[b_{\bb{j}}^{k,\pm} = \Big\{ F_k \Big\}_{\bb{j}}^\pm, \qquad \bb{j} = (j_1, \cdots, j_d).\]
In matrix form one has
\begin{equation}\label{systemFDKvNRep}
\bb{u}^{n+1} - B \bb{u}^n = 0, \quad n = 0,1,\cdots, N_t-1.
\end{equation}
The final coefficient matrix $L$ is of the same form as in Eq. \eqref{systemQdiffLiouvilleRepresenation}.

\begin{theorem}\label{thm:KvNUpwind}
Suppose $\lambda = \Delta t/\Delta x$ satisfies the following CFL condition
\[\lambda  \sum\limits_{i=1}^d \sup_x |F_i(x) |  \le 1.\]
\begin{enumerate}[(1)]
  \item The condition number and the sparsity of $L$ satisfy $\kappa \lesssim {1}/{\Delta t}$ and $s = \mathcal{O}(d)$.
  \item For fixed spatial step $\Delta x$, let $\Delta t = \mathcal{O}(\Delta x/d)$ and $\omega = (d\Delta x)^{1/3}$. Given the error tolerance $\varepsilon$, the gate complexity of the QLSA is
\[N_{\text{Gates}} = \widetilde{\mathcal{O}}\Big( \frac{d^3}{\varepsilon^3} \log \frac{1}{\varepsilon} \Big).\]
\end{enumerate}
\end{theorem}
\begin{proof}
The proof is similar to the argument in Theorem \ref{thm:FDLiouvilleRep}, so we omit the details.
\end{proof}

\begin{remark}\label{rem:kvntruc}
For the upwind discretisation of the KvN representation, the forcing term $\frac{1}{2} ( {\rm div} F) u$  will contribute to the $l_1$ error an exponentially growing term like $\e^{t_n \|\text{div} F\|_\infty}$. In fact, one easily obtains from \eqref{FDMKvN} that
\begin{align*}
\|\bb{u}^{n+1}\|_1
& \le \Big(1 + \frac{1}{2} \D t \|\text{div} F\|_\infty \Big) \|\bb{u}^n\|_1
\le \Big(1 + \frac{1}{2} \D t \|\text{div} F\|_\infty \Big)^n \|\bb{u}^0\|_1 \\
& = \Big(1 + \frac{1}{2} \frac{t_n}{n} \|\text{div} F\|_\infty \Big)^n \|\bb{u}^0\|_1
  \le \e^{t_n \|\text{div} F\|_\infty}\|\bb{u}^0\|_1.
\end{align*}
As a comparison, if one uses the Liouville equation, the $l_1$ norm of the error is contracting, as shown in Remark \ref{Rmk2.3}.
\end{remark}

\subsubsection{The gate complexity of computing the observable}

Let $n = N_t$ for convenience. For the KvN approximation, from \eqref{Opsi} we know that the observable at $t = t_n$ is
\[\langle O_{\psi^\omega,n}\rangle = \frac{1}{M^{d/2}}(\bb{\psi}_n^\omega)^\dag G_M \bb{\psi}_n^\omega \approx \frac{1}{M^{d/2}}(\bb{u}^n)^\dag G_M \bb{u}^n
 = \frac{1}{M^{d/2}}\bb{u}^\dag \mathcal{G}_M \bb{u}, \]
where $\bb{u}$ is the solution of \eqref{systemFDKvNRep}, and $\mathcal{G}_M = \text{diag}(\bb{O}, \cdots, \bb{O}, G_M)$. Let
\[\ket{\psi} = \frac{1}{N_{\psi}} \sum\limits_{\bb{j},n} \bb{u}_{\bb{j},n} \ket{\bb{j}} \ket{n}
,\]
where the normalisation $N_{\psi} = \|\bb{u}\|$. The expectation of the observable can be defined as
\[ \langle O_{\psi^\omega,n}\rangle = \frac{1}{M^{d/2}}\langle \psi | O | \psi \rangle =:\langle O \rangle , \qquad O = \frac{N_\psi^2}{M^{d/2}}  \mathcal{G}_M.\]
One easily finds from \eqref{Opsi} that $\text{Var}(\mathcal{G}_M)$ is bounded since
\[\text{Var}(\mathcal{G}_M) = \langle \mathcal{G}_M^2 \rangle - \langle \mathcal{G}_M \rangle ^2 \le  \|\mathcal{G}_M \psi\|^2 \lesssim 1.\]

Following the similar analysis in Subsect.~\ref{subsubsect:obsLiouvilleRep} for the Liouville representation, one obtains the multiplicative factor in the time complexity can be given by $n_K^4/\varepsilon^2$, where $n_K = \|\bb{u}^0\|/M^{d/4} = \|(\bb{\psi}^\omega)^0\|/M^{d/4}$, where we have omitted the parameter $g$ which characterises the decay of the final state relative to the initial state.
The arguments for computing the observable of the Liouville representation also apply to the KvN representation. The corresponding result is described in the following theorem.

\begin{theorem} \label{thm:obsFDKvNRep}
Suppose the condition of Theorem \ref{thm:KvNUpwind} is satisfied and $\sup_x |F_i(x) | = \mathcal{O}(1)$ for $i = 1,\cdots, d$. Given the error tolerance $\varepsilon$, if the QLSA for the upwind finite difference discretisation is used, then the observable of the KvN representation can be computed with gate complexity given by
\[N_{\text{Gates}}( \langle O \rangle) = \widetilde{\mathcal{O}}\Big( \frac{n_K^4 d^3}{\varepsilon^5} \log \frac{1}{\varepsilon}\Big),\]
where $n_K = \|(\bb{\psi}^\omega)^0\|/M^{d/4}$.
\end{theorem}
\begin{proof}
The KvN representation has the same error estimate as the Liouville representation, which leads to the same mesh strategy. From Theorem \ref{thm:KvNUpwind}, we also observe the same condition number and sparsity for the associated coefficient matrix. We therefore obtain the same gate complexity with the multiplicative factor replaced by $n_K^4/\varepsilon^2$.
\end{proof}

\begin{remark}
In view of the relation \eqref{rhopsi2}, one easily finds that
\[n_L = \frac{\|(\bb{\rho}^\omega)^0\|}{M^{d/2}} = \frac{\|(\bb{\psi}^\omega)^0\|^2}{M^{d/2}} =  n_K^2.\]
\end{remark}

\subsection{Spectral discretisation for the KvN representation}

The KvN representation can be solved by quantum Hamiltonian simulation directly since the evolutionary operator is Hermitian, where the Hamiltonian simulation can be realised by the quantum version of the classical Fourier spectral method. On the other hand, we can also develop the QLSA based method for the spectral discretisation.

\subsubsection{The notations} \label{subsubsect:notation}

We consider the Fourier spectral discretisation. To this end, we first introduce some notations frequently used in this article.

For one-dimensional problems we choose a uniform spatial mesh size $\Delta x = 1/M$ for $M=2N = 2^m$ with $m$
an positive integer and the time step $\Delta t$, and we let the grid points and the time step be
\[x_j = j \Delta x, ~~ t_n = n \Delta t, \quad j = 0,1,\cdots, N,~~ n = 0,1,\cdots.\]
We consider the periodic boundary conditions.
For $x\in [0,1]$, the 1-D basis functions for the Fourier spectral method are usually chosen as
\[\phi_l(x) = \e ^{\i \mu_l x} , \quad \mu_l = 2\pi l, \quad  1 = -N,\cdots, N-1.\]
For convenience, we adjust the index as
\[\phi_l(x) = \e ^{\i \mu_l x} , \quad \mu_l = 2\pi (l-N-1), \quad 1 \le l \le M = 2N.\]
The approximation in the 1-D space is
\begin{equation}\label{Fexpand}
u(t,x) = \sum\limits_{l=1}^M c_l(t) \phi_l(x), \quad x = x_j, ~~ j = 0,1,\cdots, M-1.
\end{equation}
 which can be  written in vector form,
$\bb{u}(t) = \Phi \bb{c}(t)$,
where
\[\bb{u}(t) = (u(t,x_j))_{M\times 1}, \quad \bb{c} = (c_l)_{M\times 1}, \quad
\Phi = (\phi_{jl})_{M\times M} = (\phi_l(x_j))_{M\times M}.\]

The $d$-dimensional grid points are then given by ${x}_{\bb{j}} = (x_{j_1}, \cdots, x_{j_d})$, where $\bb{j} = (j_1,\cdots,j_d)$, and
\[x_{j_i} = j_i \Delta x, \quad j_i = 0,1,\cdots, M-1, \quad i = 1,\cdots,d.\]
We use the notation $ 1\le \bb{j} \le M$ to indicate $1\le j_i \le M$ for every component of $\bb{j}$.
The multi-dimensional basis functions are written as
$\phi_{\bb{l}}({x}) = \phi_{l_1}(x_1)\cdots \phi_{l_d}(x_d)$,
where $\bb{l} = (l_1,\cdots, l_d)$ and $1 \le \bb{l}\le M$. The corresponding approximate solution is
$u(t,{x}) = \sum\nolimits_{\bb{l}} c_{\bb{l}}(t) \phi_{\bb{l}}({x})$,
with the coefficients determined by the exact values at the grid or collocation points ${x}_{\bb{j}} $. These collocation values will be arranged as a column vector:
\[\bb{u}(t) = \sum\limits_{\bb{j}}u(t,{x}_{\bb{j}}) \ket{j_1} \otimes \cdots \otimes \ket{j_d}.\]
That is, the $n_{\bb{j}}$-th entry of $\bb{u}$ is $u(t,{x}_{\bb{j}})$, with the global index given by
\[n_{\bb{j}}: = j_12^{d-1} + \cdots + j_d2^0, \qquad \bb{j} = (j_1,\cdots,j_d). \]
Similarly $c_{\bb{l}}$ is written in a column vector as $\bb{c} = \sum\nolimits_{\bb{l}} c_{\bb{l}} \ket{l_1} \otimes \cdots \otimes \ket{l_d}$.

To determine the transformation matrix between $\bb{u}$ and $\bb{c}$, let $c_{\bb{l}} = c_{l_1}\cdots c_{l_d}$.
Then
\begin{equation}\label{Fexpandhigh}
u(t,{x}_{\bb{j}}) = \sum\limits_{\bb{l}} c_{l_1}\cdots c_{l_d} \phi_{l_1}(x_{j_1})\cdots\phi_{l_d}(x_{j_d}).
\end{equation}
The direct calculation gives
\begin{align*}
 \sum\limits_{\bb{j}}u(t,{x}_{\bb{j}}) \ket{j_1} \otimes \cdots \otimes \ket{j_d}
= \Big(\sum\limits_{j_1,l_1} c_{l_1}\phi_{l_1} (x_{j_1}) \ket{j_1} \Big) \otimes  \cdots \otimes \Big(\sum\limits_{j_d, l_d} c_{l_d}\phi_{l_d} (x_{j_1}) \ket{j_d} \Big),
\end{align*}
which implies
\begin{align*}
\bb{u}
 = ( \Phi \bb{c}^{(1)}) \otimes \cdots \otimes ( \Phi \bb{c}^{(d)})
  = ( \Phi   \otimes \cdots \otimes \Phi ) ( \bb{c}^{(1)} \otimes \cdots \otimes \bb{c}^{(d)} )
 = \Phi^{\otimes ^d } \bb{c},
 \end{align*}
where
\[\Phi^{\otimes ^d } = \underbrace{\Phi \otimes \cdots \otimes \Phi}_{\text{$d$ matrices}},\qquad
\bb{c}^{(i)} = ( c_{l_i} )_{M\times 1},\]
\begin{equation}\label{cdef}
\bb{c} = \bb{c}^{(1)} \otimes \cdots \otimes \bb{c}^{(d)} = \sum\limits_{\bb{l}} c_{\bb{l}} \ket{l_1} \otimes \cdots \otimes \ket{l_d}.
\end{equation}
This shows that by arranging ${x}_{\bb{j}}$ in the order of $\ket{j_1} \otimes \cdots \otimes \ket{j_d}$, and ${c}_{\bb{l}}$ in the order of $\ket{l_1} \otimes \cdots \otimes \ket{l_d}$, the corresponding coefficient matrix is exactly the tensor product of the matrices in one dimension.

For later use, we next determine the transitions between the position operator $\hat{x}_j$ and the momentum operator $\hat{P}_j = -\i \frac{\partial}{\partial x_j}$ in discrete settings.

We first consider the one-dimensional case.
Let $u(x)$ be a function in one dimension and $\bb{u} = [u(x_0),\cdots,u(x_{M-1})]^T$ be the mesh function with $M=2N$. The discrete position operator $\hat{x}^{\rm d}$ of $\hat{x}$ can be defined as
\[\hat{x}^{\rm d}: \bb{u} = \Big( u(x_i) \Big) \quad \to \quad
  \Big( x_i u(x_i) \Big) = D_x \bb{u}  \qquad \mbox{or} \qquad
  \hat{x}^{\rm d}\bb{u} = D_x \bb{u},\]
where $D_x = \text{diag} ( x_0, x_1, \cdots, x_{M-1} )$ is the matrix representation of the position operator in $x$-space. By the discrete Fourier expansion in \eqref{Fexpand}, the momentum operator can be discretised as
\begin{align*}
\hat{P}u(x)
& \approx \hat{P} \sum\limits_{l=1}^M c_l \phi_l(x) = \sum\limits_{l=1}^M c_l \hat{P} \phi_l(x)
  = \sum\limits_{l=1}^M c_l (-\i \partial_x \phi_l(x)) \\
& = \sum\limits_{l=1}^M c_l \mu_l \phi_l(x),  \quad \mu_l = 2\pi (l-N-1)
\end{align*}
for $x = x_j$, $j = 0,1,\cdots, M-1$, which is written in matrix form as
\[\hat{P}^{\rm d} \bb{u} =  \Phi D_\mu \Phi^{-1} \bb{u} =: P_x \bb{u}, \qquad
D_\mu = \text{diag} ( \mu_1, \cdots, \mu_M ),\]
where $\hat{P}^{\rm d}$ is the discrete momentum operator. The matrices $D_\mu$ and $P_x$ can be referred to as the matrix representation of the momentum operator in $p$-space and $x$-space, respectively, and are related by the discrete Fourier transform.

For $d$ dimensions, we still denote
$\bb{u} = \sum\nolimits_{\bb{j}} u(x_{\bb{j}}) \ket{j_1} \cdots \ket{j_d}$.
Let
\[u(x_{\bb{j}}) = u(x_{j_1},\cdots,x_{j_d}) = u^{(1)}(x_{j_1})\cdots  u^{(d)}(x_{j_d}),\]
where $\bb{u}^{(l)} = \Phi \bb{c}^{(l)}$. One has $\bb{u} = \bb{u}^{(1)} \otimes \cdots \otimes \bb{u}^{(d)}$.
The discrete position operator $\hat{x}_l^{\rm d}$ is defined as
\[\hat{x}_l^{\rm d}: \bb{u} = \bb{u}^{(1)} \otimes \cdots \otimes \bb{u}^{(d)} \quad \to \quad
\bb{u}^{(1)} \otimes \cdots \otimes \tilde{\bb{u}}^{(l)} \otimes \cdots \otimes \bb{u}^{(d)}, \]
where
\[\tilde{\bb{u}}^{(l)}: = \Big( x_{j_{l_i}} u^{(l)} (x_{j_{l_i}}) \Big) = D_x \bb{u}^{(l)}.\]
Then,
\[\hat{x}_l^{\rm d} \bb{u} = (I^{\otimes^{l-1}} \otimes D_x \otimes I^{\otimes^{d-l}}) \bb{u} =: \bb{D}_l \bb{u}. \]
Using the expansion in \eqref{Fexpandhigh}, one easily finds that
\[\hat{P}_l^{\rm d} \bb{u} = (I^{\otimes^{l-1}} \otimes P_x \otimes I^{\otimes^{d-l}}) \bb{u} =: \bb{P}_l \bb{u}. \]
Note that
\begin{equation}\label{PlDl}
(\Phi^{\otimes^d})^{-1} \bb{P}_l \Phi^{\otimes^d} = I^{\otimes^{l-1}} \otimes D_\mu \otimes I^{\otimes^{d-l}}=:\bb{D}^\mu _l.
\end{equation}

\subsubsection{The QLSA for the spectral discretisation} \label{subsubsect:QLSAKvNSpectral}

Let us consider the matrix representation of the operator $\hat{H}_j$ in \eqref{KvNOperator}, where
\[\hat{H}_j =\frac{1}{2} (  F_j(\hat{x}) \hat{P}_j  + \hat{P}_j  F_j(\hat{x}) ).\]
For clarity, we still use $\bb{u}$ to denote the mesh function of $\psi$ (see the notations in Subsect.~\ref{subsubsect:notation}). When performing series expansion on $F$, one has
\begin{align*}
F_j(\hat{x}^{\rm d}) \bb{u}
& : = \sum\limits_{\bb{l}} a_{\bb{l}} (\hat{x}_1^{\rm d})^{l_1} \cdots (\hat{x}_d^{\rm d})^{l_d} \bb{u}
 =  \sum\limits_{\bb{l}} a_{\bb{l}}(\bb{D}_1^{l_1} \cdots \bb{D}_d^{l_d} ) \bb{u} \\
& = \sum\limits_{\bb{l}} a_{\bb{l}} (D_x^{l_1}\otimes \cdots \otimes D_x^{l_d}) \bb{u} =: \bb{F}_j \bb{u}
\end{align*}
in the discrete setting, where $\bb{F}_j$ is clearly a diagonal matrix. We assume that the series expansion is accurate enough to simplify the discussion. Then one has
\begin{align*}
\hat{H}_j^{\rm d} \bb{u}
& = \frac{1}{2} (  F_j(\hat{x}^{\rm d} ) \hat{P}_j^{\rm d}  + \hat{P}_j^{\rm d}  F_j(\hat{x}^{\rm d}) ) \bb{u}
  = \frac{1}{2} (  \bb{F}_j \bb{P}_j  + \bb{P}_j  \bb{F}_j ) \bb{u} 
 =: \bb{H}_j \bb{u}.
\end{align*}
One easily finds that the sparsity of $\bb{H}_j$ is $\mathcal{O}(M)$. The resulting system of ordinary differential equations is
\begin{equation}\label{SpectralSystemKvNRep}
\begin{cases}
\frac{\d }{\d t} \bb{u}(t) = A \bb{u}(t), \qquad A = - \i \sum\limits_{j=1}^d \bb{H}_j,\\
\bb{u}(0) = ( \psi^\omega(0, x_{\bb{j}}) ).
\end{cases}
\end{equation}
The analytic solution is obviously given by
\[\bb{u}(t) = \e^{A t} \bb{u}(0) = \exp \Big( - \i \sum\limits_{j=1}^d \bb{H}_j \Big) \bb{u}(0),\]
which implies $\|\bb{u}(t)\| = \|\bb{u}(0)\|$ for any $t\ge 0$ since $\bb{H}_j$ are real symmetric matrices.  Let $n = N_t$ and denote
\[\ket{\psi} =  \frac{1}{N_\psi} \sum\limits_{\bb{j}} \bb{u}_{\bb{j}}^n \ket{\bb{j}}, \qquad N_\psi = \|\bb{u}^n\|.\]
Since $N_\psi = \|\bb{u}^n\| = \|\bb{u}^0\|= : N_{\psi_0}$, the observable can be reformulated as
\[\langle O_{\psi^\omega,n}\rangle = \frac{1}{M^{d/2}}(\bb{\psi}_n^\omega)^\dag G_M \bb{\psi}_n^\omega \approx \frac{1}{M^{d/2}}(\bb{u}^n)^\dag G_M \bb{u}^n
 = \frac{N_{\psi_0}^2}{M^{d/2}} \langle \psi | G_M | \psi \rangle  =: \langle \psi | O | \psi \rangle. \]

The ODEs in \eqref{SpectralSystemKvNRep} can be solved by the quantum differential equations solver reported in \cite{Berry-2014,BerryChilds2017ODE,Childs-Liu-2020}. Here we consider the one in \cite{BerryChilds2017ODE} with the result described below. For convenience, we still refer  it to as the QLSA based method since the approach in \cite{BerryChilds2017ODE} applies the QLSA.

\begin{lemma} \label{lem:BCO2017}
Suppose $A = V^{-1}DV$ is an $N\times N$ diagonalizable matrix, where $D = \text{diag}(\lambda_1,\cdots,\lambda_N)$ satisfies $\text{Re} (\lambda_j) \le 0$ for any $j \in \{1,\cdots,N\}$. In addition, suppose $A$ has at most s nonzero entries in any row and column, and we have an oracle $O_A$ that computes these entries. Suppose $x_{in}$ and $b$ are $N$-dimensional vectors with known norms and that we have two controlled oracles, $O_x$ and $O_b$, that prepare the states proportional to $x_{in}$ and $b$, respectively. Let $x$ evolve according to the differential equation
\[\frac{\d x}{\d t} = A x + b\]
with the initial condition $x(0) = x_{in}$. Let $T>0$ and $g = \max_{t \in [0,T]} \| x(t) \| /  \| x(T) \|$.
Then there exists a quantum algorithm that produces a state $\epsilon$-close to $x(T)/\|x(T)\|$ in $l^2$ norm, succeeding with probability $\Omega(1)$, with a flag indicating success, using
\[\mathcal{O}\Big( s \kappa_V \|A\| gT \cdot \text{Poly}( \log (s \kappa_V \|A\| gT \beta/\varepsilon) )\Big)\]
queries to $O_A$, $O_x$ , and $O_b$,  where $\kappa_V$ is the condition number of the transformation matrix $V$,
$g$ characterises the decay of the final state relative to the initial state, and  $\beta = ( \| x_{in}\| + T \|b\| )/ \|x(T)\|$. The gate complexity of this algorithm is larger than its query complexity by a factor of $\text{Poly}( \log (s N \kappa_V \|A\| gT \beta/\varepsilon) )$.
\end{lemma}

Note that the parameter $g$ can be dropped if we only output the quantum state $\ket{x}$, not the projection $\ket{x(T)}$.
For the approximate evolutionary operator in \eqref{Wdelta}, we are ready to quantify the gate complexity for the QLSA.

\begin{remark} \label{rem:ODEs}
The authors in \cite{BerryChilds2017ODE} utilised the matrix exponential to construct a linear system for the ODEs and solved the linear system by using the QLSA proposed in \cite{Costa2021QLSA}. As claimed in \cite{Costa2021QLSA}, the gate complexity exceeds the query complexity by a multiplicative factor $\mathcal{O}(\log N +\log^{2.5}(s\kappa/\varepsilon))$, where $N = \mathcal{O}(M^d)$ is the order of the matrix $A$. This implies the linear dependence of the dimension $d$ when considering the gate complexity with respect to the matrix order.
\end{remark}

\begin{theorem} \label{thm:QLSAKvN}
Assume that $\max_{1\le j \le d} \|\bb{F}_j\| = \mathcal{O}(1)$ and $T = \mathcal{O}(1)$.
\begin{enumerate}[(1)]
  \item There exists a quantum algorithm that produces a state $\varepsilon$-close to $\bb{u}(T)/\|\bb{u}(T)\|$ with the gate complexity given by
\[N_{Gates} = \widetilde{\mathcal{O}}\Big( \frac{d^{2+2/\ell}}{\varepsilon^{2+4/\ell}} \Big).\]
  \item The observable of the KvN representation can be computed with gate complexity given by
\[N_{\text{Gates}}( \langle O \rangle) =  \widetilde{\mathcal{O}}\Big( \frac{n_K^4 d^{2+2/\ell}}{\varepsilon^{4+4/\ell}} \Big),\]
where $n_K = \|(\bb{\psi}^\omega)^0\|/M^{d/4}$.
\end{enumerate}

\end{theorem}
\begin{proof}
(1) Let $A = V^{-1} D V$ with $D = \text{diag}(\lambda_1,\cdots,\lambda_N)$, where $N = \mathcal{O}(M^d)$ is the order of $A$. Since $\bb{H}_j$ are real symmetric matrices, the matrix $\sum_{j=1}^d \bb{H}_j$ has only real eigenvalues and the transformation matrix $V$ can be chosen as an orthogonal matrix. This implies $\text{Re} (\lambda_j) = 0$ for any $j \in \{1,\cdots,N\}$ and $\kappa_V = 1$.
According to Lemma \ref{lem:BCO2017}, there exists a quantum algorithm that produces a state $\varepsilon$-close to $\bb{u}(T)/\|\bb{u}(T)\|$ with the gate complexity given by
\[N_{Gates} = \mathcal{O}\Big( s \kappa_V \|A\| T \cdot \text{Poly}( \log (s  N \kappa_V \|A\| T/\varepsilon) )\Big),\]
where we have omitted the parameter $g$ that characterises the decay of the final state relative to the initial state.

It is evident that the sparsity of $A$ is $\mathcal{O}(dM) = \mathcal{O}(d/\D x)$. The norm of $A$ satisfies
\begin{align*}
\|A\|
& \le  \sum\limits_{j=1}^d \| \bb{H}_j\| \le \sum\limits_{j=1}^d  \|\bb{F}_j\| \| \bb{P}_j \|
   \le \sum\limits_{j=1}^d  \|\bb{F}_j\| \| \bb{D}_j^\mu \|  \\
& \le  M \sum\limits_{j=1}^d  \|\bb{F}_j\|  \le  dM \max_{1\le j \le d} \|\bb{F}_j\| = d /\D x \cdot \max_{1\le j \le d} \|\bb{F}_j\|.
\end{align*}

 According to Lemma \ref{lem:errKvN}, the error of the spectral discretisation is $\mathcal{O}(\omega + \D t^\alpha/\omega^{\alpha} + d\Delta x^{\ell}/\omega^{\ell+1})$, where $\alpha$ is for the precision of the temporal discretisation which has been considered in the quantum algorithm in \cite{BerryChilds2017ODE}.
To reach a precision of $\varepsilon$, one just needs to set $\omega \sim d\Delta x^{\ell}/\omega^{\ell+1} \sim \varepsilon$, and gets $\Delta x \sim \varepsilon^{1 + 2/\ell}/d ^{1/\ell}$. Therefore, we have
\[N_{Gates} = \widetilde{\mathcal{O}}\Big( \frac{d^2}{\D x^2}  \Big)
= \widetilde{\mathcal{O}}\Big( \frac{d^{3+2/\ell}}{\varepsilon^{2+4/\ell}}  \Big).\]
where in the last equal sign we have included the additional factor $d$ arising from the matrix order (see Remark \ref{rem:ODEs}).

(2) For the spectral discretisation, the constant $N_{\psi}$ is known. The desired estimate follows from the general sampling law.
\end{proof}

\subsubsection{The quantum simulation for the spectral discretisation}

The ODEs \eqref{SpectralSystemKvNRep} can also be solved by quantum Hamiltonian simulations.

\begin{theorem} \label{thm:KvNsimulation}
Given the error tolerance $\varepsilon$, assume that $\max_{1\le j \le d} \|\bb{F}_j\|_{\infty} = \mathcal{O}(1)$ and the simulation time $t = \mathcal{O}(1)$.
\begin{enumerate}[(1)]
  \item  The semi-discrete problem \eqref{SpectralSystemKvNRep} obtained from the spectral discretisation of the KvN representation can be simulated with gate complexity given by
      \[N_{Gates} = \widetilde{\mathcal{O}}\Big( \frac{d^{2+2/\ell}}{\varepsilon^{3+4/\ell}} \Big).\]
  \item The observable of the KvN representation can be computed with gate complexity given by
\[N_{\text{Gates}}( \langle O \rangle) =  \widetilde{\mathcal{O}}\Big( \frac{n_K^4 d^{3+2/\ell}}{\varepsilon^{4+4/\ell}} \Big),\]
where $n_K = \|(\bb{\psi}^\omega)^0\|/M^{d/4}$.
\end{enumerate}
\end{theorem}
\begin{proof}
(1) Let $H = \sum_{j=1}^d \bb{H}_j$. Then the evolution of \eqref{SpectralSystemKvNRep} can be written as $\ket{\psi(t)} = \e^{-\i H   t } \ket{\psi(0)}$.
According to Theorem 1 in \cite{Berry-Childs-Kothari-2015}, $\e^{-\i H t}$ can be simulated within error $\varepsilon$ with
\[\mathcal{O}\Big(  \tau ( m_d + \log^{2.5}(\tau/\varepsilon) )\frac{\log (\tau/\varepsilon) }{\log\log (\tau/\varepsilon)} \Big)
= \mathcal{O}( \tau m_d \cdot \text{polylog})\]
2-qubits gates, where $\tau = s \|H\|_{\max} t$, $s$ is the sparsity of $H$ and $\|H\|_{\max}$ denotes the largest entry of $H$ in absolute value, and
\[ \text{polylog}  \equiv  \log^{2.5}(\tau/\varepsilon) \frac{\log (\tau/\varepsilon) }{\log\log (\tau/\varepsilon)}.\]
 This result is near-optimal by Theorem 2 therein.

The sparsity of $H$ is $s = \mathcal{O}(dM)$. According to the proof of Theorem \ref{thm:QLSAKvN}, the mesh strategy is
$M = 1/\Delta x = d ^{1/\ell}/\varepsilon^{1 + 2/\ell}$, and hence the number of qubits per dimension is
\[m = \mathcal{O}(\log M) = \mathcal{O} \Big( \log \frac{d^{1/\ell}}{\varepsilon^{1+2/\ell}} \Big). \]
The total number of qubits is $m_d = d m$. With these settings, noting that
\[H_{\max} \le \sum \limits_{j=1}^d \|\bb{H}_j\|_{\infty} \lesssim  M \sum \limits_{j=1}^d \|\bb{F}_j\|_{\infty}
\le d M \cdot \max_{1\le j \le d} \|\bb{F}_j\|_{\infty}, \]
one has
\[\tau = \mathcal{O} \Big(\frac{d^{2 + 2/\ell}}{\varepsilon^{2+4/\ell}} \Big), \qquad
\tau/\varepsilon = \mathcal{O} \Big(\frac{d^{2+2/\ell}}{\varepsilon^{3+4/\ell}} \Big).\]
The gate complexity for solving the ODEs is then given by
\[N_{Gates} = \widetilde{\mathcal{O}}\Big( \frac{d^{3+2/\ell}}{\varepsilon^{2+4/\ell}} \Big).\]

(2) The gate complexity for computing the observable is obtained from the general sampling law.
\end{proof}

One can also run the simulation along each direction by using the Trotter based approximation.
The evolution of \eqref{KvNOperator} can be written as
\[\ket{\psi(t+\Delta t)} = \e^{-\i (\hat{H}_1 + \cdots + \hat{H}_d) \Delta t } \ket{\psi(t)}.\]
Let
\begin{equation}\label{Udelta}
U_{\Delta t} = \e^{-\i \hat{H}_d \Delta t}\cdots \e^{-\i \hat{H}_1 \Delta t}.
\end{equation}
One has \cite{Nielsen2010,Childs2021Trotter}
\begin{equation}\label{CF0}
\e^{-\i (\hat{H}_1 + \cdots + \hat{H}_d) \Delta t } = U_{\Delta t} + C_H \Delta t^2,
\end{equation}
where $C_H$ depends on the operator $\hat{H}= \hat{H}_1 + \cdots + \hat{H}_d$ or the matrix $H$, considered as $\mathcal{O}(1)$ in the following. Therefore, the problem is reduced to the simulation of each $\hat{H}_j$.

\begin{remark}
One can clearly make the time discretization second order by using Strang's splitting. Since other methods use first order time discretization, in order to compare the time complexities on equal footing we also use first order time discretization, namely the simple splitting, here.
\end{remark}

First, we determine the mesh strategy. According to the error estimate in Lemma \ref{lem:errKvN} and noting Eq.~\eqref{CF0}, one has the error estimate
\[e_{\psi}  \le C(\omega + \Delta t/\omega +  d\Delta x^{\ell}/\omega^{\ell+1}).\]
The above error bounds suggest the following mesh strategy:
\begin{equation}\label{meshKvN}
M = 1/\Delta x =  \mathcal{O}(d ^{1/\ell}/\varepsilon^{1 + 2/\ell} ), \qquad  \Delta t  \sim \varepsilon^2.
\end{equation}

Second, we quantify the number of gates used in the quantum simulation.  According to Theorem 1 in \cite{Berry-Childs-Kothari-2015}, $\e^{-\i \hat{H}_j \Delta t}$ can be simulated within error $\eta$ with
\[\mathcal{O}\Big(  \tau ( m_d + \log^{2.5}(\tau/\eta) )\frac{\log (\tau/\eta) }{\log\log (\tau/\eta)} \Big)\]
2-qubits gates, where $\tau = s \|\bb{H}_j\|_{\max} \Delta t$, $s$ is the sparsity of $\bb{H}_j$ and $\|\bb{H}_j\|_{\max}$ denotes the largest entry of $\bb{H}_j$ in absolute value. One can check that the sparsity of $\bb{H}_j$ is $s = \mathcal{O}(M)$. Therefore, $U_{\Delta t}$ defined in \eqref{Udelta} can be simulated within error $\mathcal{O}( d \eta)$ \cite[Proposition 1.12]{Lin2022Notes} with
\[N_{\text{Gates}}(U_{\Delta t}) = \mathcal{O}(d \tau m_d \cdot \text{polylog}),\]
where
\[\tau = M \tilde{H}_{\max} \Delta t,  \qquad \tilde{H}_{\max} = \max_j \|\bb{H}_j\|_{\max} \lesssim M \max_j \|\bb{F}_j\|_{\max},\]
\[ \text{polylog} = \log^{2.5}(\tau/\eta) \frac{\log (\tau/\eta) }{\log\log (\tau/\eta)}.\]
We also need $d \eta = \mathcal{O}(\Delta t^2)$ or $\eta = \mathcal{O}(\varepsilon^4/d)$, and the number of qubits per dimension is
$m = \mathcal{O} ( \log (d^{1/\ell}/\varepsilon^{1+2/\ell}))$. The total number of qubits is $m_d = d m$. With these settings, we obtain
\[\tau = \mathcal{O} \Big(\D t \frac{d^{2/\ell}}{\varepsilon^{2+4/\ell}} \Big), \qquad
\tau/\eta = \mathcal{O} \Big(\frac{d^{1+2/\ell}}{\varepsilon^{6+4/\ell}} \Big),\]
and the total number of gates required to iterate to the $n$-th step is
\begin{align*}
N_{\text{Gates}}
= n N_{\text{Gates}}(U_{\Delta t})
= \widetilde{\mathcal{O}}\Big(\frac{d^{2+2/\ell}}{\varepsilon^{2+4/\ell}} \Big).
\end{align*}
Compared with the one-step implementation in Theorem \ref{thm:KvNsimulation}, the complexity is improved in sparsity.

\begin{remark}
From time $t=t_n$ to time $t = t_{n+1}$, we solve
$\bb{u}^{n+1} = U_{\Delta t}^{\rm d} \bb{u}^n$,
where $U_{\Delta t}^{\rm d}$ is the discrete version of $U_{\Delta t}$ in \eqref{Udelta}, given by
$U_{\Delta t}^{\rm d} = \e^{-\i \hat{H}_d^{\rm d} \Delta t}\cdots \e^{-\i \hat{H}_1^{\rm d} \Delta t}$.
According to the previous discussions, one has $U_{\Delta t}^{\rm d} \bb{u}^n = B \bb{u}^n$,
with $ B = \e^{-\i \bb{H}_d \Delta t}\cdots \e^{-\i \bb{H}_1 \Delta t} $.
One can alternatively solve a linear system $LU = F$ in the form of Eq.~\eqref{systemQdiffLiouvilleRepresenation}, which, however, is not a suitable algorithm because the sparsity of $L$ grows exponentially with the number of dimensions, i.e., $s(L) = \mathcal{O}(M^d)$.
\end{remark}

\section{Liouville representation for nonlinear Hamilton-Jacobi PDEs: finite difference vs. spectral approximations} \label{sect:semiclassical}

In \cite{JinLiu2022nonlinear}, the level set method was used to map the nonlinear Hamilton-Jacobi equation into linear Liouville equation in the phase space, based on which quantum algorithms were then constructed.

Hamilton-Jacobi equations take the following general form
\begin{align}\label{H-J}
& \partial_t S+ H(\nabla S, x) = 0,\\
& S(0,x)=S_0(x)\nonumber
\end{align}
with  $t\in \mathbb{R}^+$,  $x \in \mathbb{R}^d$, $S(t,x)\in \mathbb{R}$. Define $u=\nabla S \in \mathbb{R}^d$. Then $u$ solves a hyperbolic system of conservation laws in gradient form:
\begin{align}\label{forced-Burgers}
 & \partial_t u + \nabla H(u, x) = 0,\\
 &u(0,x)=\nabla S_0(x).\nonumber
\end{align}

The level set function $\phi_i(t,x,p)$ can be defined by
\begin{align}\label{LS-def}
\phi_i(t, x,p=u(t,x))=0, \nonumber
\end{align}
where $i=1, \cdots, d$ and $\, x, p\in \mathbb{R}^d$, and $u(t,x)$ is the solution of Eq.~\eqref{forced-Burgers}. The \textit{zero level set} of $\phi$ is the set  $\{(t,x,p)|\phi_i(t,x,p)=0\}$. Since $u(t,x)$ solves Eq.~\eqref{forced-Burgers}, one can show that $\phi=(\phi_1, \cdots, \phi_d)\in \mathbb{R}^d$ solves a (linear!) Liouville equation \cite{JinOsher2003levelset}
\begin{equation} \label{LS-Liouville}
\partial_t \phi + \nabla_p H \cdot \nabla_x \phi - \nabla_x H \cdot \nabla_p \phi=0.
\end{equation}
 The initial data can be chosen as
 \begin{equation}\label{Liou-IC}
 \phi_i(0, x,p)=p_i - u_i(0,x), \quad i=1, \cdots, d.
 \end{equation}
 Then $u$ can be recovered from the intersection of the zero level
sets of $\phi_i \, (i=1, \cdots, d)$, namely
\[u(t,x)=\{p(t,x)| \,\phi_i(t,x,p)=0, \, i=1,\cdots, d \}.\]

 To retrieve physical observables (and to avoid finding the zero level set of $\phi$ which is challenging) later, \cite{JinLiu2022nonlinear} proposed to  solve for $\psi$, defined by the following problem
\begin{align} \label{Liouville-delta}
& \partial_t \psi + \nabla_p H \cdot \nabla_x \psi - \nabla_x H \cdot \nabla_p \psi=0, \\ \nonumber
& \psi(0, x,p)=\prod_{i=1}^d\delta(p_i - u_i(0,x)),
\end{align}
whose analytical solution is $\psi(t,x,p) = \delta(\phi(t,x,p))$. We have thus transformed a $(d+1)$-dimensional nonlinear Hamilton-Jacobi PDE  to a $(2d+1)$-dimensional \textit{linear} PDE~--~the Liouville equation, without \textit{any} approximations or constraints on the nonlinearity. The mapping is {\it exact}, but  at the expense of doubling the spatial dimension. In the following, we denote $w(t,x,p)$ to be the solution to \eqref{Liouville-delta} with smoothed initial data, i.e., $w = \psi^\omega$.

\subsection{Finite difference discretisation for the Liouville equation} \label{subsect:HJFD}

\subsubsection{The QLSA for the finite difference discretisation}

Consider $x,p$ together as a new variable, and write $y = (x,p) = (x_1,\cdots,x_d, p_1,\cdots, p_d) = (y_1,\cdots, y_{2d})$. Use the same uniform mesh in each $y_i$ direction. Let $\bb{j} = (j_1,\cdots, j_d, j_{d+1},\cdots, j_{2d})$. Then the upwind discretisations for each term of the equation in \eqref{Liouville-delta} are
\begin{align*}
\partial_t w & \longrightarrow  \frac{w_{\bb{j}}^{n+1} - w_{\bb{j}}^n}{\D t},   \\[1mm]
\frac{\partial H}{\partial p_i} \frac{\partial w}{\partial x_i}
 &  \longrightarrow
  \frac{1}{\Delta x}\Big\{\frac{\partial H}{\partial p_i}\Big\}_{\bb{j}}^-  (w^n_{\bb{j}+\bb{e}_{d+i}} - w^n_{\bb{j}})  + \frac{1}{\Delta x}\Big\{\frac{\partial H}{\partial p_i}\Big\}_{\bb{j}}^+  (w^n_{\bb{j}} - w^n_{\bb{j}-\bb{e}_{d+i}} ),\\[1mm]
 -\frac{\partial H}{\partial x_k} \frac{\partial w}{\partial p_k}
 & \longrightarrow
  -\frac{1}{\Delta x}\Big\{\frac{\partial H}{\partial x_k}\Big\}_{\bb{j}}^+  (w^n_{\bb{j}+\bb{e}_k} - w^n_{\bb{j}})  - \frac{1}{\Delta x}\Big\{\frac{\partial H}{\partial x_k}\Big\}_{\bb{j}}^-  (w^n_{\bb{j}} - w^n_{\bb{j}-\bb{e}_k} ),
\end{align*}
where
\begin{equation*}\label{alphaplusminus}
\alpha^+ = \max \{ \alpha, 0\} = \frac{\alpha + |\alpha|}{2}, \quad
\alpha^- = \min \{ \alpha, 0\} = \frac{\alpha-|\alpha|}{2}.
\end{equation*}
For convenience we introduce the following notation
\[a_{\bb{j}}^{k,\pm} = \Big\{\frac{\partial H}{\partial x_k}\Big\}_{\bb{j}}^\pm, \qquad
b_{\bb{j}}^{i,\pm} = \Big\{\frac{\partial H}{\partial p_i}\Big\}_{\bb{j}}^\pm.\]
The discrete scheme can be written as
\begin{align*}
w_{\bb{j}}^{n+1} - w_{\bb{j}}^n
& + \lambda \sum\limits_{i=1}^d \Big[ b_{\bb{j}}^{i,-}  (w^n_{\bb{j}+\bb{e}_{d+i}} - w^n_{\bb{j}})  + b_{\bb{j}}^{i,+}  (w^n_{\bb{j}} - w^n_{\bb{j}-\bb{e}_{d+i}} ) \Big]  \\
& - \lambda \sum\limits_{k=1}^d \Big[ a_{\bb{j}}^{k,+}  (w^n_{\bb{j}+\bb{e}_k} - w^n_{\bb{j}})  + a_{\bb{j}}^{k,-}  (w^n_{\bb{j}} - w^n_{\bb{j}-\bb{e}_k} )\Big] = 0,
\end{align*}
or
\begin{align}
w_{\bb{j}}^{n+1}
& - \Big[ 1 - \lambda \sum\limits_{\ell=1}^d(b_{\bb{j}}^{\ell,+} - b_{\bb{j}}^{\ell,-} + a_{\bb{j}}^{\ell,+} - a_{\bb{j}}^{\ell,-} )  \Big] w_{\bb{j}}^n \nonumber\\
& + \lambda \sum\limits_{\ell=1}^d \Big[ b_{\bb{j}}^{\ell,-}  w^n_{\bb{j}+\bb{e}_{d+\ell}}  - b_{\bb{j}}^{\ell,+}  w^n_{\bb{j}-\bb{e}_{d+\ell}}
 - a_{\bb{j}}^{\ell,+} w^n_{\bb{j}+\bb{e}_\ell} + a_{\bb{j}}^{\ell,-} w^n_{\bb{j}-\bb{e}_\ell} \Big] = 0. \label{variWj}
\end{align}
In matrix form one has
\[\bb{w}^{n+1} - B \bb{w}^n = \bb{f}^{n+1}, \quad n = 0,1,\cdots, N_t-1,\]
with $\bb{f}^i$ being the terms resulting from the initial and boundary conditions, where the nodal values at $t=t_n$ are arranged as
\[\bb{w}^n = \sum\limits_{\bb{j}}w_{\bb{j}}^n \ket{j_1} \otimes \cdots \otimes \ket{j_d} \otimes \cdots \otimes \ket{j_{2d}}.\]
That is, the $n_{\bb{j}}$-th entry of $W^n$ is $w_{\bb{j}}^n$, with the global index given by
$n_{\bb{j}}: = j_12^{2d-1} + \cdots + j_{2d}2^0$. The non-zero entries of $B$ can be provided by using the global index as before.
The resulting linear system is
\begin{equation}\label{systemQdiffLiouville}
L \bb{w} = F,
\end{equation}
where
\[\bb{w} = [\bb{w}^1;\cdots; \bb{w}^{N_t}], \qquad F = [\bb{f}^1; \bb{f}^2; \cdots; \bb{f}^{N_t}].\]
The coefficient matrix $L$ is of the same form as in Eq. \eqref{systemQdiffLiouvilleRepresenation}. Note that we consider the Liouville equation with the smoothed initial data, and for periodic boundary conditions, one has
$\bb{f}^1 = B\bb{w}^0$ and $\bb{f}^i = \bb{0}$ for $i\ge 2$.

\begin{theorem}\label{thm:QdiffLiouville}
Suppose $\lambda = \Delta t/\Delta x$ satisfies the following CFL condition
\[\lambda   \sum\limits_{i=1}^d \Big ( \sup_{x,p} |\partial_{x_i} H|  +  \sup_{x,p} |\partial_{p_i} H| \Big ) \le 1,\]
and assume
\[\sup_{x,p} |\partial_{x_i} H|  +  \sup_{x,p} |\partial_{p_i} H| = \mathcal{O}(1), \quad i = 1,\cdots,d.\]
Then the condition number and the sparsity of $L$ satisfy $\kappa = \mathcal{O}(1/\Delta t)$ and $s = \mathcal{O}(d)$.
For fixed spatial step $\Delta x$, let $\Delta t = \mathcal{O}(\Delta x/d)$ and $\omega = (d\Delta x)^{1/3}$. Given the error tolerance $\varepsilon$, the gate complexity of the quantum difference method is
\[N_{\text{Gates}} = \widetilde{\mathcal{O}}\Big( \frac{d^3}{\varepsilon^3} \log \frac{1}{\varepsilon} \Big).\]
\end{theorem}
\begin{proof}
The proof is similar to the argument in Theorem \ref{thm:FDLiouvilleRep}, so we omit the details.
\end{proof}

\subsubsection{The computation of the physical observables}

In the following, we consider the computation of the physical observables for the Liouville equation
and assume the periodic boundary conditions.  Then physical observables are defined as
\begin{equation}\label{obser}
\langle G(t,x)\rangle= \int_{\mathbb{R}^d} G(p) \psi(t,x,p) \d p.
\end{equation}
For example,  $G(p) = 1, p, |p|^2/2$ yield density, momentum and kinetic energy respectively \cite{JinLiu2022nonlinear}.
As in \eqref{Grho}, one can compute the integral \eqref{obser} by using the numerical quadrature rule
 \begin{equation}\label{Gquadxp}
  \langle G(t_n,x_{\bb{j}}) \rangle
  =\int_{\mathbb{R}^d} \psi(t_n,x_{\bb{j}},p) \d p \approx \frac{1}{M^d}\sum_{\vect{l}} G_{\vect{l}} \psi^{\omega}_{ \vect{j}, \vect{l},n}
  = \frac{1}{M^d}\sum_{\vect{l}} G_{\vect{l}} w_{ \vect{j}, \vect{l},n}
  =: \langle G^\omega_{n,\vect{j}}\rangle,
 \end{equation}
 where, $G_{\vect{l}}$ are the weights, $\bb{j} = (j_1,\cdots,j_d)$ and $\bb{l} = (l_1,\cdots, l_d)$ and $M$ is the number of points in each dimension of the $2d$ phase space.

Let $\bb{w}_{\bb{j},\bb{l},n}$ be the solution of the classical spectral method or the upwind finite difference method (for the smoothed initial data). Then for the QLSA one has
\[\ket{\psi} = \frac{1}{N_{\psi}} \sum\limits_{\bb{j},\bb{l},n} \bb{w}_{\bb{j},\bb{l},n} \ket{\bb{j}} \ket{\bb{l}} \ket{n}
,\]
where the normalisation $N_{\psi} = \|\bb{w}\|$. For the quantum simulation method, one can just remove the ``time register'' (in this case $N_\psi$ is for time $t = t_n$).
With $G_{\bb{l}}$ in \eqref{Gquadxp}, we define the state
\[|G_{n, \vect{j}}\rangle := \frac{1}{N_G}\sum_{\vect{l}}G^\dagger_{\vect{l}}  |\vect{j}\rangle |\vect{l}\rangle |n\rangle \]
where $N_G=\sqrt{\sum_{\vect{l}}|G^\dagger_{\vect{l}}|^2}$ is the normalisation.
Given the density matrix $\mathcal{G} :=|G_{n, \vect{j}}\rangle \langle G_{n,\vect{j}}|$, we define
$\Upsilon := \langle \psi |\mathcal{G} | \psi \rangle$.
A simple algebra yields
\[
  \langle G(t_n,x_{\bb{j}}) \rangle \approx \langle G^{\omega}_{n,\vect{j}}\rangle  = n_{\psi}n_G|\sqrt{\Upsilon}|,
\]
where $n_G = N_G/M^{d/2} = \mathcal{O}(1)$ is known and $n_{\psi} = N_\psi /M^{d/2}$ may be unknown. We further define
\[
\langle O \rangle = \langle G^{\omega}_{n,\vect{j}}\rangle^2 = (n_Gn_{\psi})^2 \Upsilon := \langle \psi | O | \psi \rangle, \qquad
O = (n_Gn_{\psi})^2 \mathcal{G},
\]
as in Subsect.~\ref{subsubsect:obsLiouvilleRep}.

\begin{theorem} \label{thm:obsFDLouville}
Suppose the condition of Theorem \ref{thm:QdiffLiouville} is satisfied. Given the error tolerance $\varepsilon$, if the QLSA for the upwind finite difference discretisation is used, then the observable of the Liouville equation \eqref{Liouville-delta} can be computed with gate complexity given by
\[N_{\text{Gates}}( \langle O \rangle) = \widetilde{\mathcal{O}}\Big( \frac{n_H^4 d^3}{\varepsilon^5} \log \frac{1}{\varepsilon}\Big),\]
where $n_H = \|(\bb{\psi}^\omega)^0\|/M^{d/2}$.
\end{theorem}
\begin{proof}
According to Remark~\ref{rem:obsk}, one just needs to multiply the gate complexity in Theorem \ref{thm:QdiffLiouville} by the factor $n_H^4/\varepsilon^2$.
\end{proof}
\begin{remark}\label{rem:nH}
According to Lemma 14 in \cite{JinLiu2022nonlinear}, one has $n_H = \mathcal{O}(\beta^d M^{d/2})$ if we assume the initial data
has support in a box of size $\beta$.
\end{remark}

\subsection{Spectral discretisation for the Liouville equation}

For the spectral discretisation, we only consider an important case, namely $H(x,p) = \frac{1}{2} |p|^2 + V(x)$, which is the total energy (kinetic and potential energy) of classical particle. The  Liouville equation is then rewritten as
\begin{equation}\label{LiouvilleSmoothed}
\begin{cases}
\partial_t w + p \cdot \nabla_x w - \nabla_x V(x) \cdot \nabla_p w = 0, \\
w(0,x,p) = \psi^\omega(0, x,p)=\prod_{i=1}^d\delta_\omega(p_i - u_i(0,x)),
\end{cases}
\end{equation}
where we have assumed the smoothed initial data.

\subsubsection{The QLSA for the spectral discretisation}

Now we consider solving the Liouville equation in \eqref{LiouvilleSmoothed} by using the Fourier spectral methods. For simplicity, the periodic boundary conditions are used for the spectral discretisation. To this end, we introduce some notations. We always assume that $x = (x_1,\cdots,x_d) \in [0,1]^d$ and $p = (p_1,\cdots,p_d) \in [0,1]^d$. Introduce a new variable ${y} = ({x},{p})=(x_1,\cdots,x_d, p_1,\cdots,p_d)$, and set
\begin{equation}\label{wapprox}
w(t,{x},{p}) = w(t,{y}) = \sum\limits_{\bb{l}}c_{\bb{l}}(t) \phi_{\bb{l}}({y}), \quad
\bb{l} = (l_1, \cdots, l_d, l_{d+1},\cdots, l_{2d}).
\end{equation}
The collocation points are denoted by ${y}_{\bb{j}}$ with $\bb{j} = (\bb{j}_x, \bb{j}_p)$. As in \eqref{cdef}, we define $\bb{c} = \bb{c}_x \otimes \bb{c}_p$, where $\bb{c}_x = \bb{c}^{(1)} \otimes \cdots \otimes \bb{c}^{(d)}$ and
$\bb{c}_p = \bb{c}^{(d+1)} \otimes \cdots \otimes \bb{c}^{(2d)}$.
We also introduce the notation $\bb{w} = \bb{w}_x \otimes \bb{w}_p$, where
$\bb{w}_x = \bb{w}^{(1)} \otimes \cdots \otimes \bb{w}^{(d)}$, $\bb{w}_p = \bb{w}^{(d+1)} \otimes \cdots \otimes \bb{w}^{(2d)}$, and $\bb{w}^{(l)} = \Phi \bb{c}^{(l)}$ can be viewed as the approximate solution of $w$ in $y_l$ direction.

According to the discussion in Subsect.~\ref{subsubsect:notation}, the first term can be discretised as
\begin{align*}
p \cdot \nabla_x w
& = \i \sum\limits_{l=1}^d y_{l+d} (-\i \partial_{y_l}) w
  \longrightarrow \i \sum\limits_{l=1}^d \hat{y}_{l+d} \hat{P}_l^{\text{d}} ( \bb{w}_x \otimes \bb{w}_p) \\
& = \i \sum\limits_{l=1}^d ( I^{\otimes^d}\otimes \bb{D}_l ) ( \bb{P}_l \otimes I^{\otimes^d}) ( \bb{w}_x \otimes \bb{w}_p)
= \i \sum\limits_{l=1}^d ( \bb{P}_l \otimes \bb{D}_l) ( \bb{w}_x \otimes \bb{w}_p) .
\end{align*}
For the second term, one has
\begin{align*}
 \nabla_x V({x}) \cdot \nabla_p w
& = (\partial_1 V({x}), \cdots, \partial_d V({x})) \cdot \nabla_p w  =:  (v_1({x}), \cdots, v_d({x})) \cdot \nabla_p w \\
& = \sum\limits_{l=1}^d v_l (x) \partial_{y_{l+d}} w = \i \sum\limits_{l=1}^d v_l (x) (-\i \partial_{y_{l+d}}) w
 \longrightarrow \i \sum\limits_{l=1}^d v_l (\hat{x}^{\text{d}}) \hat{P}_{l+d}^{\text{d}} ( \bb{w}_x \otimes \bb{w}_p) \\
& = \i \sum\limits_{l=1}^d ( \bb{V}_l \otimes I^{\otimes^d}) ( I^{\otimes^d} \otimes \bb{P}_l) ( \bb{w}_x \otimes \bb{w}_p)
  = \i \sum\limits_{l=1}^d ( \bb{V}_l \otimes \bb{P}_l) ( \bb{w}_x \otimes \bb{w}_p),
\end{align*}
where
\[\bb{V}_k = \text{diag}( \bb{v}_k), \quad \bb{v}_k = \sum\limits_{\bb{j}_x} v_k( {x}_{\bb{j}_x}) \ket{j_1}\cdots\ket{j_d}
= \sum\limits_{\bb{j}_x} \partial_k V( {x}_{\bb{j}_x}) \ket{j_1}\cdots\ket{j_d}.\]
Let
\begin{equation}\label{AtildeA}
 A = -\i \sum\limits_{l=1}^d ( \bb{P}_l \otimes \bb{D}_l - \bb{V}_l \otimes \bb{P}_l) = : -\i \tilde{A}.
\end{equation}
Note that $\tilde{A}$ is a real symmetric matrix.  The resulting ODEs is
\begin{equation}\label{ODEsLiouvilleSpectral}
\begin{cases}
\frac{\d }{\d t} \bb{u}(t) = A \bb{u}(t),\\
\bb{u}(0) = ( \psi^\omega(0, y_{\bb{j}}) ).
\end{cases}
\end{equation}

We are ready to apply the quantum algorithm in \cite{BerryChilds2017ODE} to solve the above ODEs, with the time complexity described below.

\begin{theorem} \label{thm:QLSALiouvilleSpectral}
Assume that $\max_{1\le l \le d} \|\bb{V}_l\| = \mathcal{O}(1)$ and $T = \mathcal{O}(1)$.
\begin{enumerate}[(1)]
  \item There exists a quantum algorithm that produces a state $\varepsilon$-close to $\bb{u}(T)/\|\bb{u}(T)\|$ with the gate complexity given by
\[N_{Gates} = \widetilde{\mathcal{O}}\Big( \frac{d^{2+2/\ell}}{\varepsilon^{2+4/\ell}} \Big).\]
  \item The observable of the Liouville equation can be computed with gate complexity given by
\[N_{\text{Gates}}( \langle O \rangle) =  \widetilde{\mathcal{O}}\Big( \frac{n_H^4 d^{2+2/\ell}}{\varepsilon^{4+4/\ell}} \Big),\]
where $n_H = \|(\bb{\psi}^\omega)^0\|/M^{d/2}$.
\end{enumerate}

\end{theorem}
\begin{proof} The argument is similar to that of Theorem \ref{thm:QLSAKvN}.

(1) Let $A = V^{-1} D V$ with $D = \text{diag}(\lambda_1,\cdots,\lambda_N)$. Since the matrix $\tilde{A}$ in \eqref{AtildeA} is a real symmetric matrix,  $\text{Re} (\lambda_j) = 0$ for any $j \in \{1,\cdots,N\}$ and $\kappa_V = 1$.
According to Lemma \ref{lem:BCO2017}, there exists a quantum algorithm that produces a state $\varepsilon$-close to $\bb{u}(T)/\|\bb{u}(T)\|$ with the gate complexity given by
\[N_{Gates} = \mathcal{O}\Big( s \kappa_V \|A\| T \cdot \text{Poly}( \log (s  N \kappa_V \|A\| T/\varepsilon) )\Big),\]
where we have omitted the parameter $g$ that characterises the decay of the final state relative to the initial state.

It is evident that the sparsity of $A$ is $\mathcal{O}(M) = \mathcal{O}(1/\D x)$. The norm of $A$ satisfies
\begin{align*}
\|A\|
& \le  \sum\limits_{l=1}^d \| \bb{P}_l \otimes \bb{D}_l - \bb{V}_l \otimes \bb{P}_l \| \le \sum\limits_{l=1}^d  (\| \bb{P}_l\| \|\bb{D}_l\| + \|\bb{V}_l\| \|\bb{P}_l \|) \\
& \lesssim  M \sum\limits_{l=1}^d  (\|\bb{D}_l\| + \|\bb{V}_l\|)  \lesssim  dM \max_{1\le l \le d} (1 + \|\bb{V}_l\|) .
\end{align*}

According to Lemma \ref{lem:errKvN}, the error of the spectral discretisation is $\mathcal{O}(\omega + \D t^\alpha/\omega^{\alpha} + d\Delta x^{\ell}/\omega^{\ell+1})$, where $\alpha$ is for the precision of the temporal discretisation which has been considered in the quantum algorithm in \cite{BerryChilds2017ODE}. To reach a precision of $\varepsilon$, one just needs to set $\omega \sim d\Delta x^{\ell}/\omega^{\ell+1} \sim \varepsilon$, and gets $\Delta x \sim \varepsilon^{1 + 2/\ell}/d ^{1/\ell}$. Therefore, we have
\[N_{Gates} = \mathcal{O}\Big( \frac{d}{\D x^2} \Big)
= \mathcal{O}\Big( \frac{d^{2+2/\ell}}{\varepsilon^{2+4/\ell}} \Big),\]
where in the last equal sign we have included the additional factor $d$ arising from the matrix order (see Remark \ref{rem:ODEs}).

(2) For the spectral discretisation, the constant $n_{\psi} = N_\psi /M^{d/2}$ is known since $N_\psi = \|\bb{u}^{N_t}\| = \|\bb{u}^0\|= N_{\psi_0}$.
The desired estimate follows from the general sampling law.
\end{proof}

\subsubsection{The quantum simulation for the spectral discretisation}

One can directly solve the ODEs \eqref{ODEsLiouvilleSpectral} by using the quantum simulation since $\tilde{A}$ is real symmetric. We in the following consider the Fourier spectral methods based on the time-splitting approximations.

From time $t=t_n$ to time $t = t_{n+1}$, the Liouville equation is solved in two steps: One solves
\begin{equation}\label{LiouvilleStep1}
\partial_t w + p \cdot \nabla_x w = 0
\end{equation}
for one time step, followed by solving
\begin{equation}\label{LiouvilleStep2}
\partial_t w - \nabla_x V({x}) \cdot \nabla_p w = 0
\end{equation}
again for one time step.

\textbf{Step 1.} According to the previous discussion, one has
\begin{align*}
p \cdot \nabla_x w \longrightarrow \i \sum\limits_{l=1}^d ( \bb{P}_l \otimes \bb{D}_l) ( \bb{w}_x \otimes \bb{w}_p) .
\end{align*}
Since $\bb{P}_l \bb{w}_x =  \Phi^{\otimes^d} \bb{D}_l^\mu \bb{c}_x$, the first step \eqref{LiouvilleStep1} gives
\[\frac{\d}{\d t} (\bb{c}_{x}\otimes\bb{w}_p) + \i \sum\limits_{l=1}^d ( \bb{D}_l^\mu \otimes \bb{D}_l) ( \bb{c}_x \otimes \bb{w}_p) = \bb{0}, \]
which can be written as
\[\frac{\d}{\d t} (\bb{c}_{x}\otimes\bb{w}_p) + \i L (\bb{c}_{x}\otimes\bb{w}_p) = \bb{0}, \]
where
\[L = D_\mu \otimes I^{\otimes^{d-1}}\otimes D_p \otimes I^{\otimes^{d-1}} + \cdots +
I^{\otimes^{d-1}}\otimes D_\mu \otimes I^{\otimes^{d-1}} \otimes D_p \]
is a diagonal matrix. Therefore the intermediate solution of the first step is
\[(\bb{c}_{x}\otimes\bb{w}_p)^* = \e^{-\i L \Delta t} (\bb{c}_{x}\otimes\bb{w}_p)^n.\]

\textbf{Step 2.} The second step is to solve \eqref{LiouvilleStep2}, i.e.,
\begin{align*}
0 & = \partial_t w - \nabla_x V({x}) \cdot \nabla_p w
   =  \partial_t w - (\partial_1 V({x}), \cdots, \partial_d V({x})) \cdot \nabla_p w \\
  & =: \partial_t w - (v_1({x}), \cdots, v_d({x})) \cdot \nabla_p w.
\end{align*}
Similar to the first step, one has
\[(\bb{w}_x \otimes \bb{c}_p)^{n+1} = \e^{\i U \Delta t} (\bb{w}_x \otimes \bb{c}_p)^*,\]
where
\[U = \bb{V}_1 \otimes D_\mu \otimes I^{\otimes^{d-1}} + \cdots + \bb{V}_d \otimes I^{\otimes^{d-1}} \otimes D_\mu \]
is a diagonal matrix, and
\[\bb{V}_k = \text{diag}( \bb{v}_k), \quad \bb{v}_k = \sum\limits_{\bb{j}_x} v_k( {x}_{\bb{j}_x}) \ket{j_1}\cdots\ket{j_d}
= \sum\limits_{\bb{j}_x} \partial_k V( {x}_{\bb{j}_x}) \ket{j_1}\cdots\ket{j_d}.\]

Given the initial state of $\bb{w}^0$, applying the inverse QFT to the $x$-register, one gets $(\bb{c}_x \otimes \bb{w}_p)^0$.
At each time step, one needs to consider the following procedure
\[(\bb{c}_x \otimes \bb{w}_p)^n
\xrightarrow {\e^{-\i L \Delta t}} (\bb{c}_x \otimes \bb{w}_p)^*
\xrightarrow {F_x\otimes F_p^{-1} } (\bb{w}_x \otimes \bb{c}_p)^*
\xrightarrow {\e^{\i U \Delta t}} (\bb{w}_x \otimes \bb{c}_p)^{n+1}
 \xrightarrow {F_x^{-1}\otimes F_p } (\bb{c}_x \otimes \bb{w}_p)^{n+1},\]
where $F_x = F_p = \Phi^{\otimes^d}$.

\begin{theorem}\label{thm:ComplexityLiouville}
Given the error tolerance $\varepsilon$, assume that $S_0(x)$, $A_0(x)$ and $V(x)$ are smooth enough.
\begin{enumerate}[(1)]
  \item The Liouville equation can be simulated with gate complexity given by,
\[N_{\text{Gates}}
 = \mathcal{O}\Big( \frac{d}{\varepsilon^2} \log \frac{d^{1/\ell}}{\varepsilon^{1 + 2/\ell}} \Big).\]
  \item The observable of the Liouville equation can be computed with gate complexity given by
\[N_{\text{Gates}}( \langle O \rangle) = \mathcal{O}\Big( \frac{n_H^4 d}{\varepsilon^4} \log \frac{d^{1/\ell}}{\varepsilon^{1 + 2/\ell}} \Big),\]
where $n_H = \|(\bb{\psi}^\omega)^0\|/M^{d/2}$.
\end{enumerate}
\end{theorem}

\begin{proof}
When $S_0(x)$ and $V(x)$ are smooth, the time-splitting spectral method has the error estimate
\begin{equation}\label{errorw}
\|w^n(\cdot)-w(t_n, \cdot)\| \le C_{\ell}  \Big( \omega + \frac{\Delta t}{\omega} + \frac{d \Delta x^{\ell}}{\omega ^{\ell+1}} \Big),
\end{equation}
where $\omega ^{\ell+1}$ comes from the $\ell$-th order derivative of $w:=w^\omega$, and $C_{\ell}$ is an    $\mathcal{O}(1)$ constant.
Then one can implement the following meshing strategy
\begin{equation}\label{meshLiouville}
\omega \sim \varepsilon, \quad \Delta t \sim \varepsilon^2, \quad \Delta x \sim \varepsilon^{1 + 2/\ell} /d^{1/\ell}
\end{equation}
by forcing both error terms to be of order $\varepsilon$. Thus,
\[M = L/\Delta x = 2^m \quad \Longrightarrow \quad
m = \log \frac{L}{\Delta x} = \mathcal{O} \Big( \log \frac{d^{1/\ell}}{\varepsilon^{1 + 2/\ell}} \Big),\]
where $m$ is the number of qubits per dimension, and the total number of qubits is $m_{2d} = 2d m$.
The diagonal unitary operators $\e^{-\i L \Delta t}$ and $\e^{\i U \Delta t}$ can be implemented using $J(m_{2d}) = \mathcal{O}(m_{2d})$ gates, and the quantum Fourier transforms $F_x$ or $F_p$ can be implemented using $ d \mathcal{O}(m \log m)$ gates. Therefore, the gate complexity required to iterate to the $n$-th step is
\begin{align*}
N_{\text{Gates}}
& = 2 n( J(m_{2d}) +  2d \mathcal{O}(m \log m) )
  = 2 n \mathcal{O}( 2 dm +  2 d m \log m ) \\
& = \mathcal{O}( 4 n dm\log m )
 = \mathcal{O}\Big( \frac{d}{\varepsilon^2} \log \frac{d^{1/\ell}}{\varepsilon^{1 + 2/\ell}} \Big).
\end{align*}
The gate complexity for the observable is obtained from the sampling law.
\end{proof}

\begin{remark}
We remark that this section is the follow-up to \cite{JinLiu2022nonlinear}. Here we make precise statements about concrete implementations or running times of quantum algorithms. Compared to \cite{JinLiu2022nonlinear}, we present a different (and simpler) algorithm for computing the physical observables and include the discussion of the spectral discretization. For the finite difference discretization, we also give details on the estimation of the condition number.
\end{remark}

\section{The Schr\"odinger framework} \label{sect:Schrodinger}

In this section we propose another framework to solve the important example of the nonlinear Hamilton-Jacbobi equations with $H(x,p) = \frac{1}{2} |p|^2 + V(x)$ based on solving the Schr\"odinger equation, since the Liouville equation is the classical limit of the Schr\"odinger equation.
 The idea is to choose a semiclassical parameter,
still denoted by $\hbar$ here, sufficiently small, so the solution of the Schr\"odinger equation is close to that of the Liouville equation.

Since the error between the expectation of the wave function and its classical counterpart (the physical observables of the Liouville equation) is of $O(\hbar^2)$ \cite{Lasser2020Acta}, one can take $\hbar=\mathcal{O}(\sqrt{\varepsilon})$, to maintain the computational precision of $\mathcal{O}(\varepsilon)$ for this framework.

We consider the Schr\"{o}dinger equation in the semiclassical regime
\begin{equation} \label{Schrodinger}
\begin{cases}
{\rm i} \hbar \partial_t u(t,{x}) = -\frac{\hbar^2}{2} \Delta u(t,{x}) + V({x}) u(t,{x}) \quad \mbox{in} ~~ \Omega = (a,b)^d, \quad t>0,\\
u(0, {x}) = u_0({x})
\end{cases}
\end{equation}
with periodic boundary conditions, where ${x} = (x_1,x_2,\cdots, x_d) \in \mathbb{R}^d$, $u(t,x) := u^\hbar(t,x)$ is the complex-valued wave function, $V({x})$ is the external potential and $\hbar = \mathcal{O}(\sqrt{\varepsilon})$ with $\varepsilon \ll 1$ being the precision. Without loss of generality, we always set $a=0$ and $b=1$.
The initial condition in \eqref{Schrodinger} is chosen in a WKB form,
\begin{equation}\label{WKB0}
u_0({x}) = A_0({x}) {\rm e}^{{\rm i} \frac{S_0({x})}{\hbar}},
\end{equation}
with $A_0$ and $S_0$ independent of $\hbar$, real-valued and smooth. The periodic boundary conditions, for example, in one-dimensional case can be written as
\[u(t,a) = u(t,b), \quad u_x(t,a) = u_x(t,b), \quad  t\ge 0.\]

The problem \eqref{Schrodinger} will be solved by the classical time-splitting Fourier spectral method \cite{BJM2002splitting}, which, as described below, can be interpreted as the Trotter based Hamiltonian simulation \cite{Jin2022quantumSchrodinger,Lin2022Notes}.

\subsection{The semiclassical approximation}

We first recall the WKB analysis, which assumes that the solution remains the same form as the initial data at later time:
\[u = A(t,x) {\rm e}^{{\rm i} \frac{S(t,x)}{\hbar}},\]
where $A(t,x)$ and $S(t,x)$ are the amplitude and phase respectively. Substituting this into \eqref{Schrodinger}, and separating the real and imaginary parts, one gets
\begin{align*}
A\partial_tS  + \frac{1}{2} | \nabla_x S|^2 + AV = \frac{\hbar^2}{2} \Delta A, \\
\partial_t A + \nabla_x A \cdot \nabla_x S + \frac{1}{2} A \Delta S = 0.
\end{align*}
Ignoring the $\mathcal{O}(\hbar^2)$ terms, and multiplying the second equation by $A$, one gets
\begin{align}
\partial_t |A|^2 + \nabla_x ( |A|^2 \nabla_x S) = 0,  \label{WKB1}\\
\partial_tS  + \frac{1}{2} | \nabla_x S|^2 + V = 0. \label{WKB2}
\end{align}
The first equation \eqref{WKB1} is a transport equation, and the second one \eqref{WKB2} is the eikonal equation, which is exactly the Hamilton-Jacobi equation \eqref{H-J}. Note that the eikonal equation admits solutions $S$ with discontinuous derivatives (usually referred to as the caustic) even if the initial data of $u$ is smooth. Thus the WKB analysis is only valid up to the time when the first caustic forms.
Beyond caustics, the solution becomes multi-valued \cite{Jin2022Acta,Jin2011Acta}.

In contrast to that, the Wigner transform technique yields the Liouville equation on phase space, in the semiclassical limit $\hbar \to 0$, whose solution does not exhibit caustics, hence is valid globally in time. The Wigner transform of $u$ is defined as \cite{Jin2022Acta,Jin2011Acta}
\begin{equation}\label{Wigner}
w^{\hbar}(t,x,p) = w^{\hbar}[u](t,x,p): = \frac{1}{(2\pi)^d} \int_{\mathbb{R}^d} u \Big( x + \frac{\hbar}{2} \eta\Big)
\overline{u} \Big( x - \frac{\hbar}{2} \eta\Big) {\rm e}^{{\rm i} p\cdot \eta}{\rm d} \eta.
\end{equation}
Applying this transformation on the Schr\"{o}dinger equation \eqref{Schrodinger}, one obtains the Wigner equation (also called the quantum Liouville equation):
\[\partial_t w^{\hbar} + p\cdot\nabla_x w^{\hbar} - H_V w^{\hbar} =0, \quad w^{\hbar}(0,x,p) = w_{\rm in}(x,p),\]
where
\[H_V w^{\hbar} = \frac{{\rm i}}{(2\pi)^d} \iint_{\mathbb{R}^d \times \mathbb{R}^d} \delta V(x,y)w^\hbar (x,p'){\rm e}^{{\rm i} \eta(p-p')}{\rm d}\eta{\rm d}p',\]
\[\delta V = \frac{1}{\hbar} \Big( V(x-\frac{\hbar}{2}y) - V(x + \frac{\hbar}{2}y) \Big).\]
When $\hbar \to 0$, the Wigner equation becomes the classical Liouville equation on the phase space:
\[
\partial_t w + p \cdot \nabla_x w - \nabla_x V(x) \cdot \nabla_p w = 0,
\]
which corresponds to the important case $H(x,p) = \frac{1}{2} |p|^2 + V(x)$. One easily finds that the Liouville equation can be written as $\partial_t w + \{ w, H \} = 0$, where $\{\cdot, \cdot\}$ is the Poisson bracket, defined as
\begin{equation*}
 \{w, H\} = \nabla_p H \cdot \nabla_x w - \nabla_x H \cdot \nabla_p w.
\end{equation*}

 When $u_0$ is given in WKB form \eqref{WKB0}, the corresponding Wigner measure is found to be
\[w^\hbar[u_0] \xrightarrow{\hbar \to 0} w_0 = |A_0(x)|^2 \delta(p-\nabla S_0(x)),\]
see \cite[Eq.~(3.9)]{Jin2011Acta}.
It should be pointed out that Eqs.~\eqref{WKB1} and \eqref{WKB2} can be deduced from the moment-closure
of the Liouville problem
\begin{equation}\label{LiouvillePro}
\begin{cases}
\partial_t w + \nabla_p H \cdot \nabla_x w - \nabla_x H \cdot \nabla_p w = 0, \\
w(0,x,p) = |A_0(x)|^2 \delta(p-\nabla S_0(x))
\end{cases}
\end{equation}
with mono-kinetic ansatz $w(t,x,p) = |A(t,x)|^2 \delta(p-\nabla_x S(t,x))$, but are not valid beyond caustics since $\delta(p-\nabla_x S(t,x))$ is not well-defined when $\nabla_x S(t,x)$ becomes discontinuous.
However, the equation in \eqref{LiouvillePro} is valid globally in time, since it unfolds the caustics in the phase space \cite{Jin2022Acta,Jin2011Acta,JinLiu2022nonlinear}. For this reason, we instead solve \eqref{LiouvillePro} in the semiclassical regime.

Clearly, \eqref{LiouvillePro} is the level set formulation
\eqref{Liouville-delta} if $A_0(x)\equiv 1$. Here we leave the general $A_0(x)$.
One can solve the problems in \eqref{LiouvillePro} by upwind finite difference methods or spectral methods as shown in the previous section.


\subsection{Quantum simulations for the spectral discretisation}

\subsubsection{The time-splitting spectral approximations} \label{subsubsect:Schrodingersimulation}

From time $t=t_n$ to time $t = t_{n+1}$, the Schr\"odinger equation is solved in two steps \cite{BJM2002splitting}:
One solves
\begin{equation}\label{udStep1}
\hbar u_t - \i \frac{\hbar^2}{2} \Delta u = 0
\end{equation}
for one time step, followed by solving
\begin{equation}\label{udStep2}
\hbar u_t  + \i V({x}) u = 0
\end{equation}
again for one time step.
Equation \eqref{udStep1} will be discretised in space by the Fourier pseudo-spectral method and integrated in time exactly. The ODE \eqref{udStep2} will then be solved exactly.

\begin{remark}
We remark that usually the Trotter or Strang splitting is used for quantum simulation of the Schr\"odinger equation, which is second order in time rather than first order in the above simple splitting. We use the first order one in order to compare, on the equal footing, with other methods since all other methods use the first order approximation in time.  For time complexity of the  Trotter or Strang splitting see \cite{Jin2022quantumSchrodinger}.
\end{remark}

Let $u_{\bb{j}}^n$ be the numerical solution at $t=t_n$ and $u_{\bb{j}}^*$ the solution given by the first step for $0\le \bb{j} \le M-1$.
\begin{itemize}
  \item For the first step, according to the previous discussion one easily obtains
  \[  \frac{\d }{\d t} \bb{u}(t) + \i \hbar/2 \cdot (\bb{P}_1^2 + \cdots + \bb{P}_d^2)   \bb{u}(t)  = \bb{0},\]
  or
  \[  \frac{\d }{\d t} \bb{c}(t) + \i \hbar/2 \cdot (( \bb{D}_1^\mu)^2 + \cdots + ( \bb{D}_d^\mu)^2)   \bb{c}(t)  = \bb{0},\]
  where the relation \eqref{PlDl} is used, which gives
  \[\bb{c}^* =  \Big( \e^{-\i \hbar \Delta t D_\mu^2 /2 } \Big)^{\otimes ^d}\bb{c}^n, \qquad
D_\mu = {\rm diag}(\mu_1, \cdots,  \mu_M).\]
  \item The updated numerical solution for the second step is $u_{\bb{j}}^{n+1} = \e^{-\i V({x}_{\bb{j}}) \Delta t/\hbar} u_{\bb{j}}^*$, which can be written in vector form as $\bb{u} = \e^{-\i \bb{V} \Delta t/\hbar}  \bb{u}^*$,
where $\bb{V}$ is a diagonal matrix with
\[\bb{V}_{n_{\bb{j}}, n_{\bb{j}} } = V({x}_{\bb{j}}), \quad
n_{\bb{j}} = j_12^{d-1} + \cdots +  j_d2^0. \]
\end{itemize}

\subsubsection{The quantum simulation of the Schr\"odinger equation}

We only consider the 1-D case since it is straightforward to extend the arguments to high-dimensional cases by using tensor products.
According to the previous discussion, we have the following algorithm (Algorithm \ref{alg:SchrodingerSplitting}) represented by the matrix-vector multiplication.
\begin{algorithm}[H]
\caption{Time splitting approximations for the Schr\"odinger equation \label{alg:SchrodingerSplitting}}
\begin{enumerate}
  \item Given the initial data $\bb{u}^0$ and $n=0$, compute the discrete Fourier coefficients
  $\bb{c}^n = \Phi^{-1} \bb{u}^n$.
  \item Calculate the intermediate variables
 $\bb{c}^* = \e^{- \i \hbar D_\mu^2 \Delta t/2} \bb{c}^n$ and $\bb{u}^* = \Phi \bb{c}^*$.
  \item Update the numerical solution
  $\bb{u}^{n+1} = \e^{-\i \bb{V} \Delta t/ \hbar } \bb{u}^*$.
\end{enumerate}
\end{algorithm}

In the above algorithm, the matrix $\Phi$ plays the role of the discrete Fourier transform (DFT), where given a set of numbers $x_0, x_1, \cdots, x_{M-1}$, the DFT and the inverse DFT are defined by
\[y_k = \frac{1}{\sqrt{M}}\sum\limits_{j=0}^{M-1} \e^{2\pi{\rm i} jk/M}x_j, \quad k = 0,\cdots, M-1\]
and
\[x_j = \frac{1}{\sqrt{M}}\sum\limits_{k=0}^{M-1} \e^{-2\pi{\rm i} jk/M}y_k, \quad j = 0,\cdots, M-1,\]
respectively. Denote the transformation matrix of DFT by $F$. It is easy to find the transformation matrix in Algorithm~\ref{alg:SchrodingerSplitting} satisfies $\Phi = \sqrt{M} SF$, where $S$ is the diagonal matrix
\[S = {\rm diag} \Big( [1, -1, \cdots, 1,-1]_{M\times 1} \Big),\]
which in turn gives
\[\bb{u}^{n+1} = \e^{-\i \bb{V} \Delta t/ \hbar } S F \e^{- \i \hbar D_\mu^2 \Delta t/2} F^{-1} S \bb{u}^n \]
since $S^{-1} = S$. For convenience, the above two diagonal matrices are denoted by $D_1$ and $D_2$, respectively, and then one has
\begin{align*}
\bb{u}^{n+1}
 = D_1 S F D_2 F^{-1} S\bb{u}^n
 = S D_1 F D_2 F^{-1} S \bb{u}^n,
\end{align*}
or
\[\bb{v}^{n+1} = (D_1 F D_2 F^{-1})  \bb{v}^n, \quad \bb{v}^n = S \bb{u}^n,\]
where we have used the fact that $S D_1 = D_1 S$.

Therefore, when preparing the variable $\bb{v}$ in the computational basis, the implementation in each iteration involves one application of an inverse quantum Fourier transform (QFT), followed by a multiplication of a diagonal unitary operator $D_2$, and a QFT and another diagonal unitary operator $D_1$, since the QFT is exactly the quantum version of the DFT.

According to the above discussion, the quantum simulation algorithm to find $\bb{v}^{n}: = S\bb{u}^{n}$ is described as follows (see Algorithm~\ref{alg:qsimSchrodingerNew}).
\begin{algorithm}[H]
\caption{Quantum simulation of the Schr\"odinger equation \label{alg:qsimSchrodingerNew}}
\begin{enumerate} \setlength{\itemindent}{1em}
  \item [Step 0.] Initialization of the quantum state:  Given $v_j^0$ encode it as
  \[\ket{\psi^0} = \frac{1}{\mathcal{N}}\sum\limits_{j=0}^{M-1} \psi_j^0 \ket{j},  \quad \psi_j^0= v_j^0,\]
  where $\mathcal{N}$ is the normalization constant. Let $n=0$.
  \item [Step 1.] Performing inverse QFT on $\ket{\psi^n}$ yields $\ket{\tilde{\psi}}$.
  \item [Step 2.] Perform a diagonal unitary operator $ \Big( \e^{- \i \hbar \mu_l^2 \Delta t/2} \Big)_{1\le l \le M}$ for $\ket{\tilde{\psi}}$, and the resulting state is denoted as  $\ket{\psi^\mu}$.

  \item [Step 3.] Perform QFT on $\ket{\psi^\mu}$ with the output denoted by $\ket{\psi^*}$.
  \item [Step 4.] Apply a diagonal unitary operator $\Big( \e^{-\i V(x_j) \Delta t/ \hbar } \Big)_{0\le j \le M-1}$ to $\ket{\psi^*}$. The output is denoted by $\ket{\psi^{n+1}}$.
  \item[Step 5.] Let $n \leftarrow n+1$ and go back to Step 1.
\end{enumerate}
\end{algorithm}

\begin{figure}[!htb]
  \centering
  \subfigure[$\hbar=0.0256, h = 1/16$]{\includegraphics[scale=0.34]{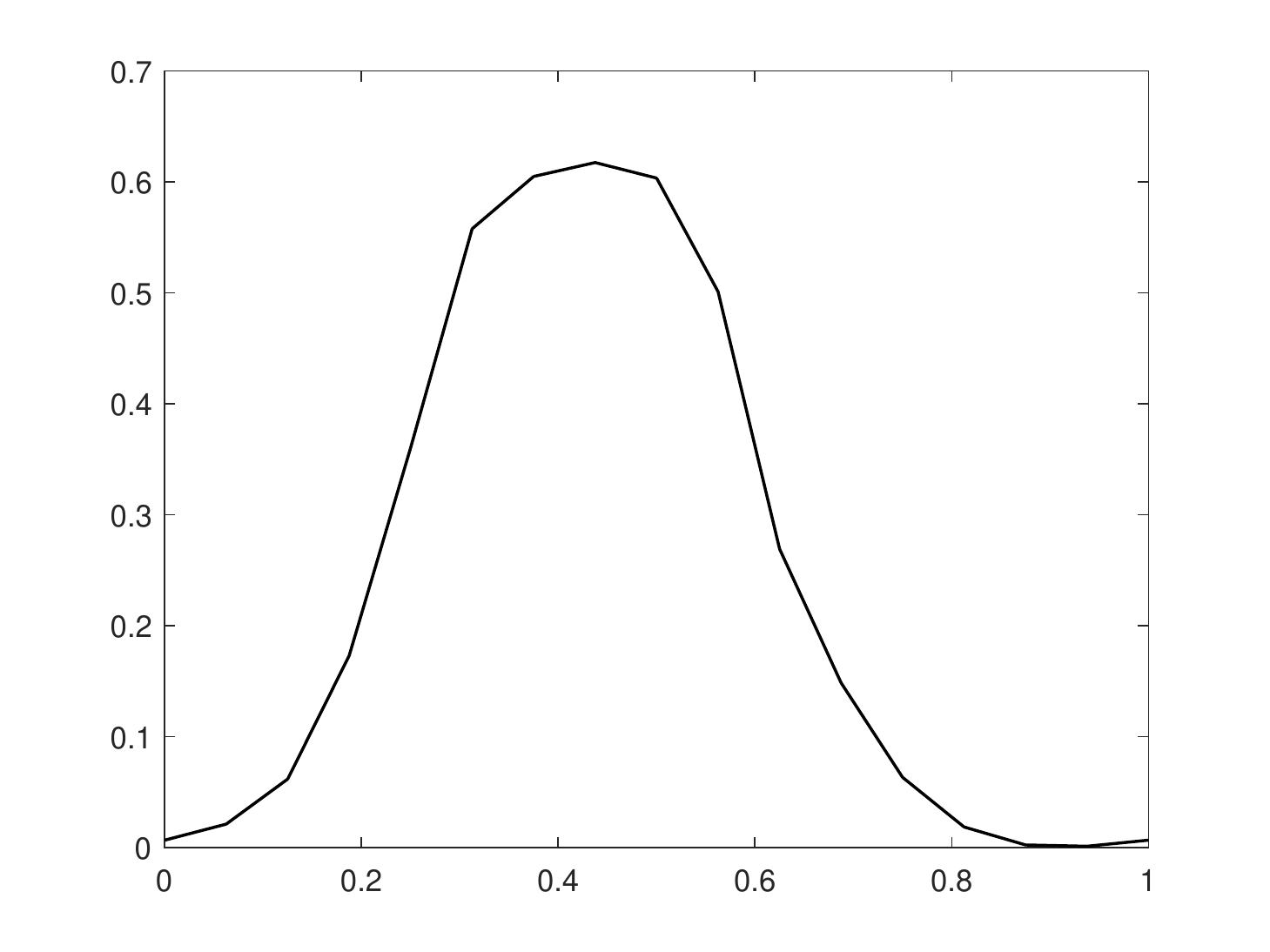}}
  \subfigure[$\hbar=0.0064, h = 1/64$]{\includegraphics[scale=0.34]{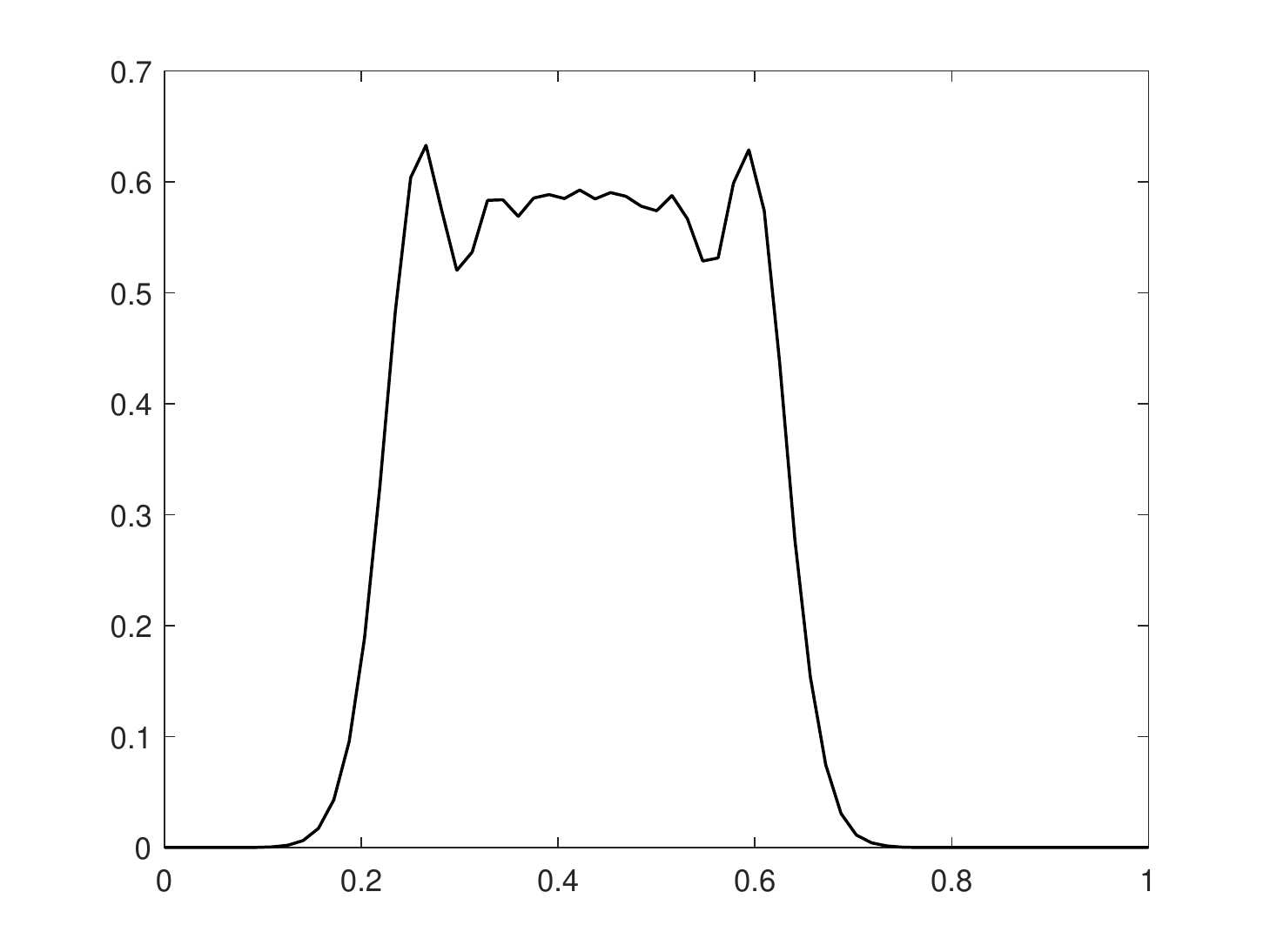}}
  \subfigure[$\hbar=0.0008, h = 1/512$]{\includegraphics[scale=0.34]{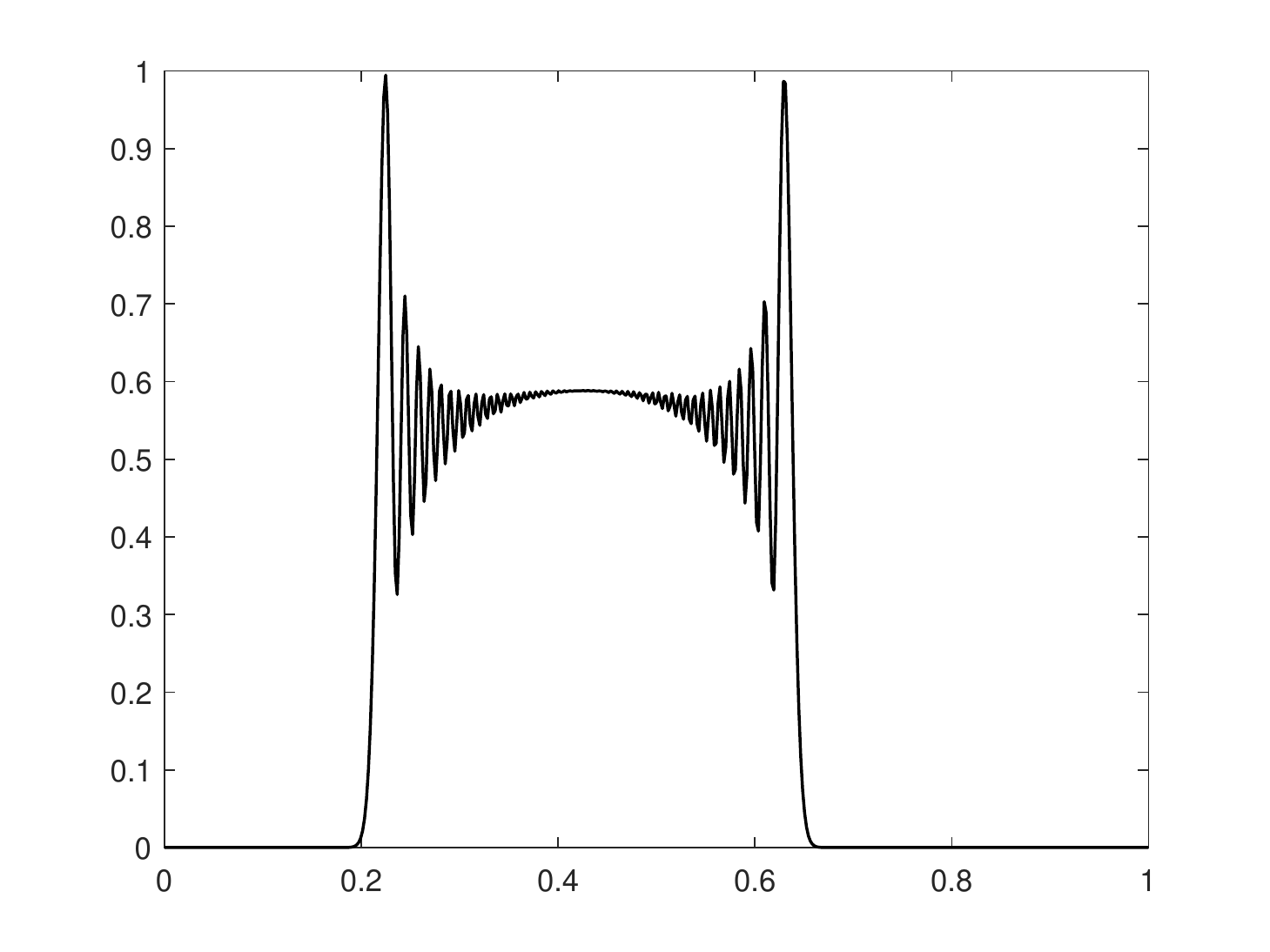}}\\
  \subfigure[$\hbar=0.0001, h = 1/4096$]{\includegraphics[scale=0.34]{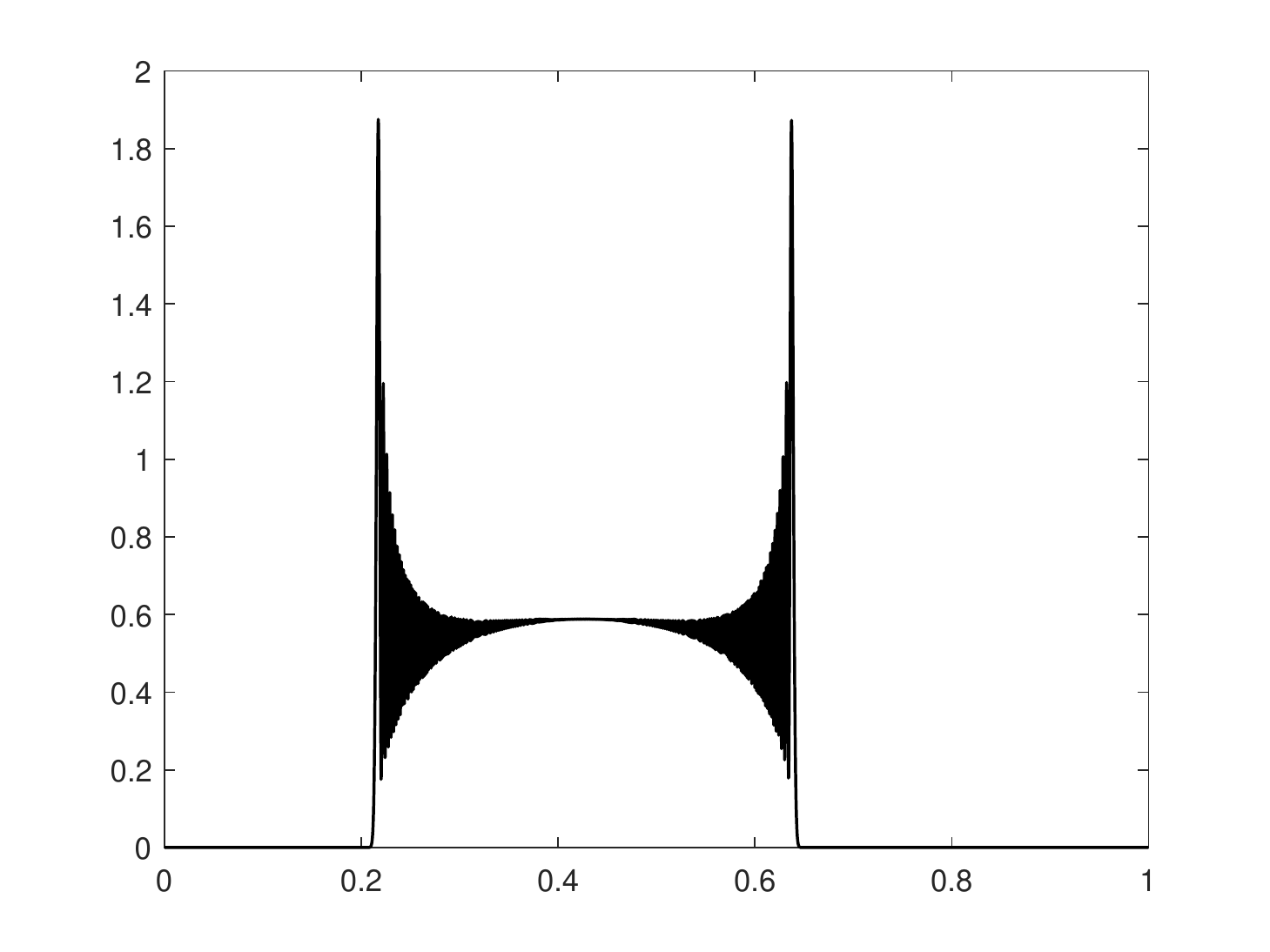}}
  \subfigure[$\hbar=0.000025, h = 1/16384$]{\includegraphics[scale=0.34]{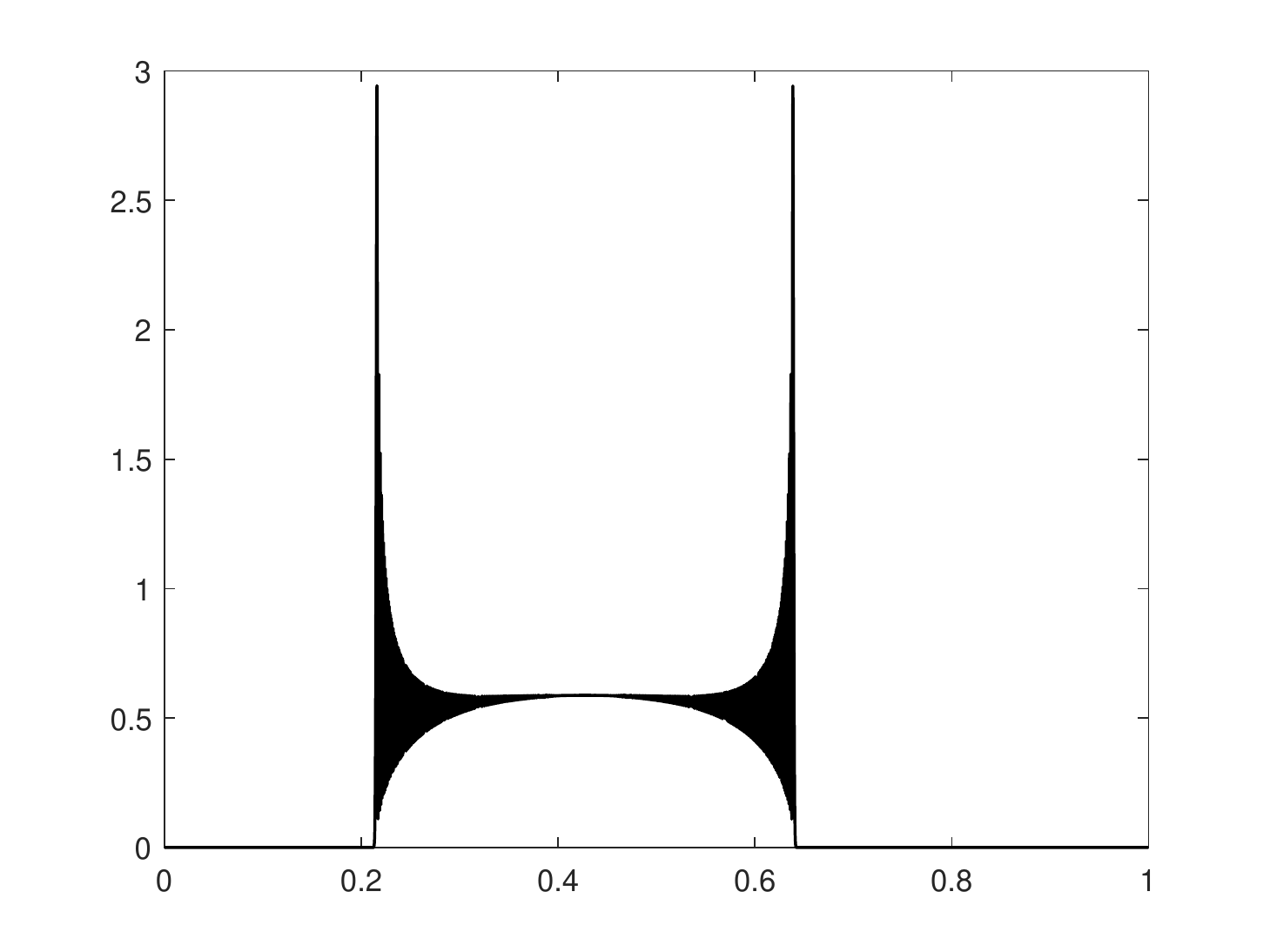}}
  \subfigure[$\hbar=0.0000125, h = 1/32768$]{\includegraphics[scale=0.34]{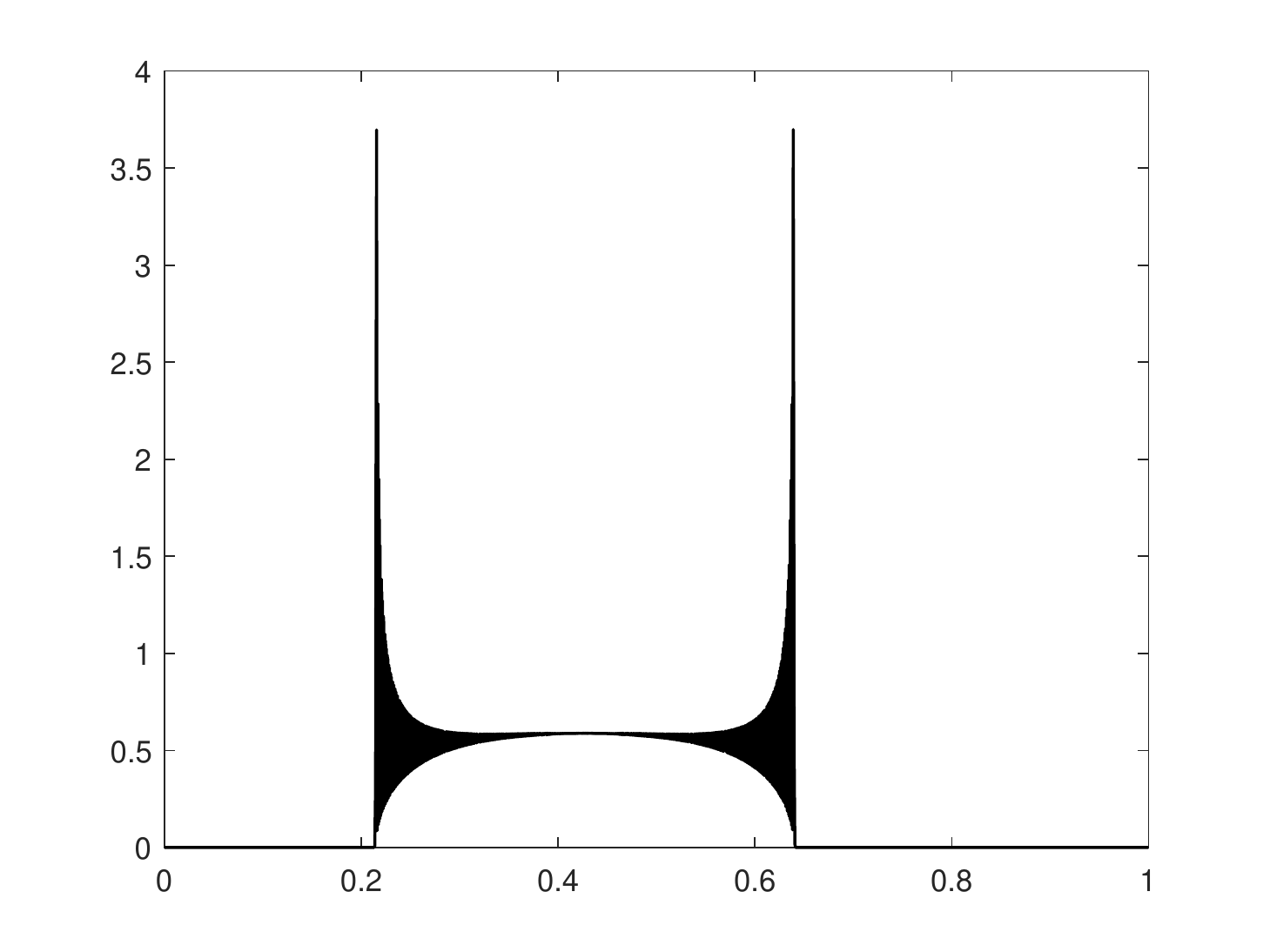}}\\
  \caption{The position density $\rho(t,x) = |u(t,x)|^2$ at $t=0.54$.}\label{fig:Pden}
\end{figure}

The following example is taken from Example 1 in \cite{BJM2002splitting}.
\begin{example}\label{ex:WKB}
The initial data is $u(x,0) = A_0(x) \e ^ {\i S_0(x)/ \hbar}$, where
\[A_0(x) = \e^{-25(x-0.5)^2}, \quad S_0(x) = -\frac{1}{5} \ln \Big( \e^{5(x-0.5)} + \e^{-5(x-0.5)}  \Big).\]
We take $[a,b]= [0,1]$ and $V(x) = 10$. The position density $\rho(t,x) = |u(t,x)|^2$ is shown in Fig.~\ref{fig:Pden}. One can see that the solution is {\bf oscillatory} for small $\hbar$. For numerical descriptions, please refer to \cite{BJM2002splitting}.
\end{example}

\subsubsection{Gate counts for the computation of wave functions}

To simplify the discussion, we set $L = b-a = 1$ and the evolution time $t = 1$ throughout the paper.
The time-splitting scheme involves only diagonal operators and QFTs whose complexities depend on the number of qubits $m$ per dimension. Since the meshing satisfies $\Delta x = L/M$ and $M = 2^m$,
we can determine $\Delta x$ and thus $m$ by the expected precision of the algorithm.

\begin{theorem}\label{thm:ComplexitySchrodinger}
Given the error tolerance $\varepsilon$, assume $V({x})$ is sufficiently smooth and $\hbar = \mathcal{O}(\sqrt{\varepsilon})$. Then the Schr\"odinger equation \eqref{Schrodinger} can be simulated using $m_d$
qubits with gate complexity $N_{\text{Gates}}$, given respectively by,
\[m_d = \mathcal{O}\Big( d \log \frac{d^{1/\ell}}{\varepsilon^{1/2+5/(2\ell)} } \Big), \qquad
N_{\text{Gates}}
 = \mathcal{O}\Big( \frac{d}{\varepsilon^{3/2}} \log \frac{d^{1/\ell}}{\varepsilon^{1/2+5/(2\ell)} }  \Big).\]
\end{theorem}

\begin{proof}
According to Refs.~\cite{BJM2002splitting,Jin2022quantumSchrodinger}, if $\Delta x/\hbar = \mathcal{O}(1)$ and $\Delta t/\hbar = \mathcal{O}(1)$, then for each $\ell$, the error of the Fourier spectral method is bounded by,
\begin{equation}\label{usp1}
 \| u^n - u(t_n, \cdot) \| \le C_\ell   n \Big(   d \Big(\frac{\Delta x}{\hbar }\Big)^
 \ell +  \frac{ \Delta t^2}\hbar\Big).
\end{equation}
Here $C_\ell$ is considered as $\mathcal{O}(1)$.  The mesh strategy is
\begin{equation*}
\frac{\Delta t}{\hbar} \sim \varepsilon,  \qquad  \frac{\Delta x}{\hbar} \sim \Big(\frac{\varepsilon}{nd} \Big)^{1/\ell}, \qquad \hbar \sim \sqrt{\varepsilon},
\end{equation*}
or equivalently,
 \begin{equation}\label{meshstrategySchr}
\Delta t \sim \varepsilon^{3/2},  \qquad  \Delta x \sim \varepsilon^{1/2 + 5/(2\ell)} / d^{1/\ell} ,
\end{equation}
which is obtained by forcing both error terms to be of order $\varepsilon$.
With this choice of $\Delta x$ and $\Delta t$, the number of qubits is
\[m_d = d m, \quad m = \log \frac{1}{\Delta x} = \mathcal{O}\Big( \log \frac{d^{1/\ell}}{\varepsilon^{1/2 + 5/(2\ell)} }\Big).\]

The diagonal unitary operators can be implemented using $J(m_d) = \mathcal{O}(m_d)$ gates \cite{Kassal2008Diagonal,Jin2022quantumSchrodinger}. Thus the gate complexity required to iterate to the $n$-th step is
\begin{align}
N_{\text{Gates}}
 \sim n  ( 2d m\log m + 2  J(m_d)  ) \sim  2 n d m\log m
  = \mathcal{O}\Big( \frac{d}{\varepsilon^{3/2}} \log \frac{d^{1/\ell}}{\varepsilon^{1/2+5/(2\ell)} }  \Big), \label{gatesSchrsim}
\end{align}
as required.
\end{proof}

\subsection{The QLSA for the spectral discretisation}

According to the discussion in Subsect.~\ref{subsubsect:Schrodingersimulation}, one has the following ODEs
\[  \frac{\d }{\d t} \bb{u}(t) = A \bb{u}(t),\]
where
\begin{equation}\label{AtildeASchr}
A = -\i \Big( \frac{\hbar}{2}  (\bb{P}_1^2 + \cdots + \bb{P}_d^2) + \frac{1}{\hbar} \bb{V} \Big) =:-\i \tilde{A}.
\end{equation}
Obviously, $\tilde{A}$ is a real symmetric matrix.

As in Theorem \ref{thm:QLSALiouvilleSpectral}, we can apply the quantum algorithm in \cite{BerryChilds2017ODE} to solve the above ODEs.

\begin{theorem} \label{thm:QLSASchrodinger}
Assume that $\|\bb{V}\| = \mathcal{O}(1)$ and $T = \mathcal{O}(1)$. Then there exists a quantum algorithm that produces a state $\varepsilon$-close to $\bb{u}(T)/\|\bb{u}(T)\|$ with the gate complexity given by
\[N_{Gates} = \widetilde{\mathcal{O}}\Big(\frac{d^{2 + 2/\ell} }{\varepsilon^{1+5/\ell}}\Big).\]
\end{theorem}
\begin{proof}
Let $A = V^{-1} D V$ with $D = \text{diag}(\lambda_1,\cdots,\lambda_N)$, where $N = \mathcal{O}(M^d)$ is the order of $A$. Since the matrix $\tilde{A}$ in \eqref{AtildeASchr} is a real symmetric matrix,  $\text{Re} (\lambda_j) = 0$ for any $j \in \{1,\cdots,N\}$ and $\kappa_V = 1$.
According to Lemma \ref{lem:BCO2017}, there exists a quantum algorithm that produces a state $\varepsilon$-close to $\bb{u}(T)/\|\bb{u}(T)\|$ with the gate complexity given by
\[N_{Gates} = \mathcal{O}\Big( s \|A\| T \cdot \text{Poly}( \log (s  N  \|A\| T/\varepsilon) )\Big).\]

It is evident that the sparsity of $A$ is $\mathcal{O}(M) = \mathcal{O}(1/\D x)$. For the mesh strategy in \eqref{meshstrategySchr}, the norm of $A$ satisfies
\begin{align*}
\|A\|
 \le  \frac{\hbar}{2} \sum\limits_{l=1}^d \| \bb{P}_l \| + \frac{1}{\hbar} \|\bb{V}\|
 \lesssim  \hbar d M + \frac{1}{\hbar} \lesssim
  \frac{d^{1+1/\ell}}{ \varepsilon ^{5/(2\ell)}} + \frac{1}{\varepsilon^{1/2}}
 \le  \frac{d^{1+1/\ell}}{\varepsilon^{1/2 + 5/(2\ell)}}.
\end{align*}
Then one has
\[N_{Gates} = \widetilde{\mathcal{O}}\Big(\frac{d^{2 + 2/\ell} }{\varepsilon^{1+5/\ell}}\Big),\]
where in the last equal sign we have included the additional factor $d$ arising from the matrix order (see Remark \ref{rem:ODEs}).
This completes the proof.
\end{proof}

\subsection{The computation of physical observables} \label{Subsect:observablesSchr}

The quantum mechanical wave function $u(t,x)$ can be considered as an auxiliary quantity used to compute physical
quantities. The most basic quadratic observables \cite{BJM2002splitting,Jin2011Acta,Jin2022Acta} include the position density $\rho(t,x):=|u(t,x)|^2$, the current density
\[J(t,x) = \hbar {\rm Im} (\overline{u}(t,x) \nabla u(t,x) ) =
 \frac{\hbar}{2\i}(\overline{u}\nabla u - u \nabla \overline{u} ),\]
and the kinetic or total energy
\[E(t,x) = \frac{\hbar^2}{2}|\nabla u(t,x)|^2  \qquad \mbox{or} \qquad
\frac{\hbar^2}{2}|\nabla u(t,x)|^2 + V(x)|\nabla u(t,x)|^2. \]

\subsubsection{The expectation of observables for the quantum simulation}

The observables $\rho(t_n,x_{\bb{i}})$, $J(t_n,x_{\bb{i}})$ and $E(t_n,x_{\bb{i}})$ can be expressed as the standard form of $\langle O \rangle = \langle \widetilde{\bb{u}} | O |  \widetilde{\bb{u}} \rangle$, which is the expectation of the observable $O$. Here, $\widetilde{\bb{u}}$ is normalized to $1$ since the output of the quantum algorithms is the normalized state. For the spectral discretisation of the Schr\"odinger equation,  one has
\begin{equation}\label{condUn}
\|\bb{u}^{N_t}\| = \cdots = \|\bb{u}^0\| =: N_{u_0},
\end{equation}
and $\widetilde{\bb{u}} = \bb{u}^{N_t}/N_{u_0}$.
The output state of the quantum simulation is
\[\ket{\widetilde{\bb{u}}} = \frac{1}{N_{u_0}} \sum\limits_{\bb{j}} \bb{u}^{N_t} \ket{\bb{j}}
= \sum\limits_{\bb{j}} \widetilde{\bb{u}} \ket{\bb{j}}, \qquad \bb{j} = (j_1,\cdots,j_d).\]

For the position density $\rho(t_n,x_{\bb{i}})$, it is obvious that one can measure the magnitude of the wave function using multiple shots in the computational basis $\ket{\bb{i}}$ as was done in the numerical experiments in \cite{Jin2022quantumSchrodinger},
so we choose
\[O_{\rho} =: N_{u_0}^2 \ket{\bb{i}}\bra{\bb{i}} \quad \mbox{for the position density}.\]
Let $p = |\widetilde{\bb{u}}_{\bb{i}}|^2\le 1$. Then ${\rm Var}(O_{\rho}) = N_{u_0}^4 p(1-p) \le N_{u_0}^4$, hence the number of samples  is $n_\rho = N_{u_0}^4 /\varepsilon^2$ for the position density.

For the current density, we consider $d=1$ for simplicity and denote $\bb{e}_i = \ket{i}$.
 Let $\bb{u} = \bb{u}^{N_t}$ for simplicity. Using the previous notations, one has
\[\overline{u}(x_i) =  \overline{\bb{u}}^T \bb{e}_i, \qquad
(\partial_x u) (x_i) = (\Phi D_\mu\Phi^{-1} \bb{u})^T \bb{e}_i = \bb{e}_i^T(\Phi D_\mu \Phi^{-1} \bb{u}),\]
which gives
\[(\overline{u}\partial_x u)(x_i) = \overline{\bb{u}}^T (\bb{e}_i \bb{e}_i^T \Phi D_\mu \Phi^{-1}) \bb{u} =
\bb{u}^\dag (\bb{e}_i \bb{e}_i^T \Phi D_\mu \Phi^{-1}) \bb{u} =: \bb{u}^\dag A \bb{u},\]
and
\[(u\partial_x \overline{u})(x_i) = \Big( (\overline{u}\partial_x u)(x_i) \Big)^\dag
= \bb{u}^\dag A^\dag \bb{u} .\]
Therefore,
\[J(t_n,x_i) = \frac{\hbar}{2\i} \bb{u}^\dag (A-A^\dag)\bb{u}
= \frac{\hbar N_{u_0}^2}{2\i } \widetilde{\bb{u}}^\dag (A-A^\dag) \widetilde{\bb{u}},\]
and we can choose
\[O_J := \frac{\hbar N_{u_0}^2}{2\i }( A - A^\dag) \quad \mbox{for the current density}.  \]
One can check that
\[ {\rm Var}(O_J)\le \|O_J \widetilde{\bb{u}} \|^2 \lesssim \Big( \frac{\hbar M N_{u_0}^2}{4 } \Big)^2,\]
 where $M$ comes from $D_\mu = {\rm diag}(-N, \cdots, N-1)$ with $N=M/2$, which implies a multiplicative factor $n_J = (\hbar M N_{u_0}^2 )^2 / \varepsilon^2 = M^2 N_{u_0}^4/\varepsilon$ for the current density since $\hbar = \mathcal{O}(\sqrt{\varepsilon})$.

For the kinetic energy, we similarly obtain
\[E(t_n,x_i) = \frac{\hbar^2}{2} \bb{u}^\dag (\Phi D_\mu \Phi^{-1} \bb{e}_i \bb{e}_i^T  \Phi D_\mu \Phi^{-1}) \bb{u}
=: \frac{\hbar^2}{2} \bb{u}^\dag B \bb{u} = \frac{\hbar^2 N_{u_0}^2}{2 } \widetilde{\bb{u}}^\dag B \widetilde{\bb{u}}\]
for the one-dimensional case. We can choose
\[O_E := \frac{\hbar^2 N_{u_0}^2 }{2 } B \quad \mbox{for the kinetic energy}.  \]
It is obvious that
\[ {\rm Var}(O_E)\le \|O_E \widetilde{\bb{u}} \|^2 \lesssim \Big( \frac{\hbar^2 M^2 N_{u_0}^2}{4 } \Big)^2,\]
which implies a multiplicative factor $n_E = ( \hbar^2 M^2N_{u_0}^2 )^2/\varepsilon^2 = M^4 N_{u_0}^4 $ for the kinetic energy.

\begin{remark}\label{rem:Nu0}
 By definition, $\rho(t,x):=|u(t,x)|^2$, where
 \[\rho(t,x) = \int w(t,x,p) \d p \approx \frac{1}{M^d}\sum_{\vect{l}} G_{\vect{l}} w^\omega_{ \vect{j}, \vect{l},n},\]
 where $w$ is the solution to \eqref{LiouvillePro}. Then
 \[N_{u_0}^2 = \|\bb{u}^0\|^2 = \|\bb{\rho}^0\| \lesssim \|\bb{G}\|\|(\bb{w}^\omega)^0\|/M^d \lesssim \|(\bb{w}^\omega)^0\|/M^{d/2} = n_H.\]
 Here, $(\bb{w}^\omega)^0$ is exactly the $(\bb{\psi}^\omega)^0$ in Theorem \ref{thm:obsFDLouville} when considering the problem \eqref{LiouvillePro}.
\end{remark}

\subsubsection{The expectation of observables for the QLSA} \label{subsubsect:obsQLSASchr}

Unlike the quantum simulation, the solution of the QLSA is a quantum state that is a superposition of the solution at all temporal and spatial points, denoted as
\[\ket{\widetilde{\bb{u}}} = [\widetilde{\bb{u}}^1; \cdots; \widetilde{\bb{u}}^{N_t}], \qquad
\widetilde{\bb{u}}^n = \frac{1}{N_u}\bb{u}^n,\]
where the normalization constant is
\[N_u = \|\bb{u}\| = (\|\bb{u}^1\|^2+ \cdots + \|\bb{u}^{N_t}\|^2)^{1/2} = \sqrt{N_t} \|\bb{u}^0\|
=: \sqrt{N_t} N_{u_0}.\]
Here we have used \eqref{condUn}.
Let $O_{\bb{i}} = \ket{\bb{i}}\bra{\bb{i}}$, $O_n = \ket{n}\bra{n}$ and
\[O_{\bb{i}}^n = O_n \otimes O_{\bb{i}} = \ket{n,\bb{i}}\bra{n,\bb{i}},\]
where $\ket{n}$ is of size $N_t$.
Then the position density
\[\rho(t = t_n, x_{\bb{i}}) = (\bb{u}^n)^\dag  O_{\bb{i}} \bb{u}^n
= \bb{u}^\dag (O_n \otimes O_{\bb{i}}) \bb{u}
= N_t N_{u_0}^2 \cdot \langle \widetilde{\bb{u}} | O_{\bb{i}}^n | \widetilde{\bb{u}} \rangle.\]
The expectation $\langle O_{\bb{i}}^n \rangle := \langle \widetilde{\bb{u}} | O_{\bb{i}}^n | \widetilde{\bb{u}} \rangle$ satisfies the condition that ${\rm Var}( O_{\bb{i}}^n)$ is bounded.
In this case, however, we must evaluate $\langle O_{\bb{i}}^n \rangle$ to precision $\mathcal{O}(\varepsilon/(N_tN_{u_0}^2))$, which increases the number of samples by another factor $(N_tN_{u_0}^2)^2$ when considering the general sampling law. We remark that the factor $N_t^2$ can be removed by using the dilation procedure. In this case, the multiplicative factor is still given by $n_\rho = N_{u_0}^4/\varepsilon^2$.

For the current density, one easily finds that ($d=1$)
\[J(t_n,x_i) = \frac{\hbar}{2\i} (\bb{u}^n)^\dag (A - A^\dag) \bb{u}^n
 = N_t N_{u_0}^2\cdot \frac{\hbar}{2\i} \widetilde{\bb{u}}^\dag (A - A^\dag) \widetilde{\bb{u}}.\]
We still need to apply the dilation procedure \eqref{dilation} to remove the unexpected multiplicative factor $N_t^2$, and still obtain $n_J = M^2 N_{u_0}^4/\varepsilon$ for the current density.

The kinetic energy can be analysed similarly, with the factor given by $n_E = M^4 N_{u_0}^4 $.

\subsection{Gate complexity for the computation of the observables}

It is worth pointing out that the time step $\Delta t$ can be chosen independently of the small parameter $\hbar$ if one is only concerned with the computation of the physical quantities. This observation was interpreted by using the Wigner transformation approach in
\cite{BJM2002splitting}, and mathematically rigorously investigated in \cite{GJP2021convergenceSplitting,Caroline2020Acta}.
For instance, the first-order time splitting spectral method gives \cite{Jin2022quantumSchrodinger,Caroline2020Acta}
\begin{equation}\label{obsp1}
\|\langle O \rangle_{u^n} - \langle O \rangle_{u(t_n,\cdot)}\|
\le C_\ell  n \Big(  d \Big( \frac{\Delta x}{\hbar} \Big)^\ell  +  \Delta t^2 + \Delta t \hbar^2 \Big),
\end{equation}
where $\hbar = \mathcal{O}(\sqrt{\varepsilon})$ implies the above observation since $n \Delta t \hbar^2 = \mathcal{O}(\varepsilon)$. Note that the term $ n \Delta t \hbar^2 = t \hbar^2$ is the error between the classically evolved Wigner function and the expectation value of the Schr\"odinger solution  \cite{Caroline2020Acta}.

\subsubsection{The gate counts of the quantum simulation method}

\begin{theorem}\label{thm:SchrObs}
Given the error tolerance $\varepsilon$, suppose that the estimate \eqref{obsp1} holds with $C_\ell$ considered as $\mathcal{O}(1)$.
The gate complexities for the observables from the Schr\"odinger equation \eqref{Schrodinger} are given by
   \[N_{Gates} ( \langle O \rangle )  = \mathcal{O}\Big(  c_O\frac{N_{u_0}^4 d}{\varepsilon^3} \log \frac{d^{1/\ell}}{\varepsilon^{1/2+2/\ell}}  \Big),\]
   where
  \begin{equation}\label{nO}
   c_O = \begin{cases}
   1,  & \text{for the position density},\\
    d^{2/\ell}/\varepsilon^{4/\ell}, & \text{for the current density},\\
   d^{4/\ell}/\varepsilon^{8/\ell}, & \text{for the kinetic energy}.
   \end{cases}
  \end{equation}
\end{theorem}
\begin{proof}
For the observables, one can implement the mesh strategy according to \eqref{obsp1}, given by
  \begin{equation}\label{meshS2}
\Delta t = \mathcal{O}(\varepsilon), \quad \Delta x = \mathcal{O}(\varepsilon^{1/2+2/\ell}/d^{1/\ell}).
  \end{equation}
Then the number of gates for outputting the quantum sate is
\[\mathcal{O}\Big( \frac{d}{\varepsilon} \log \frac{d^{1/\ell}}{\varepsilon^{1/2+2/\ell}}  \Big).\]
For computing the observables, we must add a multiplicative factor $\text{Var}(O)/\varepsilon^2$ if the general sampling law is used. According to the previous discussion, one has
\begin{equation}\label{nObser}
n_\rho = \frac{N_{u_0}^4}{\varepsilon^2}, \qquad
n_J = \frac{M^2 N_{u_0}^4}{\varepsilon} \sim \frac{N_{u_0}^4 d^{2/\ell}}{\varepsilon^{2+4/\ell}}, \qquad
n_E = M^4 N_{u_0}^4 \sim \frac{N_{u_0}^4 d^{4/\ell}}{\varepsilon^{2 + 8/\ell}}.
\end{equation}
This completes the proof.
\end{proof}

\subsubsection{The gate counts of the quantum linear systems algorithm}

\begin{theorem}\label{thm:SchrObsQLSA}
Given the error tolerance $\varepsilon$, suppose that the estimate \eqref{obsp1} holds with $C_\ell$ considered as $\mathcal{O}(1)$.
The gate complexities of the QLSA for the observables from the Schr\"odinger equation \eqref{Schrodinger} are given by
   \[N_{Gates} ( \langle O \rangle )  = \widetilde{\mathcal{O}}\Big( c_O \frac{N_{u_0}^4 d^{2+2/\ell}}{\varepsilon^{3+4/\ell}} \Big) ,\]
   where $c_O$ is defined by \eqref{nO}.
\end{theorem}
\begin{proof}
For the mesh strategy in \eqref{meshS2}, according to the proof of Theorem \ref{thm:QLSASchrodinger}, one easily finds that the number of gates for approximating the wave function is
\[N_{\text{Gates}}  = \widetilde{\mathcal{O}}\Big( \frac{d^{2+2/\ell}}{\varepsilon^{1+4/\ell}} \Big),\]
where in the last equal sign we have included the additional factor $d$ arising from the matrix order.
For computing the observables, one just need to include the multiplicative factor $\text{Var}(O)/\varepsilon^2$ as given in \eqref{nObser}.
\end{proof}

\section{Linear representation approach for scalar nonlinear hyperbolic PDEs}\label{Sec: Hyp}

We consider the linear representations for the spatially $d$-dimensional scalar nonlinear hyperbolic PDE
\begin{equation}\label{scalarPDEs}
\begin{cases}
\partial_t u + F(u)\cdot \nabla_x u + Q(x,u) = 0, \\
u(0,x) = u_0(x),
\end{cases}
\end{equation}
where $x\in \mathbb{R}^d$ and $u\in \mathbb{R}$.

\subsection{The Liouville representation}

For this general scalar hyperbolic equation, one can still use the Liouville representation but the Schr\"odinger representation is not available.
In fact, the Liouville representation has been considered in \cite{JinLiu2022nonlinear} by using the level set formalism. To this end, we first review the construction.

Let $\phi(t,x,p)$ be the level set function in $(d+1)+1 = d+2$ dimensions, where $p \in \mathbb{R}$. The zero level set of $\phi$ gives solution $u$:
\[\phi(t,x,p) = 0 \quad \mbox{at} \quad p = u(t,x).\]
One easily finds that $\phi$ satisfies
\[\begin{cases}
\partial_t \phi + F(p)\cdot \nabla_x \phi - Q(x,p)\partial_p \phi = 0,  \\
\phi(0,x,p) = p - u_0(x).
\end{cases}\]
Like for the Hamilton-Jacobi PDEs, we can similarly define a function $\varphi$ such that
\begin{equation}\label{Liouvillescalar}
\begin{cases}
\partial_t \varphi + F(p)\cdot \nabla_x \varphi - Q(x,p)\partial_p \varphi = 0,  \\
\varphi(0,x,p) = \delta(p - u_0(x)),
\end{cases}
\end{equation}
with the solution given by $\varphi(t,x,p) = \delta(\phi(t,x,p))$.  Eq.~\eqref{Liouvillescalar} is referred to as the Liouville representation of \eqref{scalarPDEs}.

One can apply the quantum difference method to solve \eqref{Liouvillescalar} and compute the physical observables as in Subsect.~\ref{subsect:HJFD}. For the spectral discretisation, one can utilize the Trotter based technique in Subsect.~\ref{subsubsect:spectralLiouoville}. The similar numerical performance can be deduced, so we omit the detailed discussions in view of the length of the article and the similarity of the numerical implementation.

\subsection{The KvN representation}

For scalar nonlinear hyperbolic PDEs, it is not very clear how to formulate the KvN representation. Here we offer an idea, which is inspired by the evolution of the phase factor of the KvN wave function in \cite{Joseph2020KvN}.

Let $\bb{x} = (x,p)$ and denote $\bb{v} = [F(p), -Q(x,p)]$. Then problem \eqref{Liouvillescalar} can be written as
\begin{equation}\label{Liouvillescalar-1}
\begin{cases}
\partial_t \varphi + \bb{v} \cdot \nabla_{\bb{x}} \varphi = 0,  \\
\varphi_0(\bb{x}) = \varphi(0,\bb{x}) = \varphi(0,x,p) = \delta(p - u_0(x)).
\end{cases}
\end{equation}
Let $\psi$ be the complex-valued KvN wave function and $f = \psi^\dag\psi$ be the probability distribution function. Inspired by the discussion in Section II(B) of \cite{Joseph2020KvN}, we define $\psi = f^{1/2} \e^{\i \varphi}$:
\begin{itemize}
  \item The amplitude $f$ satisfies the Liouville equation in conservative form:
  \begin{equation}\label{amplitude}
\begin{cases}
\partial_t f + \nabla_{\bb{x}} \cdot  (\bb{v} f) = 0,  \\
f(0,\bb{x}) = f_0(\bb{x}),
\end{cases}
\end{equation}
   where $f_0(\bb{x}) = \delta(\bb{x}- \bb{q}_0)$ and $\bb{q}_0$ is an arbitrarily given vector.
  \item The phase factor $\varphi$ satisfies the non-conservative Liouville equation in \eqref{Liouvillescalar-1}.
\end{itemize}
As in \cite{Joseph2020KvN}, one can check that the KvN wave function $\psi$ is governed by the KvN equation
\begin{equation}\label{KvNscalar}
\i \partial_t \psi = \mathcal{H}_{\text{KvN}} \psi = -\i \Big(\bb{v}\cdot \nabla_{\bb{x}} + \frac{1}{2} \nabla_{\bb{x}} \cdot \bb{v} \Big) \psi,
\end{equation}
with the initial data given by
\[\psi_0(\bb{x}) = \psi(0,\bb{x}) = f_0^{1/2}(\bb{x})\e^{\i \varphi_0(\bb{x})}.\]

The original intention of \cite{Joseph2020KvN} is to introduce the linear representation for the following nonlinear ODEs
\begin{equation}\label{vtx}
\frac{\d \bb{q}}{\d t} = \bb{v}(t, \bb{q}).
\end{equation}
Like in Sect.~\ref{sect:linearrep}, one can first obtain the Liouville representation \eqref{amplitude} corresponding to \eqref{vtx}. The KvN representation is then derived by assuming the non-conservative hyperbolic equation \eqref{Liouvillescalar} for the phase factor. In view of the evolution of the phase factor, we therefore propose the KvN representation for the scalar nonlinear hyperbolic PDEs.

It should be pointed out that one cannot get the phase factor $\varphi$ from $\psi = f^{1/2} \e^{\i \varphi}$ since $\e^{\i \varphi}$ is periodic with respect to $\varphi$. For this reason, it may be impossible to define the physical observables
\[\langle g(t,x) \rangle = \int g(p) \varphi(t,x,p) \d p\]
as in \cite{JinLiu2022nonlinear}. One may instead define
\[\langle g_\chi(t,x) \rangle = \int g(p) \chi(\varphi(t,x,p)) \d p\]
to remove the unexpected period, where $\chi$ is a function with period $2\pi$. However, $\langle g_{\chi}(t, x)\rangle$ is actually very tricky to compute. It's not clear if this is possible to do. Suppose we have $\chi(\varphi)=\sin(\varphi)$ or $\cos(\varphi)$. If we have access to a Hamiltonian diagonal in $\varphi$, i.e. $H=\sum_j \varphi_j|j\rangle \langle j|$, then we can create a control unitary $U=|0\rangle \langle 0| \otimes \mathbf{1}+|1\rangle \langle 1|\otimes \e^{\i H}$. Then a standard Hadamard test can be employed to compute $\langle g_{\chi}(t, x)\rangle =\text{Im} \langle \sqrt{g}| U|\sqrt{g}\rangle$  or $\text{Re} \langle \sqrt{g}| U|\sqrt{g}\rangle$ where $|\sqrt{g}\rangle$ has amplitudes $\sqrt{g_j}$. However, by solving Eq.~\eqref{KvNscalar} alone, one cannot create access to $H$ unless every $\varphi_j$ is individually computed, which defeats the purpose of a quantum algorithm.

Instead we can try to create the state $|\sin(\varphi)\rangle$ or $|\cos(\varphi)\rangle$ whose amplitudes are proportional to $\sin(\varphi_j)$ or $\cos(\varphi_j)$. Note that
\[\sin \varphi_j=\text{Im}(\psi_j/\sqrt{f_j}) = \text{Im}(\psi_j/\sqrt{\psi_j\psi_j^\dag})
= \frac{1}{2\i} \frac{\psi_j-\psi_j^\dag}{\sqrt{\psi_j\psi_j^\dag}}, \qquad \cos(\varphi_j)=\frac{\psi_j+\psi_j^\dag}{2\sqrt{\psi_j \psi_j^\dag}}.\]
The question is if we can create a quantum state with amplitudes proportional to $\text{Im}(\psi_j/\sqrt{f_j})$ or $\text{Re}(\psi_j/\sqrt{f_j})$. If we solve Eq.~\eqref{KvNscalar} alone to obtain $|\psi\rangle$, then we cannot easily access these states.

An alternative to solving Eq.~\eqref{KvNscalar} alone is to include also its complex conjugate and we instead solve for $\widetilde{\psi}$ obeying
\[
\i \frac{\partial}{\partial t} \widetilde{\psi}=\widetilde{\mathcal{H}}_{\text{KvN}}\widetilde{\psi}
\]
where
\[
    \widetilde{\psi}=
\begin{bmatrix}
\psi-\psi^{\dagger} \\
\psi+\psi^{\dagger}
\end{bmatrix}, \qquad \widetilde{\mathcal{H}}_{\text{KvN}}=\begin{bmatrix}
\mathcal{H}_{\text{KvN}}& 0 \\
0 & \mathcal{H}_{\text{KvN}}
\end{bmatrix}.
\]
We now perform quantum simulation of the state  $|\widetilde{\psi}\rangle \propto |\psi-\psi^{\dagger}\rangle+|\psi+\psi^{\dagger}\rangle$ with the new Hamiltonian $\widetilde{\mathcal{H}}_{\text{KvN}}$ and we can choose to post-select either the state $|\psi-\psi^{\dagger}\rangle$ or $|\psi+\psi^{\dagger}\rangle$. Given $|\psi\pm\psi^{\dagger}\rangle$, we can compute the inner product $|\langle g|\psi\pm\psi^{\dagger}\rangle|^2$ to obtain the observable $\langle g_{\chi}(t,x)\rangle$ for $\chi(\varphi)=|\psi\pm\psi^{\dagger}|$ using standard methods. This differs from $\chi(\varphi)=\sin(\varphi)$ or $\cos(\varphi)$ up to norm factors $\sqrt{\psi \psi^{\dagger}}$. We leave it as an open question on how to design algorithms for more general $\chi(\varphi)$ functions.

\section{Summary and discussion} \label{summary}

In this paper, we systematically studied the quantum difference methods and the quantum spectral methods for solving the linear representations of nonlinear ODEs and nonlinear PDEs. Since our studies involve many different methods,   we summarize the results for computing the physical observables in Tab.~\ref{tab:summaryObs}, from which we clearly observe that the quantum simulation methods give the best performance in the computational cost. On the other hand, for the classical treatment the order of the iterative matrix in $\bb{u}^{n+1} = B \bb{u}^n$ scales as $\mathcal{O}(N_x^d)$, leading to the classical run time scaling as $C = \mathcal{O}( d^{d+3}/\varepsilon^{d+1} )$  for the first-order hyperbolic equation discussed in \cite{JLY2022multiscale}, while the quantum cost is $Q = \mathcal{O}( d^3\log(1/\varepsilon))$,  where $\varepsilon = \mathcal{O}(d/N_x)$ and the quantum difference method is employed. This implies the exponential speedup with respect to $d$ and $\varepsilon$ even for producing the observables.

\begin{table}[!htb]
\begin{threeparttable}
  \centering
  \caption{Time complexities for the computation of physical observables}  \label{tab:summaryObs}
  \begin{tabular}[\textwidth]{|c|cc|cc|cc|}
  \hline
  \multicolumn{7}{|c|}{\textbf{Nonlinear Ordinary Differential Equations}}    \\
  \hline
  \multirow{2}{*}{Problem} & \multicolumn{2}{c|}{Quantum simulation}  & \multicolumn{2}{c|}{Spectral QLSA}
  & \multicolumn{2}{c|}{FD QLSA~($\alpha\ge 4$)}\\ \cline{2-7}
         &  Subroutine   & Observable   & Subroutine  & Observable  & Subroutine   & Observable \\
  \hline
  \makecell[c]{Liouville \\ representation}
  & $\dfrac{d^{2+2/\ell}}{\varepsilon^{2+4/\ell}} ^*$   & $\dfrac{ n_L^4 d^{2+2/\ell}}{\varepsilon^{4+4/\ell}} ^*$
  & $\dfrac{d^{3+2/\ell}}{\varepsilon^{4+4/\ell}} ^*$   & $\dfrac{ n_L^4 d^{3+2/\ell}}{\varepsilon^{6+4/\ell}} ^*$
  & $\dfrac{d^\alpha}{\varepsilon^3}$  & $\dfrac{n_L^4d^\alpha}{\varepsilon^5}$   \\
  \hline
  \makecell[c]{KvN \\ representation}
  & $\dfrac{d^{2+2/\ell}}{\varepsilon^{2+4/\ell}}$  & $\dfrac{n_L^2 d^{2+2/\ell}}{\varepsilon^{4+4/\ell}}$
  & $\dfrac{d^{3+2/\ell}}{\varepsilon^{2+4/\ell}}$ & $\dfrac{n_L^2d^{3+2/\ell}}{\varepsilon^{4+4/\ell}}$
  & $\dfrac{d^\alpha}{\varepsilon^3}$  & $\dfrac{n_L^2d^\alpha}{\varepsilon^5}$   \\
  \hline
  \multicolumn{7}{|c|}{\textbf{Nonlinear Hamilton-Jacobi PDEs$^\sharp$}}    \\
  \hline
  \multirow{2}{*}{Problem} & \multicolumn{2}{c|}{Quantum simulation}  & \multicolumn{2}{c|}{Spectral QLSA}
  & \multicolumn{2}{c|}{FD QLSA~($\alpha\ge 4$)}\\ \cline{2-7}
         &  Subroutine   & Observable   & Subroutine  & Observable  & Subroutine   & Observable \\
  \hline
  \makecell[c]{Liouville \\ equation}
  & $\dfrac{d}{\varepsilon^2}$   & $\dfrac{n_H^4 d}{\varepsilon^4}$
  & $\dfrac{d^{2+2/\ell}}{\varepsilon^{2+4/\ell}}$  & $\dfrac{n_H^4d^{2+2/\ell}}{\varepsilon^{4+4/\ell}}$
  &  $\dfrac{d^\alpha}{\varepsilon^3}$  & $\dfrac{n_H^4d^\alpha}{\varepsilon^5}$  \\
  \hline
  \makecell[c]{Schr\"odinger \\ equation}
  & $\dfrac{d}{\varepsilon}$  & $\dfrac{c_ON_{u_0}^4 d}{\varepsilon^3}$
  & $\dfrac{d^{2+2/\ell}}{\varepsilon^{1+4/\ell}}$  & $\dfrac{c_O N_{u_0}^4 d^{2+2/\ell}}{\varepsilon^{3+4/\ell}}$
  &    & \\
  \hline
  \end{tabular}
   \begin{tablenotes}
  \item[a)] The notations $\mathcal{O}$ for quantum simulations and $\widetilde{\mathcal{O}}$ for QLSA based methods are omitted. Since the dependence on the matrix order is not explicitly presented in \cite{Costa2021QLSA}, we just include the multiplicative factor $d^{\alpha-3}~(\alpha\ge 4)$ in the gate complexity for the finite difference discretisations (note that the order of the matrix grows exponentially with respect to the dimension).
  \item[b)] $n_L$ and $n_H$ are the sampling factors for the Liouville equation, given respectively in Theorems \ref{thm:obsFDLouvilleRep} and \ref{thm:obsFDLouville}.  $N_{u_0}$ and $c_O$ are defined by \eqref{condUn} and \eqref{nO}, respectively. Note that $c_O$ depends on $d, \varepsilon$ when computing the current density and kinetic energy. For the Hamilton-Jacobi equations, $c_O$ may be neglected, and $N_{u_0} \lesssim n_H^{1/2}$ (see Remark \ref{rem:Nu0}).
  \item[*)] Despite the absence of the unitary structure, we still proposed a ``quantum simulation'' algorithm for the Liouville representation in Appendix~\ref{subsect:spectralLiouvillerep}, where non-unitary procedures are involved. The results are presented in Theorem \ref{thm:QLSAspectralLiouvilleRep} and Theorem \ref{thm:simulationLiouvillerep} when the cost arising from multiple copies of initial quantum states is ignored.
  \item[$\sharp$)] For the spectral discretisation of the Liouville/Schr\"odinger equation, we only consider a prototype case $H(x,p) = \frac{1}{2}|p|^2 + V(x)$.
  \end{tablenotes}
  \end{threeparttable}
\end{table}

\begin{figure}[!htb]
  \centering
  \includegraphics[scale=0.7]{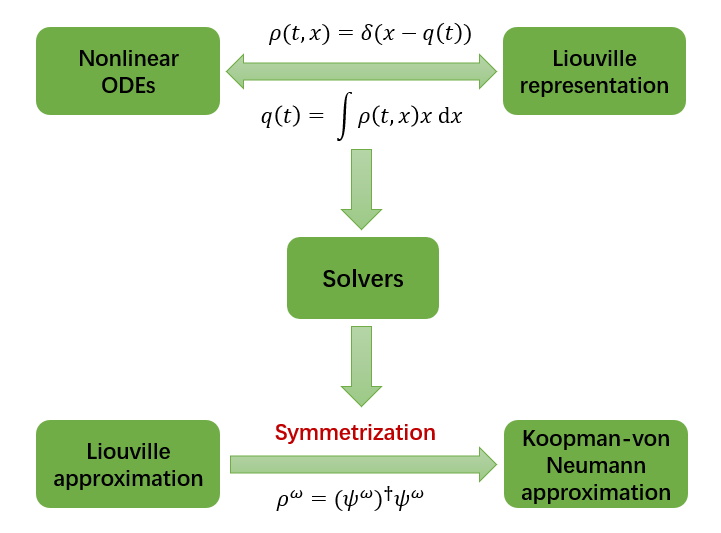}\\
  \caption{Schematic diagram of linear representations}\label{fig:SchematicLinearRep}
\end{figure}

Let us summarise the several parts of the article as follows.

(1) Motivated by the idea in \cite{Dobin2021plasma,JinLiu2022nonlinear}, we established the correspondence between the nonlinear dynamic system and the Liouville representation via a simple ansatz, which relates the ODE solution and the density distribution by the Dirac delta function. In this case, the ODE solution can be recast as a physical observable of the Liouville equation, which provides an efficient way to solve the nonlinear ODEs by using quantum algorithms for the resulting linear Liouville equation.  We introduced the Liouville approximation and the KvN approximation
from the perspective of quantum differential equations solvers for the Liouville equation with smoothed initial data,  while the KvN approximation can be regarded as the ``symmetrized'' counterpart of the Liouville approximation (See Fig.~\ref{fig:SchematicLinearRep} for illustration).
For both linear representation approaches, we proposed the upwind difference discretisations and Fourier spectral discretisations and provided in detail the time complexity analysis for the QLSA based methods and the quantum simulation methods, including the output of the quantum states and the post-processing for the computation of physical observables. The KvN mechanism allows direct quantum Hamiltonian simulations, while the Liouville method can also be translated into the evolution of symmetric propagators with additional non-unitary procedures at every time step after an appropriate Trotter based approximation.

(2) For the nonlinear PDEs, more specifically the Hamilton-Jacobi equations, by using the level set mechanism as proposed in \cite{JinLiu2022nonlinear}, one can map the nonlinear PDEs of $(d+1)$-dimension to a linear $(2d+1)$-dimensional Liouville equation, referred to as the Liouville representation for nonlinear PDEs. For a classical device, doubling the dimension of the problem may seem too costly because the cost increases exponentially with dimension. However, for quantum algorithms, the relative overhead in doubling the dimension can be up to exponentially smaller, which has been verified by the proposed quantum algorithms since no exponential terms in dimension like $d^d$ and $(1/\varepsilon)^d$ for the classical cost are included.

(3) It is well-known that the Schr\"odinger equation can be transformed into the quantum Liouville equation via the Wigner transform, which in turn leads to the Liouville equation when taking the semiclassical limit. In view of the close relations between their physical observables, we introduced the Schr\"odinger framework for solving the Liouville equation. We studied the quantum interpretation of the classical time-splitting Fourier spectral method proposed in \cite{BJM2002splitting} for the Schr\"odinger equation, and presented a comprehensive discussion in the correspondence between the time-splitting spectral method and the Trotter based Hamiltonian simulation, although this issue has been addressed (in less detail) in some earlier literature. The time-splitting spectral discretisation for the Schr\"odinger equation generates a discretised Hamiltonian system, which can be handled by standard Hamiltonian simulation algorithms or quantum linear systems algorithms. We analysed in detail the gate complexity of these two approaches for numerically resolving the wave function and the physical observables. Despite the advantages in terms of time complexity, it should  be pointed out that the solution to the Schr\"odiner equatin is {\it oscillatory}, hence if one wants high-resolution (oscillation-free) numerical results the Liouville framework is preferred.


\subsection*{Declaration of competing interest}

The authors declare that they have no known competing financial interests or personal relationships that could have
appeared to influence the work reported in this paper.

\section*{Acknowledgement}
SJ was partially supported by the NSFC grant No.~12031013, the Shanghai Municipal Science and Technology Major Project (2021SHZDZX0102), and the Innovation Program of Shanghai Municipal Education Commission (No. 2021-01-07-00-02-E00087).  NL acknowledges funding from the Science and Technology Program of Shanghai, China (21JC1402900), the Shanghai Pujiang Talent Grant (no. 20PJ1408400) and the NSFC International Young Scientists Project (no. 12050410230). YY was partially supported by China Postdoctoral Science Foundation (no. 2022M712080).

\appendix

\addcontentsline{toc}{section}{Appendix}

\section{Spectral discretisation for the Liouville representation of nonlinear ODEs} \label{subsect:spectralLiouvillerep}

\subsection{The Trotter based spectral discretisation} \label{subsubsect:spectralLiouoville}

Consider the problem \eqref{LiouvilleExplicit}. Let $u = F_i(x) w(t,x)$. According to the above discussion, one has
\[ -\i \frac{\partial}{\partial x_i} u \longrightarrow  \hat{P}_i^{\rm d}\bb{u} = \bb{P}_i \bb{u} = \bb{P}_i \Lambda_{F_i} \bb{w},\]
where $\Lambda_{F_i} = \text{diag}(\bb{F}_i)$ is a diagonal matrix, where the vector $\bb{F}_i= \sum_{\bb{j}} F_i(x_{\bb{j}}) \ket{\bb{j}}$. The resulting system of ordinary differential equations is
\begin{equation}\label{SpectralSystemLiouvilleRep}
\begin{cases}
\frac{\d }{\d t} \bb{w}(t) = -\i A \bb{w}(t),\\
\bb{w}(0) = \bb{w}^0 = ( w_0(x_{\bb{j}}) ),
\end{cases}
\end{equation}
where $A = \sum_{i=1}^d A_i$ with $A_i = \bb{P}_i \Lambda_{F_i}$. One can check from \eqref{PlDl} that $\bb{P}_i$ are Hermitian matrices. However, we note that this does not mean that $A$ is Hermitian, since in general $(\bb{P}_i \Lambda_{F_i})^\dag = \Lambda_{F_i} \bb{P}_i \ne \bb{P}_i \Lambda_{F_i}$.

The evolution of \eqref{SpectralSystemLiouvilleRep} can be {\it formally} written as
\[\ket{\psi(t+\Delta t)} = \e^{-\i (A_1 + \cdots + A_d) \Delta t } \ket{\psi(t)},\]
where $\ket{\psi(t)}$ is a quantum state whose amplitudes are proportional to $\bb{w}(t)$ and the evolution operator $\exp(-\i A_i\Delta t)$ is not necessarily unitary.
Let us consider the first-order product formula
\begin{equation}\label{Wdelta}
U_{\Delta t} = \e^{-\i A_d \Delta t}\cdots \e^{-\i A_1 \Delta t} .
\end{equation}
One obtains from \cite{Nielsen2010,Childs2021Trotter} that
\begin{equation}\label{CF0A}
\e^{-\i (A_1 + \cdots + A_d) \Delta t } = U_{\Delta t} + C_A \Delta t^2,
\end{equation}
where $C_A$ depends on the matrix $A$, considered as $\mathcal{O}(1)$ in the following. Therefore, the problem is reduced to the simulation of each $A_j$, where $A_j$ is not necessarily symmetric. Take $j=1$ as an example and consider the decomposition $\Lambda_{F_1}  = \Lambda_{F_1}^+ - \Lambda_{F_1}^-$, where $\Lambda_{F_1}^\pm = \text{diag}(d_1^\pm,\cdots,d_n^\pm)$ are diagonal matrices with $d_j^\pm > \alpha$ for some positive constant $\alpha$. One can further require that $\|\Lambda_{F_i}^\pm\| \lesssim \max_{1\le j \le d} \|\bb{F}_j\|$. Let $A_1^\pm = \bb{P}_1 \Lambda_{F_1}^\pm$. The Strang splitting gives
\begin{equation}\label{strangApF}
\e^{-\i A_1 \D t} = \e^{-\i A_1^+ \D t/2}\e^{\i A_1^- \D t}\e^{-\i A_1^+ \D t/2} + C_1 \D t^3,
\end{equation}
where $\e^{\i A_1^- \D t}$ and $\e^{-\i A_1^+ \D t/2}$ can be evolved in the similar way (note that one can also use the first-order approximation). To this end, we consider the second one as an example.

The simulation of $\e^{-\i A_1^+ \Delta t/2}$ is related to the following ODEs:
\[\frac{\d }{\d t} \bb{w}(t) = - \frac{\i}{2}A_1^+ \bb{w}(t) = - \frac{\i}{2} \bb{P}_1 \Lambda_{F_1}^+\bb{w}(t),  \qquad 0\le t \le \D t.\]
Let $\widetilde{\bb{w}} = \sqrt{\Lambda_{F_1}^+} \bb{w}$, where
$\sqrt{\Lambda_{F_1}^+} = \text{diag}(\sqrt{d_1^+}, \cdots, \sqrt{d_n^+})$. Then the above ODEs can be reformulated as
\begin{equation}\label{ODEtw}
\frac{\d }{\d t} \widetilde{\bb{w}}(t) = - \frac{\i}{2} \widetilde{A}_1^+ \widetilde{\bb{w}}(t),   \qquad 0\le t \le \D t,
\end{equation}
where $\widetilde{A}_1^+  =  \sqrt{\Lambda_{F_1}^+} \bb{P}_1 \sqrt{\Lambda_{F_1}^+}$ is a Hermitian matrix.
The one-step simulation gives
\begin{equation}\label{tpsi1}
\ket{\widetilde{\psi}(t+\Delta t)} = \e^{-\i \widetilde{A}_1^+ \Delta t/2}\ket{\widetilde{\psi}(t)},
\end{equation}
where $\ket{\widetilde{\psi}(t)}$ corresponds to $\widetilde{\bb{w}}(t)$.
Since $d_j^\pm > \alpha>0$, $\widetilde{\bb{w}} = \sqrt{\Lambda_{F_1}^+} \bb{w}$ can be viewed as a linear systems problem:
\begin{equation}\label{w2tw}
\widetilde{\bb{w}} = D_1^{-1} \bb{w}, \qquad D_1 = (\sqrt{\Lambda_{F_1}^+})^{-1}.
\end{equation}
Similarly,
\begin{equation}\label{tw2w}
\bb{w} = D_2^{-1}  \widetilde{\bb{w}}, \qquad D_2 = \sqrt{\Lambda_{F_1}^+} = D_1^{-1}.
\end{equation}

Given the initial state of $\bb{w}^0$,  denoted by $\ket{\psi^0}$. At each time step, one needs to consider the procedure
\[\ket{\psi^0}
\xrightarrow { \eqref{w2tw} } \ket{\widetilde{\psi}^0}
\xrightarrow { \eqref{tpsi1} } \ket{\widetilde{\psi}^1}
\xrightarrow { \eqref{tw2w} } \ket{\psi^1}\]
for $\e^{-\i A_1^+ \Delta t/2}$, followed by the similar procedures for $\e^{\i A_1^- \Delta t}$ and the first $\e^{-\i A_1^+ \Delta t/2}$ in \eqref{strangApF}, where \eqref{tpsi1} can be solved by quantum Hamiltonian simulations or quantum differential equations solvers.

\begin{remark} \label{rem:multiplecopies}
The transition between $\widetilde{\bb{w}}$ and $\bb{w}$ in $\eqref{w2tw}$ and \eqref{tw2w} may be implemented in a simpler manner, for example, the LCU method, which decomposes the diagonal (and Hermitian) matrices $D_i=\text{diag}(d_{i,1},\cdots,d_{i,n})$, $i=1,2$ into a sum of two unitary operations. In fact, it is always possible to write $D_i=(U_i+V_i)/2$, where $U_i=D_i+\i\sqrt{\mathbf{1}-D^2_i}$, and $V_i=D_i-\i\sqrt{\mathbf{1}-D^2_i}$ (One can assume $\|D_i\|\leq 1$ after an adjustment). These unitaries are also diagonal matrices with diagonal entries $\text{diag}(d_{i,1} \pm \i\sqrt{1-d^2_{i,1}},\cdots,d_{i,n} \pm \i\sqrt{1-d^2_{i,n}})$. Applying the operation $D_i$ onto a quantum state can be done by a straightforward application of the LCU method (e.g. Lemma 6 in \cite{Childs2017QLSA}). Here we can assume access to the control operation $|0\rangle \langle 0| \otimes U_i+|1\rangle \langle 1|\otimes V_i$.

However, multiple copies are needed at every time step for both the QLSA and the LCU since they are not unitary procedures, hence the cost (i.e. number of copies needed of the initial state) will increase exponentially with $N_t$. We can see the last statement more explicitly.
\begin{itemize}
  \item For the QLSA, when solving $D_1 \ket{\widetilde{w}} = \ket{w}$ with the quantum state $\ket{w}$ given, one must prepare unitary operations to query the entries of $w$. This needs post-processing or multiple uses of $\ket{w}$, hence multiple copies of $\ket{w}$.
  \item For the LCU, to obtain the state $D_i |w\rangle$ for some state $|w\rangle$, one can construct a unitary procedure acting on $|w\rangle$ and an ancilla that will output a state $|w'\rangle=\alpha_i D_i |w\rangle+\beta_i|v\rangle$, where $|v\rangle$ is some state we don't want. One can then obtain a single copy of $D_i|w\rangle$ upon post-selection of $|w'\rangle$ with $\mathcal{O}(1/\|\alpha_i D_i |w\rangle\|^2)$ number of measurements, and hence multiple copies of $|w\rangle$, where $\|\alpha_i D_i |w\rangle\| \le 1$.
\end{itemize}
We are not only interested in the cost at every time step: we want to evolve for a long time, to $N_t$ time-steps. Suppose we wanted to repeat this procedure $N_t$ times with at least $C>1$ copies at every time step. At the last $N_t^{\text{th}}$ step if we need only a single copy of the desired state, then $C$ copies of the state in the previous $N_t-1$ time-step is required, which means $C^2$ copies of the state in the $(N_t-2)^{\text{th}}$ time-step is needed. Ultimately, this requires $C^{N_t}$ copies of the original $|w\rangle$ state in the initial time-step.
\end{remark}

In the following, we ignore the cost of multiple copies at each time step.

\subsubsection{The QLSA for the spectral discretisation}

\begin{theorem}\label{thm:QLSAspectralLiouvilleRep}
Assume further that $\max_{1\le j \le d} \|\bb{F}_j\| = \mathcal{O}(1)$ and $T = \mathcal{O}(1)$.
\begin{enumerate}[(1)]
  \item There exists a quantum algorithm that produces a state $\varepsilon$-close to $\bb{w}(T)/\|\bb{w}(T)\|$ with the gate complexity given by
\[N_{Gates} = \widetilde{\mathcal{O}}\Big(\frac{d^{3+2/\ell}}{\varepsilon^{4+4/\ell}}\Big).\]
  \item The observable of the Liouville representation can be computed with gate complexity given by
\[N_{\text{Gates}}( \langle O \rangle) = \widetilde{\mathcal{O}}\Big(\frac{ n_L^4 d^{3+2/\ell}}{\varepsilon^{6+4/\ell}}\Big),\]
where $n_L = \|(\bb{\rho}^\omega)^0\|/M^{d/2}$.
\end{enumerate}
\end{theorem}

\begin{proof}
For simplicity, we omit the discussion of the cost and the error resulting from \eqref{w2tw} and \eqref{tw2w}.

(1) Since $\widetilde{A}_1^+$ is hermitian, the eigenvalues of $\widetilde{A}_1^+$ are real and $\kappa_V = 1$, where $V$ is the transformation matrix associated with $\widetilde{A}_1^+$. At each time step, by Lemma \ref{lem:BCO2017}, the gate complexity of solving \eqref{ODEtw} within error $\eta$ is
\[Q_{\D t} = \widetilde{\mathcal{O}}( s \kappa_V \|\widetilde{A}_1^+\| ) = \widetilde{\mathcal{O}}( M^2 )= \mathcal{O}( M^2 \text{Polylog}(M^{d+2}/\eta) ),\]
where $s = \mathcal{O}(M) = \mathcal{O}(1/\D x)$, $\|\widetilde{A}_1^+\| = \mathcal{O}(M)$. Therefore, $U_{\Delta t}$ defined in \eqref{Wdelta} can be evolved within error $\mathcal{O}( d (3\eta+\D t^2) )$ \cite[Proposition 1.12]{Lin2022Notes} with
\begin{equation}\label{NgatesUdelta}
N_{\text{Gates}}(U_{\Delta t}) = \mathcal{O}(d Q_{\D t}) = \widetilde{\mathcal{O}}( d M^2).
\end{equation}
In the following, we choose $\eta \sim \D t^2$, and hence $U_{\Delta t}$ can be evolved within error $\mathcal{O}( d \D t^2 )$.
According to the error estimate \eqref{rhoErr} and noting Eq.~\eqref{CF0A}, one may have
\[e_{\rho} \le C( \omega +  d \D t/\omega + d\Delta x^{\ell}/\omega^{\ell+1} ),\]
which is also true for the computation of the observable. The above error bounds suggest the following mesh strategy:
\begin{equation}\label{meshSpectralLiouville}
\omega \sim d \D t/\omega \sim d\Delta x^{\ell}/\omega^{\ell+1} \sim \varepsilon,
\end{equation}
or equivalently,
\begin{equation}\label{meshSLiouvilleSpec}
M \sim d^{1/\ell}/\varepsilon^{1+2/\ell}, \qquad \D t = \varepsilon^2/d.
\end{equation}
The total number of gates required to iterate to the $n$-th step is
\begin{align*}
N_{\text{Gates}}
= n N_{\text{Gates}}(U_{\Delta t})
= \widetilde{\mathcal{O}}\Big(\frac{d^{2+2/\ell}}{\varepsilon^{4+4/\ell}}\Big)
= \widetilde{\mathcal{O}}\Big(\frac{d^{3+2/\ell}}{\varepsilon^{4+4/\ell}}\Big),
\end{align*}
where in the last equal sign we have included the additional factor $d$ arising from the matrix order (see Remark \ref{rem:ODEs}).

(2) For the computation of the observable, according to Remark \ref{rem:obsk}, one just needs to multiply the original gate complexity by the sampling factor $k = \mathcal{O}(n_L^4/\varepsilon^2)$. This completes the proof.
\end{proof}

\subsubsection{Quantum simulation for the spectral discretisation}

The approximate operator in \eqref{Wdelta} can also evolved by the quantum simulation.

\begin{theorem}\label{thm:simulationLiouvillerep}
Assume further that $\max_{1\le j \le d} \|\bb{F}_j\| = \mathcal{O}(1)$ and $T = \mathcal{O}(1)$.
\begin{enumerate}[(1)]
  \item There exists a quantum algorithm that produces a state $\varepsilon$-close to $\bb{w}(T)/\|\bb{w}(T)\|$ with the gate complexity given by
\[N_{Gates} = \widetilde{\mathcal{O}}\Big(\frac{d^{2+2/\ell}}{\varepsilon^{2+4/\ell}}\Big).\]
  \item The observable of the Liouville representation can be simulated with gate complexity given by
\[N_{\text{Gates}}( \langle O \rangle) = \widetilde{\mathcal{O}}\Big(\frac{ n_L^4 d^{2+2/\ell}}{\varepsilon^{4+4/\ell}}\Big),\]
where $n_L = \|(\bb{\rho}^\omega)^0\|/M^{d/2}$.
\end{enumerate}
\end{theorem}
\begin{proof}
We first quantify the number of gates used in the quantum simulation.  According to Theorem 1 in \cite{Berry-Childs-Kothari-2015}, $\e^{-\i \widetilde{A}_j^+ \Delta t/2}$ in \eqref{tpsi1} can be simulated within error $\eta$ with
\[\mathcal{O}\Big(  \tau ( m_d + \log^{2.5}(\tau/\eta) )\frac{\log (\tau/\eta) }{\log\log (\tau/\eta)} \Big)\]
2-qubits gates, where $\tau = s \|\widetilde{A}_j^+\|_{\max} \Delta t/2$, $s$ is the sparsity of $\widetilde{A}_j^+$ and $\|\widetilde{A}_j^+\|_{\max}$ denotes the largest entry of $\widetilde{A}_j^+$ in absolute value. This result is near-optimal by Theorem 2 therein. One can check that the sparsity of $\widetilde{A}_j^+$ is $s = \mathcal{O}(M)$. Therefore, $U_{\Delta t}$ defined in \eqref{Wdelta} can be simulated within error $\mathcal{O}( d (3\eta + \D t^2) )$ \cite[Proposition 1.12]{Lin2022Notes} with
\[N_{\text{Gates}}(U_{\Delta t}) = \mathcal{O}(d \tau m_d \cdot \text{polylog}),\]
where
\[\tau = M \widetilde{A}_{\max} \Delta t,  \qquad \widetilde{A}_{\max} = \max_j \|\widetilde{A}_j^\pm\|_{\max} = \mathcal{O}(M),\]
\[ \text{polylog} = \log^{2.5}(\tau/\eta) \frac{\log (\tau/\eta) }{\log\log (\tau/\eta)}.\]
In the following, we choose $\eta \sim \D t^2$, and hence obtain the same mesh strategy for the QLSA.

With the mesh strategy given in \eqref{meshSLiouvilleSpec}, one has the number of qubits per dimension is
\[m = \mathcal{O}(\log M) = \mathcal{O} \Big( \log \frac{d^{1/\ell}}{\varepsilon^{1+2/\ell}} \Big). \]
The total number of qubits is $m_d = d m$. With these settings, we obtain
\[\tau = \mathcal{O} \Big(\frac{d^{2/\ell}}{\varepsilon^{2+4/\ell}} \D t\Big), \qquad
\tau/\eta = \mathcal{O} \Big(\frac{d^{1 + 2/\ell}}{\varepsilon^{4+4/\ell}} \Big),\]
and the total number of gates required to iterate to the $n$-th step is
\begin{align*}
N_{\text{Gates}}
= n N_{\text{Gates}}(U_{\Delta t})
= \mathcal{O}\Big(\frac{d^{2+2/\ell}}{\varepsilon^{2+4/\ell}} \log \frac{d^{1/\ell}}{\varepsilon^{1+2/\ell}} \cdot \text{polylog}\Big).
\end{align*}

(2) For the computation of the observable, one just needs to add the multiplicative factor $\mathcal{O}(n_L^4/\varepsilon^2)$. This completes the proof.
\end{proof}

\begin{remark}
We emphasise that here only the total cost of the Hamiltonian simulation component at each time-step is included. The simulation protocol here is different from the traditional time-marching Hamiltonian simulation since non-unitary procedures are involved at each time step, leading to exponential increase of the cost as discussed in Remark~\ref{rem:multiplecopies}. It is worth pointing out that this obstacle can be overcome by our recently proposed technique~-~the Schr\"odingerisation approach \cite{JinLiuYu2022Schr,JinLiuYu2022Schrdetail}. Please refer to Section 4.5 of \cite{JinLiuYu2022Schrdetail} for details.
\end{remark}

\end{document}